\documentclass[11pt,a4paper]{article}
\usepackage{indentfirst}
\usepackage{setspace}
\usepackage[font=small,  margin=2cm]{caption}
\usepackage{amsthm,amsmath,amssymb}
\usepackage{natbib}
\usepackage{multirow}
\usepackage[pdftex]{graphicx}
\usepackage{subfigure}
\usepackage{makecell}
\usepackage{booktabs}
\usepackage{array}
\usepackage{fullpage}
\usepackage{url}
\usepackage{algorithm}
\usepackage{algpseudocode}
\usepackage{bm}
\usepackage{smile}
\usepackage{mathtools}
\usepackage{wrapfig}
\usepackage{lipsum}
\usepackage{mathrsfs}
\usepackage{dsfont}
\usepackage{rotating} 
\usepackage{color}
\usepackage{natbib}
\usepackage{multirow}
\usepackage{apalike}
\usepackage{appendix}

\def\P{{\mathbb P}}
\def\E{{\mathbb E}}

\def\supp{\mathop{\text{supp}\kern.2ex}}
\def\argmin{\mathop{\text{\rm arg\,min}}}
\def\argmax{\mathop{\text{\rm arg\,max}}}

\def\supp{\mathop{\text{supp}}}

\DeclarePairedDelimiter\floor{\lfloor}{\rfloor}

\usepackage[usenames,dvipsnames,svgnames,table]{xcolor}
\usepackage[colorlinks=true,
            linkcolor=blue,
            urlcolor=blue,
            citecolor=blue]{hyperref}

\numberwithin{equation}{section}
\numberwithin{theorem}{section}
\numberwithin{corollary}{section}
\numberwithin{asmp}{section}
\numberwithin{definition}{section}

\renewcommand{\baselinestretch}{1.5}
\begin{document}

\title{Simultaneous Change Point Detection and Identification  for High Dimensional Linear Models}

\author{
Bin Liu\footnotemark[1],
Xinsheng Zhang\footnotemark[1],
Yufeng Liu\footnotemark[2]
}
\renewcommand{\thefootnote}{\fnsymbol{footnote}}
 
\footnotetext[1]{Department of Statistics and Data Science, School of Management at Fudan University, Shanghai, China; E-mail:{\tt liubin0145@gmail.com; xszhang@fudan.edu.cn }}
\footnotetext[2]{Department of Statistics and Operations Research, Department of Genetics, and Department of Biostatistics, Carolina Center for Genome Sciences, Linberger Comprehensive Cancer Center, University of North Carolina at Chapel Hill, U.S.A; E-mail:{\tt yfliu@email.unc.edu }}

\maketitle\vspace{-0.4in}
\begin{abstract}
In this article, we consider change point inference for high dimensional linear models. For change point detection, given any subgroup of variables, 
we propose a new method for testing the homogeneity of corresponding regression coefficients across the observations. 
Under some regularity conditions, the proposed new testing procedure controls the type I error asymptotically and is powerful against sparse alternatives and enjoys certain optimality.
For change point identification,  an ``argmax" based change point estimator is  proposed which  is shown to be consistent  for the true change point location. 
Moreover, combining with the binary segmentation technique, we further extend our new method for detecting and identifying multiple change points.
Extensive numerical studies justify the validity of our new method and an  application to the Alzheimer’s disease data analysis further
demonstrate its competitive performance.
\end{abstract}


\section{Introduction}		
Driven by the great improvement of data collection and storage capacity, 
high dimensional linear regression models have attracted a lot of attentions because of its simplicity for interpreting the  effect of different variables in predicting the response. Specifically, we are interested in the following  model:
\begin{equation*}\label{equation: linear model}
	Y=\bX^\top \bbeta+\epsilon,
\end{equation*}
where $Y\in\RR$ is the response variable, $\bX=(X_1,\ldots,X_p)\in \RR^{p}$ is the covariate vector, $\bbeta=(\beta_1,\ldots,\beta_p)^\top$ is a $p$-dimensional unknown vector of coefficients, and $\epsilon\in\RR$ is the error term. 

For high dimensional linear regression, the $L_1$-penalized technique lasso (\cite{tibshirani1996regression}) is
a popular method for estimating $\bbeta$. In the past decades, lots of research attentions {both in machine learning and statistics}  have been focused   on studying theoretical properties of lasso  and other penalized methods. 
Most of the existing literature on high dimensional linear regression focuses on the case with a homogeneous linear model, where the regression coefficients are assumed invariant  across  the observations. With many modern complex datasets for analysis in practice, data heterogeneity is a common challenge in many real applications such as economy and genetics. In some applications, the regression coefficients may have a sudden change at some unknown time point, which is called a change point. Typical examples include racial segregation  and crime prediction  in sociology, and financial contagion in economy. For these problems, methods and theories designed for independently and identically ($i.i.d.$) distributed settings are no longer applicable. As a result, ignoring these structural breaks in {machine learning applications may lead to  misleading results and wrong decision making.} For the regression change point problem, a fundamental question is whether the underlying regression model remains homogenous across the observations. To address this issue, in this article, we investigate change point inference for high dimensional linear models. Specifically, let $(Y_i,\bX_i)_{i=1}^{n}$ be $n$  ordered independent realizations of $(Y,\bX)$. We aim to detect whether the regression coefficients have a change point during the observations. In particular, let $\bbeta^{(1)}$ and $\bbeta^{(2)}$ be two $p$-dimensional vectors of coefficients with $\bbeta^{(1)}=(\beta^{(1)}_1,\ldots,\beta^{(1)}_p)^\top$ and $\bbeta^{(2)}=(\beta^{(2)}_1,\ldots,\beta^{(2)}_p)^\top$. We consider the following linear regression model with a possible change point:
\begin{equation}\label{equation: linear model with change point}
	\begin{array}{ll}
		Y_i=\bX_i^\top \bbeta^{(1)}\mathbf{1}\{1\leq i\leq k_*\}+\bX_i^\top \bbeta^{(2)}\mathbf{1}\{  k_*+1\leq i\leq n\} +  \epsilon_i,
	\end{array}
\end{equation}
where $k_* $ is the possible but unknown change point location and $(\epsilon_i)_{i=1}^n$ are the error terms. In this paper, we assume $k_*=\lfloor nt_0 \rfloor$ for some $t_0\in(0,1)$.
For any given subgroup $\cG\subset\{1,\ldots,p\}$,  the first goal is to test 
\begin{equation}\label{hypothesis: H0}
	\begin{array}{ll}
		\Hb_{0,\cG}:\beta_{s}^{(1)}=\beta_{s}^{(2)}~~\text{for~all}~ s\in\cG~~\text{v.s.}~~\\
		\Hb_{1,\cG}: \text{There exist}~ s\in\cG ~\text{and}~k_*\in\{1,\dots,n-1\}~\text{s.t.}~ \beta_{s}^{(1)}\neq \beta_{s}^{(2)}.
	\end{array}
\end{equation}
In other words, under $\Hb_{0,\cG}$, the regression coefficients in each subgroup $\cG$ are homogeneous across the observations, and under $\Hb_{1,\cG}$ there is a  change point at an unknown time point $k_*$ such that the  regression coefficients have a sudden change after $k_*$.
Our second goal of the paper is to identify the change point location once we reject $\Hb_{0,\cG}$ in (\ref{hypothesis: H0}). In this paper, we assume that the number of coefficients can be much larger than the number of observations, i.e., $p\succeq n$, which is known as a high dimensional problem.

For the  low dimensional setting  with a fixed $p$ and $p<n$, change point inference for linear regression models has been well-studied. For example, \cite{quandt1960tests} considered testing (\ref{hypothesis: H0}) for a simple regression model with $p=2$. Based on that, several techniques were proposed in the literature. Among them are maximum likelihood ratio tests \citep{horvath1995detecting}, partial sums of regression residuals \citep{bai1998estimating},  and the union intersection test \citep{horvath1995limit}. Moreover, as a special case of linear regression models, \cite{chan2014group} considered change point detection for the autoregressive model. 
As compared to the broad literature in the low dimensional setting, methods and theory  for high dimensional change point inference of  (\ref{equation: linear model with change point}) have not been investigated much until recently. For instance,  \cite{lee2016lasso} considered  a high dimensional regression model with a possible change point due to a covariate threshold. Based on the $L_1/L_0$ regularization,  \cite{kaul2019efficient} proposed  a two-step algorithm for the detection and estimation of parameters in a high-dimensional change point regression model.
As extensions to multiple structural breaks  in high dimensional linear  models, \cite{leonardi2016computationally} proposed fast algorithms for multiple change point estimation  based on dynamic programming and binary search algorithms.  In addition,
\cite{zhang2015change-point} developed an approach for estimating multiple change points  based on sparse group {l}asso. \cite{Wangdaren2021} proposed a  projection-based algorithm for estimating multiple change points. 
{Recently, \cite{cho2022high,bai2022unified}  constructed estimates for the  multiple change points in high-dimensional regression models based on methods of moving window  and blocked fused lasso}. {\cite{kaul2021inference, xu2022change} respectively considered the problem of constructing confidence interval for the change point  in the context of high dimensional mean vector-based models and linear regression models.} {\cite{chen2023data} proposed a new method for determing the number of change points with false discovery rate controls. Other related papers include \cite{he2023multiple,wang2022denoising}.} 

It is worth noting that the majority of  above mentioned papers mainly focus on the estimation of regression coefficients as well as the change point locations by assuming a pre-existing change point in the model. To our best knowledge, the testing problem of  (\ref{hypothesis: H0}) has not been considered yet. How to make effective change point detection remains to be an urgent but challenging task. 
To fill this gap, in this article, we consider  change point inference in the context of  high dimensional linear models.  

The main contributions of this paper are as follows.
For any pre-specified subgroup $\cG\subset\{1,\ldots,p\}$, we propose a new method for testing the homogeneity of corresponding regression coefficients across the observations. For change point detection, the proposed test statistic $T_{\cG}$ is constructed based on a weighted $L_{\infty}$ aggregation, both temporally and spatially, 
of the process $\{Z_j(\lfloor nt \rfloor)\}_{j\in\cG,t\in[\tau_0,1-\tau_0]}$, where $Z_j(\lfloor nt \rfloor)=\breve{\beta}_j^{(0,t)}-\breve{\beta}_j^{(t,1)}$ with $\breve{\beta}_j^{(0,t)}$ and $\breve{\beta}_j^{(t,1)}$ denoting the de-biased lasso estimators for coordinate $j$ before and after time point $\lfloor nt \rfloor$, respectively. It is shown that $T_{\cG}$ is powerful against sparse alternatives with only a few entries in $\cG$ having a change point. To approximate its limiting null distribution, a multiplier bootstrap procedure is introduced. The proposed bootstrap can automatically account for the dependence structures of $\{Z_j(\lfloor nt \rfloor)\}_{j\in\cG,t\in[\tau_0,1-\tau_0]}$ and allow the group size $|\cG|$ {to} grow exponentially with the sample size $n$. Furthermore, to identify the change point location, for each time point $\lfloor nt \rfloor$, we first aggregate the coordinates with the $L_\infty$-norm, then a change point estimator $\hat{t}_{0,\cG}$ is obtained by taking ``argmax" with respect to $t$ of the above aggregated process with some proper weights. 
In addition to single change point detection, by combining with the binary segmentation technique \citep{Vostrikovadetecting}, we extend our new algorithm for detecting multiple change points which enjoys better performance than the existing methods.

In terms of theoretical investigation, with mild moment conditions on the covariates and errors in the regression model, we justify the validity of our proposed  method in terms of change point detection and identification. In particular, our bootstrap procedure consistently approximates the  limiting null distribution of $T_{\cG}$, which implies that the proposed new test preserves the pre-specified significance level asymptotically. Furthermore, under $\Hb_{1,\cG}$,  our new method is sensitive to sparse alternatives and can reject the null hypothesis with probability tending to one. {It is worth mentioning that  \cite{xia2018two-sample} considered  two sample tests for high dimensional linear regression models. They derived some conditions for consistently distinguishing two sample regression models, which are shown to be minimax optimal. Our requirement for detecting a change point under $\Hb_{1,\cG}$ has  the same order as the condition  derived in \cite{xia2018two-sample}.
	As for the change point estimation, we prove that  our proposed argmax-based change point estimator  is consistent for $t_{0}$ with an estimation error rate of
	$\big|\hat{t}_{0,\cG}-t_0\big|=O_p\big(\dfrac{\log(|\cG|n)}{n\|\bdelta\|_{\cG,\infty}^2}\big),$
	where $\bdelta:=\bbeta^{(1)}-\bbeta^{(2)}$ with $\|\bdelta\|_{\cG,\infty}=:\max_{j\in\cG}|\beta_j^{(1)}-\beta_j^{(2)}|$. {Hence, the above estimation result shows that our proposed change point estimator is consistent as long as $\|\bbeta^{(1)}-\bbeta^{(2)}\|_{\cG,\infty}\gg \sqrt{\log(|\cG|n)/n}$ and} 
	allows the overall sparsity of regression coefficients and the group's magnitude $|\cG|$ {to} grow simultaneously with the sample size $n$. We demonstrate that our new testing procedure is relatively simple to implement and extensive numerical studies provide strong support to our theory.  Moreover,  an R package called ``RegCpt'' is developed to implement our proposed new algorithms.
	The rest of this paper is organized as follows. In Section \ref{section: methodology}, we introduce our new methodology for Problem (\ref{hypothesis: H0}). In Section \ref{section: theoretical properties}, some theoretical results are derived in terms of change point detection and identification.
	In Section \ref{section: simulation studies},  extensive numerical studies are investigated.  
	The detailed proofs of the main theorems, additional numerical studies and  an interesting real data application 
	are given in the Appendix.

	For $\bv=(v_1,\ldots,v_p)^\top \in \mathbb{R}^p$,  let its $L_p$ norm be $\|\bv\|_p=(\sum_{j=1}^d|v_j|^p)^{1/p}$ for $1\leq p\leq \infty$. For $p=\infty$, define $\|\bv\|_\infty=\max_{1\leq j\leq d}|v_j|$. For a subset $\cG\subset\{1,\ldots,p\}$, denote $\|\bv\|_{\cG,\infty}$ by $\max_{j\in \cG}|v_j|$.
	For any set $\cS$, denote its cardinality by $|\cS|$. For two real numbered sequences $a_n$ and $b_n$,  set $a_n=O(b_n)$ if there exits a constant $C$ such that $|a_n|\leq C|b_n|$ for a sufficiently large $n$; $a_n=o(b_n)$ if $a_n/b_n\rightarrow0$ as $n\rightarrow\infty$; $a_n\asymp b_n$ if there exists constants $c$ and $C$ such that $c|b_n|\leq|a_n|\leq C|b_n|$ for a sufficiently large $n$. 
	For a sequence of random variables $\{\xi_1,\xi_2,\ldots\}$,  denote $\xi_n=o_p(1)$ if $\xi_n\xrightarrow{\P} 0$. 
	Define $\floor{x}$ as the largest integer less than or equal to $x$ for $x\geq 0$.
	\section{Methodology}\label{section: methodology}

	\subsection{New test statistic}\label{section: test statistics}
	We present our methodology for testing the existence of a change point in Model (\ref{equation: linear model with change point}). To this end, we first introduce some basic model settings. Recall  the   regression model
	\begin{equation}\label{equation: linear model2}
		Y_i=\bX_i^\top \bbeta^{(1)}\mathbf{1}\{1\leq i\leq \lfloor nt_0 \rfloor\}+\bX_i^\top \bbeta^{(2)}\mathbf{1}\{ \lfloor nt_0 \rfloor+1\leq i\leq n\} +  \epsilon_i.
	\end{equation}
	Denote $\bY=(Y_1,\ldots,Y_n)^\top$ as a $n\times 1$ response vector, $\Xb$ is a $n\times p$ design matrix with $\bX_i=(X_{i,1},\ldots,X_{i,p})^\top$ being its $i$-th row for $1\leq i\leq n$, and $\bepsilon=(\epsilon_1,\cdots,\epsilon_n)^\top$ is the error vector.   
	For the unknown $p\times 1$ regression vectors $\bbeta^{(1)}=(\beta_1^{(1)},\ldots,\beta_p^{(1)})^\top$ and $\bbeta^{(2)}=(\beta_1^{(2)},\ldots,\beta_p^{(2)})^\top$, define $\cS^{(1)}=\{1\leq j\leq p: \beta_j^{(1)}\neq 0\}$ and
	$\cS^{(2)}=\{1\leq j\leq p: \beta_j^{(2)}\neq 0\}$
	as the active sets of variables. Denote $s^{(1)}=|\cS^{(1)}| $  and $s^{(2)}=|\cS^{(2)}| $ as the cardinalities of $\cS^{(1)}$ and $\cS^{(2)}$, respectively.  Define $\bSigma=(\Sigma_{i,j})=\text{Cov}(\bX_1)$ as the covariance matrix of $\bX_1$ and $\bTheta=(\theta_{i,j})$ as the inverse of $\bSigma$. For $\bTheta$, let $s_j=|\{1\leq k\leq p: \theta_{j,k}\neq 0, k\neq j\}|$.  In addition to the above notations, we assume that the change point does not happen at the beginning or end of data observations. In other words, there exists some  $\tau_0\in(0,0.5)$ such that $t_0\in[\tau_0,1-\tau_0]$ holds. Note that the search boundary scales with $n$ by allowing $\tau_0\rightarrow 0$.

	To propose our method, we first introduce the de-sparsified (de-biased) lasso estimator, which was  proposed in \cite{van2014asymptotically} and \cite{zhang2014confidence}. Specifically, for Model (\ref{equation: linear model2}), let $\hat{\bbeta}_n$ be a lasso estimator  from
		$	\hat{\bbeta}^n=\argmin_{\bbeta\in \RR^p}\|\bY-\Xb\bbeta\|_2^2/n+2\lambda_n\|\bbeta\|_1,$
	where $\lambda_n$ is the non-negative regularization parameter. 
	Then for a homogeneous model with no change points, the de-biased lasso estimator is defined:
	\begin{equation}\label{equation: de-biased lasso}
		\breve{\bbeta}^n=\hat{\bbeta}^n+\hat{\bTheta}\Xb^\top\big(\bY-\Xb\hat{\bbeta}^n\big)/n,
	\end{equation}
	where $\hat{\bTheta}$ is some appropriate estimator for $\bTheta$. Essentially, the de-biased lasso estimator $\breve{\bbeta}_n$ is a lasso solution by plugging in a Karush-Kuhn-Tucker (KKT) condition.  It has been widely used for constructing confidence intervals and statistical tests for high dimensional parameters, and proven to be asymptotically  optimal in terms of semiparametric efficiency.
	\begin{remark}
		In this paper, we adopt the node-wise estimation for obtaining $\hat{\bTheta}$, as proposed in \cite{meinshausen2006high}. The main idea is to perform regression on each variable using the remaining ones. In particular, denote $\Xb^j$ as the $j$-th column of $\Xb$ and $\Xb^{-j}$ as the remaining columns. For each $j=1\ldots,p$, define
		\begin{equation}\label{equation: node-wise regression}
			\hat{\bgamma}_j=\argmin_{\bgamma\in \RR^{p-1}}\Big(\|\Xb^j-\Xb^{-j}\bgamma\|_2^2/n+2\lambda_{(j)}\|\bgamma\|_1\Big),
		\end{equation}
		with $\hat{\bgamma_j}=\{\hat{\gamma}_{j,k}:k=1\ldots,p,k\neq j\}$. Denote by $\hat{\Cb}=(\hat{c}_{i,j})_{i,j}^p$ with $\hat{c}_{i,i}=1$ and $\hat{c}_{i,j}=-\gamma_{i,j}$ for $i\neq j$.
		Let $\hat{\tau}_j^2=\|\Xb^j-\Xb^{-j}\hat{\bgamma}_j\|_2^2/n+\lambda_{(j)}\|\hat{\bgamma}_j\|_1$ and $\hat{\Tb}^2=\diag\{\hat{\tau}_1^2,\ldots,\hat{\tau}_p^2\}$. The node-wise lasso estimator for $\bTheta$ is defined as
		\begin{equation}\label{equation: node-wise lasso}
			\small
			\hat{\bTheta}=\hat{\Tb}^{-2}\hat{\Cb}.
		\end{equation}
		It is shown that $\hat{\bTheta}$ enjoys good properties in estimation accuracy.  More importantly, it is possible to use  parallel computation for calculating $\hat{\bTheta}$, which is more appropriate for modern statistical applications with large scale datasets.
	\end{remark}
	Since there is a possible but unknown change point in Model (\ref{equation: linear model2}), we can not use (\ref{equation: de-biased lasso}) directly to make statistical inferences on $\bbeta^{(1)}$ and $\bbeta^{(2)}$. The main challenge comes from the unknown change point $t_0$. 
	To overcome this difficulty,  instead of only calculating a single de-biased lasso estimator $\breve{\bbeta}^n$, we need to construct the de-biased lasso-based process. {To that end, we need some notations.}  For any $0\leq s< t\leq 1$, define
	\begin{equation*}
		\begin{array}{ll}
			\bY_{(s,t)}=(Y_{\floor{ns}+1},\ldots,Y_{\floor{nt}})^\top,~~\bepsilon_{(s,t)}=(\epsilon_{\floor{ns}+1},\ldots,\epsilon_{\floor{nt}})^\top,\\
			\Xb_{(s,t)}=(\bX_{\floor{ns}+1},\ldots,\bX_{\floor{nt}})^\top, 	\hat{\bSigma}_{(s,t)}=\dfrac{1}{\lfloor nt \rfloor -\lfloor ns \rfloor+1 }\sum\limits_{i=\floor{ns}+1}^{\floor{nt}}\bX_i\bX_i^\top.
		\end{array}
	\end{equation*}
	{To motivate our testing statistic, for each fixed $t\in[\tau_0,1-\tau_0]$, we define} 
	\begin{equation}\label{equation: true parameters before and after cpt}
		\small
		\begin{array}{ll}
			{\bbeta}^{(0,t)}=\argmin\limits_{\bbeta\in \RR^p}\E\big\|\bY_{(0,t)}-\Xb_{(0,t)}\bbeta\big\|_2^2,
			{\bbeta}^{(t,1)}=\argmin\limits_{\bbeta\in \RR^p}\E\big\|\bY_{(t,1)}-\Xb_{(t,1)}\bbeta\big\|_2^2.
		\end{array}
	\end{equation}
	{By definition, ${\bbeta}^{(0,t)}$ and ${\bbeta}^{(t,1)}$ are the best regression coefficients for predicting $\bY_{(0,t)}$ and $\bY_{(t,1)}$ under the squared error loss, respectively. More importantly, suppose there is a change point $t_0$ in the linear model (\ref{equation: linear model2}). According to the search location $t$ and the true change point location $t_0$, the underlying true parameters can have the following explicit form: }
	\begin{equation*}
		\bbeta^{(0,t)}=\bbeta^{(1)}\mathbf{1}\{t\in[\tau_0,t_0]\}+\big(\dfrac{\floor{nt_0}}{\floor{nt}}\bbeta^{(1)}+\dfrac{\floor{nt}-\floor{nt_0}}{\floor{nt}}\bbeta^{(2)}\big)\mathbf{1}\{t\in[t_0,1-\tau_0]\},
	\end{equation*}
	and 
	\begin{equation*}
		\bbeta^{(t,1)}=\big(\dfrac{\floor{nt_0}-\floor{nt}}{n-\floor{nt}}\bbeta^{(1)}+\dfrac{n-\floor{nt_0}}{n-\floor{nt}}\bbeta^{(2)}\big)\mathbf{1}\{t\in[\tau_0,t_0]\}+\bbeta^{(2)}\mathbf{1}\{t\in[t_0,1-\tau_0]\}.
	\end{equation*}
	{From the population level, we can define the theoretical signal jump process:}
	\begin{equation}\label{equation: signal function}
		\begin{array}{ll}
			\bdelta_n(t)&:=\sqrt{n}\dfrac{\lfloor nt \rfloor}{n}\dfrac{\lfloor nt \rfloor^* }{n}(\bbeta^{(0,t)}-\bbeta^{(t,1)})
			\\&=\sqrt{n}\dfrac{\lfloor nt \rfloor}{n}\dfrac{\lfloor nt_0\rfloor^* }{n}\big(\bbeta^{(1)}-\bbeta^{(2)}\big)\mathbf{1}\{t\in [\tau_0,t_0]\}\\
			&+\sqrt{n}\dfrac{\lfloor nt_0\rfloor }{n }\dfrac{\lfloor nt \rfloor ^*}{n}\big(\bbeta^{(1)}-\bbeta^{(2)}\big)\mathbf{1}\{t\in [t_0,1-\tau_0]\},\\
			\end{array}
		\end{equation}
			where $\floor{nt}^*:=n-\floor{nt}$.
			
			The signal function in (\ref{equation: signal function}) has some interesting properties. First, under $\Hb_{0,\cG}$ of no change points, it reduces to a vector of zeros at each time point $\lfloor nt \rfloor$.  Second, under $\Hb_{1,\cG}$,  {$\bdelta_n(t)$ is at most $(s^{(1)}+s^{(2)})$-sparse} since we require sparse regression coefficients in the model. Third, we can see that $\|\bdelta_n(t)\|_{\cG,\infty}$ with $t\in[\tau_0,1-\tau_0]$ obtains its maximum value at the true change point location $t_0$. Hence, to make change point inference for high dimensional linear models, the key point is how to propose a test statistic that can estimate $\bdelta_n(t)$ well under $\Hb_{1,\cG}$, and has some theoretically tractable limiting null distributions under $\Hb_{0,\cG}$. {A natural idea is to use the lasso estimators directly.} Specifically, for each time point, we obtain the lasso estimators $\hat{\bbeta}^{(0,t)}=(\hat{\beta}^{(0,t)}_1,\ldots,\hat{\beta}^{(0,t)}_p)^\top$  and $\hat{\bbeta}^{(t,1)}=(\hat{\beta}^{(t,1)}_1,\ldots,\hat{\beta}^{(t,1)}_p)^\top$:
			\begin{equation}\label{equation: lasso estimation before and after cpt}
				\begin{array}{l}
					\hat{\bbeta}^{(0,t)}=\argmin\limits_{\bbeta\in \RR^p}\dfrac{1}{2\floor{nt}}\big\|\bY_{(0,t)}-\Xb_{(0,t)}\bbeta\big\|_2^2+\lambda_1(t)\|\bbeta\|_1,\\
					\hat{\bbeta}^{(t,1)}=\argmin\limits_{\bbeta\in \RR^p}\dfrac{1}{2\floor{nt}^*}\big\|\bY_{(t,1)}-\Xb_{(t,1)}\bbeta\big\|_2^2+\lambda_2(t)\|\bbeta\|_1,
				\end{array}
			\end{equation}
			where $\lambda_1(t)$ and $\lambda_2(t)$ are some regularity parameters to account for the data heterogeneity. 
			{It is well known that due to the $\ell_1$ regularized penalization in (\ref{equation: lasso estimation before and after cpt}), the lasso estimators are typically biased and do not have a tractable limiting null distribution. As a result, some``de-biased'' process is needed. The main idea is to plug into the KKT conditions under both $\Hb_{0,\cG}$ and $\Hb_{1,\cG}$ for the change point model. To give an insight into the de-biased process for change point detection, in what follows, we assume $\Hb_{1,\cG}$ holds.} 
			
			{Firstly, we consider the case that the search location satisfies $t\in[\tau_0,t_0]$. Let $\hat{\bkappa}_1(t)\in \RR^p$ and $\hat{\bkappa}_2(t)\in \RR^p$ be the subdifferentials of $\|\bbeta\|_1$ for the first and second optimization problems in (\ref{equation: lasso estimation before and after cpt}), respectively. Then, by the KKT condition, we have:}
			\begin{equation}\label{equation: KKT condition}
				\begin{array}{ll}
					-\Xb^\top_{(0,t)}(\bY_{(0,t)}-\Xb_{(0,t)}\hat{\bbeta}^{(0,t)} )/\floor{nt}+\lambda_1(t)\hat{\bkappa}_1(t)=\mathbf{0},\\
					-\Xb^\top_{(t,1)}(\bY_{(t,1)}-\Xb_{(t,1)}\hat{\bbeta}^{(t,1)} )/\floor{nt}^*+\lambda_2(t)\hat{\bkappa}_2(t)=\mathbf{0}.
				\end{array}
			\end{equation}
			{Note that for $t\in[\tau_0,t_0]$, the samples $\{\bY_{(0,t)},\Xb_{(0,t)}\}$ are homogeneous with regression coefficients being $\bbeta^{(0,t)}=\bbeta^{(1)}$. Hence, similar to the analysis in \cite{van2014asymptotically}, for the first term in (\ref{equation: KKT condition}), for $t\in[\tau_0,t_0]$, we have the following decomposition:} 
			\begin{equation}\label{equation: KKT for beta1 before the cpt}
				\begin{array}{ll}
					\hat{\bbeta}^{(0,t)}+\hat{\bTheta}\lambda_1(t)\hat{\bkappa}_1(t)-\bbeta^{(1)}=\Xb^\top_{(0,t)}\bepsilon_{(0,t)}/\floor{nt}\overbrace{-( \hat{\bTheta}\hat{\bSigma}_{(0,t)}  -\Ib)(\hat{\bbeta}^{(0,t)}-\bbeta^{(1)})}^{\bDelta_{(0,t)}^I}.
				\end{array}
			\end{equation}
			{For the second term in (\ref{equation: KKT condition}), we note that the samples $\{\bY_{(t,1)},\Xb_{(t,1)}\}$ with $t\in[\tau_0,t_0]$ are heterogeneous due to the change point at $t_0$.} Observe that 
			\begin{equation*}
				\Xb^\top_{(t,1)}=(\Xb^\top_{(t,t_0)},\Xb^\top_{(t_0,1)}), ~\text{and}~\Yb_{(t,1)}=((\Xb_{(t,t_0)}\bbeta^{(1)} )^\top+\bepsilon_{(t,t_0)}^\top,(\Xb_{(t_0,1)}\bbeta^{(2)} )^\top+\bepsilon_{(t_0,1)}^\top)^\top.
			\end{equation*}
			{Then, the KKT condition for the second equation in (\ref{equation: KKT condition}) becomes:}
			\begin{equation}\label{equation: KKT1}
				\begin{array}{ll}
					\lambda_2(t)\hat{\bkappa}_2(t)\\
					=\Xb^\top_{(t,t_0)}\Xb_{(t,t_0)}(\bbeta^{(1)}-\hat{\bbeta}^{(t,1)})/\floor{nt}^*+\Xb^\top_{(t_0,1)}\Xb_{(t_0,1)}(\bbeta^{(2)}-\hat{\bbeta}^{(t,1)})/\floor{nt}^*+\Xb^\top_{(t,1)}\bepsilon_{(t,1)}/\floor{nt}^*\\
					=\hat{\bSigma}_{(t,1)}(\bbeta^{(2)}-\hat{\bbeta}^{(t,1)})+\dfrac{\floor{nt_0}-\floor{nt}}{\floor{nt}^*}\hat{\bSigma}_{(t,t_0)}(\bbeta^{(1)}-\bbeta^{(2)})+\Xb^\top_{(t,1)}\bepsilon_{(t,1)}/\floor{nt}^*.
				\end{array}
			\end{equation}
			{Multiplying $\hat{\bTheta}$ on both sides of (\ref{equation: KKT1}), for the case of $t\in[\tau_0,t_0]$, we have:}
			\begin{equation}\label{equation: KKT2}
				\small
				\begin{array}{ll}
					\hat{\bbeta}^{(t,1)}+\hat{\bTheta}\lambda_2(t)\hat{\bkappa}_2(t)-\underbrace{\big(\dfrac{\floor{nt_0}-\floor{nt}}{\floor{nt}^*}\bbeta^{(1)}+\dfrac{n-\floor{nt_0}}{\floor{nt}^*}\bbeta^{(2)}\big)}_{\bbeta^{(t,1)}}\\
					=\underbrace{-(\hat{\bTheta}\hat{\bSigma}_{(t,1)}-\Ib)(\hat{\bbeta}^{(t,1)}-\bbeta^{(2)})-\dfrac{\floor{nt_0}-\floor{nt}}{\floor{nt}^*}(\hat{\bTheta}\hat{\bSigma}_{(t,t_0)}-\Ib)(\bbeta^{(2)}-\bbeta^{(1)})}_{\bDelta_{(t,1)}^{I}}+\Xb^\top_{(t,1)}\bepsilon_{(t,1)}/\floor{nt}^*.\\
				\end{array}
			\end{equation}
			{Secondly, for the case of $t\in[t_0,1-\tau_0]$, using a very similar analysis, we can prove that:}
			\begin{equation}\label{equation: KKT second case}
				\begin{array}{ll}
					\hat{\bbeta}^{(0,t)}+\hat{\bTheta}\lambda_1(t)\hat{\bkappa}_1(t)-\underbrace{\big(\dfrac{\floor{nt_0}}{\floor{nt}}\bbeta^{(1)}+\dfrac{\floor{nt}-\floor{nt_0}}{\floor{nt}}\bbeta^{(2)}\big)}_{\bbeta^{(0,t)}}=\Xb^\top_{(0,t)}\bepsilon_{(0,t)}/\floor{nt}+\bDelta_{(0,t)}^{II},\\	
					\hat{\bbeta}^{(t,1)}+\hat{\bTheta}\lambda_2(t)\hat{\bkappa}_2(t)-\underbrace{\bbeta^{(2)}}_{\bbeta^{(t,1)}}=\Xb^\top_{(t,1)}\bepsilon_{(t,1)}/\floor{nt}^*+\bDelta_{(t,1)}^{II},\\	
				\end{array}
			\end{equation}
			{where the two terms $\bDelta_{(0,t)}^{II}$ are $\bDelta_{(t,1)}^{II}$ are defined as} 
			\begin{equation*}
				\begin{array}{ll}
					\bDelta_{(0,t)}^{II}&:=-\dfrac{\lfloor nt \rfloor-\lfloor n{t}_{0}\rfloor}{\lfloor nt \rfloor }\big(\hat{\bTheta}\hat{\bSigma}_{(t_0,t)}-\Ib\big)\big(\bbeta^{(1)}-\bbeta^{(2)} \big)-\big(\hat{\bTheta}\hat{\bSigma}_{(0,t)}-\Ib\big)\big(\hat{\bbeta}^{(0,t)}-\bbeta^{(1)} \big),\\
					\bDelta_{(t,1)}^{II}&:=-\big(\hat{\bTheta}\hat{\bSigma}_{(t,1)}-\Ib\big)\big(\hat{\bbeta}^{(t,1)}-\bbeta^{(2)} \big).
				\end{array}
			\end{equation*}
			Combining the results in (\ref{equation: KKT condition})-(\ref{equation: KKT second case}), for each $t\in[\tau_0,1-\tau_0]$, we then construct the de-biased lasso estimators 
			$\breve{\bbeta}^{(0,t)}=(\breve{\beta}^{(0,t)}_1,\ldots,\breve{\beta}^{(0,t)}_p)^\top$ and $\breve{\bbeta}^{(t,1)}=(\breve{\beta}^{(t,1)}_1,\ldots,\breve{\beta}^{(t,1)}_p)^\top$ as follows:
			\begin{equation}\label{equation: de-biased before and after cpt}
				\begin{array}{ll}
					\breve{\bbeta}^{(0,t)}=\hat{\bbeta}^{(0,t)}+
					{\hat{\bTheta}\Xb_{(0,t)}^\top}\big(\bY_{(0,t)}-\Xb_{(0,t)}\hat{\bbeta}^{(0,t)}\big)/{\floor{nt}},\\
					\breve{\bbeta}^{(t,1)}=\hat{\bbeta}^{(t,1)}+
					\hat{\bTheta}\Xb_{(t,1)}^\top\big(\bY_{(t,1)}-\Xb_{(t,1)}\hat{\bbeta}^{(t,1)}\big)/{\floor{nt}^*}.
				\end{array}
			\end{equation}
			The construction of our new test statistic comes from our important new derivation (\ref{equation: de-biased before and after cpt}). In particular,  under some regularity conditions,  the difference between $\breve{\bbeta}^{(0,t)}$ and $\breve{\bbeta}^{(t,1)}$ has the following decomposition:
			\begin{equation}\label{equation: difference of de-biased lasso based process}
				\sqrt{n}\dfrac{\floor{nt}}{n}\dfrac{\floor{nt}^* }{n}\big(\breve{\bbeta}^{(0,t)}-\breve{\bbeta}^{(t,1)}\big)=\underbrace{\bdelta_n(t)}_{\bf{Signal~function}}+\underbrace{\dfrac{1}{\sqrt{n}}\sum_{i=1}^{\floor{nt}}\hat{\bTheta}\bX_i\epsilon_i}_{\bf{Random ~noise}}+\underbrace{\sqrt{n}\dfrac{\floor{nt}}{n}\dfrac{\floor{nt}^* }{n}(\bR^{(0,t)}-\bR^{(t,1)})}_{\bf{Random ~bias}}, 
			\end{equation}
			where $\bdelta_n(t)$ is defined in (\ref{equation: signal function}), 
			and $\bR^{(0,t)}$ and $\bR^{(t,1)}$ are the residuals:
			\begin{equation*}
				\begin{array}{ll}
					\bR_{(0,t)}=\bDelta_{(0,t)}^I\mathbf{1}\{t\in[\tau_0,t_0]\}+\bDelta_{(0,t)}^{II}\mathbf{1}\{t\in[t_0,1-\tau_0]\},\\
					\bR_{(t,1)}=\bDelta_{(t,1)}^I\mathbf{1}\{t\in[\tau_0,t_0]\}+\bDelta_{(t,1)}^{II}\mathbf{1}\{t\in[t_0,1-\tau_0]\}.
				\end{array}
			\end{equation*}
			
			{The above de-biased lasso-based process enjoys several advantages for making change point inference. 
				Firstly, under $\Hb_{0,\cG}$ of no change points, it is the combination of a partial sum-based process plus a random bias term. The latter one  can be shown to be negligible. Moreover, under $\Hb_{1,\cG}$, we can see that the de-biased lasso-based process is an asymptotically unbiased estimator for the signal function defined in (\ref{equation: signal function}), allowing us to make change point detection and identification.} The derivation of (\ref{equation: difference of de-biased lasso based process}) is different from the original de-biased lasso estimator in (\ref{equation: de-biased lasso})
			and requires a fundamental modification of \cite{bickel2009simultaneous})	to account for data heterogeneity. More details can be found in the Appendix. 
			
			Motived by the above observation, for any given subgroup $\cG\subset\{1,\ldots,p\}$, a natural test statistic for the hypothesis (\ref{hypothesis: H0}) is defined as
			\begin{equation*}\label{equation: natural test statistics}
				\tilde{T}_\cG=\max_{t\in[\tau_0,1-\tau_0]}\max_{j\in \cG}\sqrt{n}\dfrac{\floor{nt}}{n}\Big(1-\dfrac{ \floor{nt}}{n}\Big) \Big| \breve{\beta}^{(0,t)}_j-\breve{\beta}^{(t,1)}_j\Big|.
			\end{equation*} 
			For any given subgroup $\cG$, the proposed new  statistic $\tilde{T}_\cG$ searches all possible locations of time points. It is demonstrated that $\tilde{T}_{\cG}$ is powerful against sparse alternatives with only a few entries in $\cG$ having a change point, and a large value of $\tilde{T}_{\cG}$ leads to a rejection of $\Hb_{0,\cG}$.

			\subsection{Weighted variance estimation}\label{section: variance estimation}
			In Section \ref{section: test statistics}, we  introduced $\tilde{T}_{\cG}$ for  the hypothesis (\ref{hypothesis: H0}). Considering the variability of the design matrix $\Xb$ and the error term $\bepsilon$, the test statistic $\tilde{T}_G$ is heterogeneous. Hence, we need to take  its variance into account and standardize it. In this paper, we adopt a weighted variance estimator. Specifically, let $\hat{\bOmega}=(\hat{\omega}_{i,j})_{i,j}^p=\hat{\bTheta}\hat{\bSigma}_n\hat{\bTheta}^\top$ with $\hat{\bSigma}_n:=\Xb^\top\Xb/n$. For each $t\in[\tau_0,1-\tau_0]$, denote
			\begin{equation}\label{equation: pooled variance estimation}
				\hat{\sigma}_{\epsilon}^2(t)=\dfrac{1}{n}\big( \big\|\bY_{(0,t)}-\Xb_{(0,t)}\hat{\bbeta}^{(0,t)}\big\|_2^2+\big\|\bY_{(t,1)}-\Xb_{(t,1)}\hat{\bbeta}^{(t,1)}\big\|_2^2\big).
			\end{equation}
			Under $\Hb_{0,\cG}$ of no change points in the model, we can prove that 
			\begin{equation*}\label{equation: variance estimator under H0}
				\max\limits_{\tau_0 \leq t\leq 1-\tau_0}\max_{1\leq j\leq p}|\hat{\sigma}_{\epsilon}^2(t)\hat{\omega}_{j,j }-\sigma_{\epsilon}^2\omega_{j,j}|=o_p(1).
			\end{equation*}
			Under $\Hb_{1,\cG}$, however, $\hat{\sigma}_{\epsilon}^2(t)$ is not a consistent estimator for $\sigma_{\epsilon}^2$ because of the unknown change point $t_0$. Furthermore, as discussed in \cite{shao2010testing}, an inappropriate variance estimator may lead to non-monotonic power performance.  In order to form a powerful test statistic, it is necessary to construct consistent variance estimation for $\Hb_{0,\cG}$ and $\Hb_{1,\cG}$. To address this issue, we need to deal with the unknown change point first. In particular, for a given subgroup $\cG$, define 
			\begin{equation*}\label{equation:H-G}
				H_{\cG}(t)=\max_{j\in \cG}\dfrac{\floor{nt}}{n}\Big(1-\dfrac{\floor{nt} }{n}\Big) \Big| \breve{\beta}^{(0,t)}_j-\breve{\beta}^{(t,1)}_j\Big|, ~\text{with}~t\in[\tau_0,1-\tau_0].
			\end{equation*}
			By maximizing $H_{\cG}(t)$, we obtain the  argmax-based change point estimator:
			\begin{equation}\label{equation: t-hat}
				\small
				\hat{t}_{0,\cG}=\argmax_{t\in[\tau_0,1-\tau_0]}H_{\cG}(t).
			\end{equation}
			Based on (\ref{equation: t-hat}), let $\hat{t}_0=\hat{t}_{0,\cG}$ with $\cG=\{1,\ldots,p\}$.
			We put $\hat{t}_{0}$ into $\hat{\sigma}_{\epsilon}^2(t)$ and get a weighted variance estimator for $\sigma^2_{\epsilon}$ as
			\begin{equation}\label{equation: variance estimator under H1}
				\hat{\sigma}_{\epsilon}^2=\dfrac{1}{n}\big( \big\|\bY_{(0,\hat{t}_{0})}-\Xb_{(0,\hat{t}_{0})}\hat{\bbeta}^{{(0,\hat{t}_{0})}}\big\|_2^2+\big\|\bY_{(\hat{t}_{0},1)}-\Xb_{(\hat{t}_{0},1)}\hat{\bbeta}^{{(\hat{t}_{0},1)}}\big\|_2^2\big).
			\end{equation}
			As shown in our theoretical analysis, the new variance estimation in (\ref{equation: variance estimator under H1}) is consistent under both $\Hb_{0,\cG}$ and $\Hb_{1,\cG}$. 
			The proof is nontrivial since we need to justify the consistency of $\hat{t}_{0,\cG}$ for $t_0$, which is known to be
			an important but difficult task for high dimensional linear models (\cite{lee2016lasso}).
			
			Using the new variance estimator in (\ref{equation: variance estimator under H1}), for any given subgroup $\cG\subset\{1,\ldots,p\}$, our new test statistic for the hypothesis (\ref{hypothesis: H0}) is finally defined as follows:
			\begin{equation}\label{equation: test statistics}
				T_\cG=\max_{t\in[\tau_0,1-\tau_0]}\max_{j\in \cG}\sqrt{n}\dfrac{\floor{nt}}{n}\Big(1-\dfrac{\floor{nt} }{n}\Big) \Big| \dfrac{\breve{\beta}^{(0,t)}_j-\breve{\beta}^{(t,1)}_j}{\sqrt{\hat{\sigma}_{\epsilon}^2\hat{\omega}_{j,j }}}\Big|.
			\end{equation} 
			
			\subsection{Multiplier bootstrap  for approximating the  null  distribution} \label{section: bootstrap}
			In Section \ref{section: variance estimation}, we have proposed the new test statistic $T_{\cG}$  for the hypothesis (\ref{hypothesis: H0}). It is challenging to directly obtain its limiting null distribution  in high dimensions.  Bootstrap has been widely used for making statistical inference on high dimensional linear models since the seminal work of \cite{chernozhukov2013gaussian}. 
			For high dimensional linear models with change points, however, existing bootstrap techniques are not applicable and it is desirable to design a new method. To overcome this problem, we investigate two types of multiplier bootstrap.
			
			\subsubsection{Bootstrap-I} 
			Recall the decomposition in (\ref{equation: difference of de-biased lasso based process}). Under $\Hb_{0,\cG}$, we have
			\begin{equation*}\label{equation: difference of de-biased lasso based process under H0}
				\small
				\sqrt{n}\dfrac{\floor{nt}}{n}\dfrac{\floor{nt}^* }{n}\big(\breve{\bbeta}^{(0,t)}-\breve{\bbeta}^{(t,1)}\big)=\dfrac{1}{\sqrt{n}}\sum_{i=1}^{\floor{nt}}\hat{\bTheta}\bX_i\epsilon_i+(\bR^{(0,t)}-\bR^{(t,1)}).
			\end{equation*}
			It is shown that under $\Hb_{0,\cG}$, the residual-based process $\{\bR^{(0,t)}-\bR^{(t,1)},t\in[\tau_0,1-\tau_0]\}$ is asymptotically negligible and the partial sum-based process $\{n^{-1/2}\sum_{i=1}^{\floor{nt}}\hat{\bTheta}\bX_i\epsilon_i,t\in[\tau_0,1-\tau_0]\}$ determines the limiting null distribution of $T_{\cG}$, which is known as the leading term. This motivates us to first  consider the following bootstrap method:\\
			\textbf{Step~1:}  For the $b$-th bootstrap, generate $i.i.d.$ random variables  $\epsilon_1^b,\ldots,\epsilon_n^b$  with  $\epsilon_i^b\sim N(0,1)$.\\
			\textbf{Step~2:}	Calculate the testing statistic for the $b$-th bootstrap by
			\begin{equation*}
				W^b_\cG=\max_{t\in[\tau_0,1-\tau_0]}\max_{ j\in \cG}\sqrt{n}\dfrac{\floor{nt}}{n}\dfrac{\floor{nt}^* }{n}{\hat{\omega}_{j,j }}^{-1/2}\Big|\dfrac{1}{\floor{nt}}\sum\limits_{i=1}^{\floor{nt}}\hat{\bTheta}_j^\top\Xb_i\epsilon_i^b-\dfrac{1}{\floor{nt}^*}\sum\limits_{i=\floor{nt}+1}^{n}\hat{\bTheta}_j^\top\Xb_i\epsilon_i^b\Big|,
			\end{equation*}
			where $\hat{\bTheta}_j^\top$ is the $j$-th row of $\hat{\bTheta}$.\\
			\textbf{Step~3:} Repeat the above process for $B$ times. \\
			\textbf{Step~4:} Based on the bootstrap samples $\{W^1_\cG,\ldots,W^B_{\cG}\}$, calculate the bootstrap sample-based critical value
			\begin{equation*}\label{equation: estimated critical values for the first bootstrap}
				\hat{w}_{\cG,\alpha}=\inf\Big\{t:{(B+1)}^{-1}\sum_{b=1}^B\mathbf{1}\{W^b_{\cG}\leq t|\Xb,\bY \}\geq 1-\alpha   \Big\}.
			\end{equation*}
			\textbf{Step~5:}  Reject $\Hb_{0,\cG}$ if and only if $T_{\cG}\geq \hat{w}_{\cG,1-\alpha}$.
			
			Note that the above bootstrap method essentially bootstraps the partial sum-based process, which has been recently used for change point detection of high dimensional mean vectors in \cite{Jirak2015Uniform,yu2021finite}. As shown in our numerical studies, Bootstrap-I suffers from serious size distortions. This phenomenon is due to large biases arising from the residual-based process $\{\bR^{(0,t)}-\bR^{(t,1)},t\in[\tau_0,1-\tau_0]\}$, which can not be ignored in finite sample performance although it is asymptotically negligible. Hence, for change point detection in high dimensional linear models, substantial modifications are needed and it is desirable to consider a new candidate bootstrap method. To overcome this problem, different from the existing methods, we choose to bootstrap the entire  de-biased lasso-based process as shown in the following  Bootstrap-II.
			
			\subsubsection{Bootstrap-II} 
			
			The key idea of this bootstrap procedure is to approximate the null limiting distribution  under both $\Hb_{0,\cG}$ and $\Hb_{1,\cG}$.  It proceeds as follows:\\
			\textbf{Step~1:} Given $\hat{\sigma}_{\epsilon}^2$ in (\ref{equation: variance estimator under H1}), for the $b$-th bootstrap, let $\epsilon_1^b,\ldots,\epsilon_n^b$ be $i.i.d.$ random variables following $N(0,\hat{\sigma}_{\epsilon}^2)$. Define the $b$-th bootstrap of response vectors $\bY^b=(Y_1^b,\ldots,Y_n^b)^\top$:			
			\begin{equation}\label{equation: bootstrap observations}
				Y_i^b=\bX_i^\top\hat{\bbeta}^{(0,\hat{t}_{0})}\mathbf{1}\{1\leq i\leq \lfloor n\hat{t}_{0}\rfloor\}+\bX_i^\top\hat{\bbeta}^{(\hat{t}_{0},1)}\mathbf{1}\{\lfloor n\hat{t}_{0}\rfloor<i\leq n\}+\epsilon_i^b,
			\end{equation}			
			where $\hat{\bbeta}^{(0,\hat{t}_{0})}$ and $\hat{\bbeta}^{(\hat{t}_{0},1)}$ are the lasso estimators before and after $\hat{t}_0$.\\						
			\textbf{Step~2:} Denote $\bY^{b}_{(0,t)}=(Y^b_1,\ldots,Y^b_{\floor{nt}})^\top$, and  $\bY^{b}_{(t,1)}=(Y^b_{\floor{nt}+1},\ldots,Y^b_{n})^\top$. We then define the $b$-th bootstrap version of the de-biased lasso estimators before and after $\floor{nt}$ as
			$\breve{\bbeta}^{b,(0,t)}=(\breve{\beta}^{b,(0,t)}_1,\ldots,\breve{\beta}^{b,(0,t)}_p)^\top$ and $\breve{\bbeta}^{b,(t,1)}=(\breve{\beta}^{b,(t,1)}_1,\ldots,\breve{\beta}^{b,(t,1)}_p)^\top$,
			where 
			\begin{equation}\label{equation: boot de-biased before and after cpt}
				\begin{array}{ll}
					\breve{\bbeta}^{b,(0,t)}:=\hat{\bbeta}^{b,(0,t)}+
					{\hat{\bTheta}\Xb_{(0,t)}^\top}\Big(\bY^{b}_{(0,t)}-\Xb_{(0,t)}\hat{\bbeta}^{b,(0,t)}\big)/\floor{nt},\\
					\breve{\bbeta}^{b,(t,1)}:=\hat{\bbeta}^{b,(t,1)}+
					{\hat{\bTheta}\Xb_{(t,1)}^\top}\Big(\bY^{b}_{(t,1)}-\Xb_{(t,1)}\hat{\bbeta}^{b,(t,1)}\big)/\floor{nt}^*,
				\end{array}
			\end{equation}
			and $\hat{\bbeta}^{b,(0,t)}$ and $\hat{\bbeta}^{b,(t,1)}$ are the lasso estimators  before and after $t$ using the bootstrap samples
			$\{\bY^{b}_{(0,t)},\Xb_{(0,t)}\}$ and 
			$\{\bY^{b}_{(t,1)},\Xb_{(t,1)}\}$. \\
			\textbf{Step~3:} Define the bootstrap sample-based signal function $\hat{\bdelta}(t) =(\hat{\delta}_1(t),\ldots,\hat{\delta}_p(t))^\top$:
			\begin{equation*}
				\hat{\bdelta}(t)=\dfrac{n-\lfloor n\hat{t}_{0}\rfloor }{n-\lfloor nt\rfloor }\big(\hat{\bbeta}^{(0,\hat{t}_{0}) }-\hat{\bbeta}^{(\hat{t}_{0},1)}\big)\mathbf{1}\{t\in [\tau_0,\hat{t}_{0}]\}+\dfrac{\lfloor n\hat{t}_{0}\rfloor }{\lfloor nt\rfloor }\big(\hat{\bbeta}^{(0,\hat{t}_{0}) }-\hat{\bbeta}^{(\hat{t}_{0},1)}\big)\mathbf{1}\{t\in[\hat{t}_{0},1-\tau_0]\}.
			\end{equation*}
			\textbf{Step~4:}
			Calculate the $b$-th bootstrap version for the test statistic $T_{\cG}$ by
			\begin{equation}\label{equation: bootstrap test statistic}
				T_\cG^b=\max_{t\in[\tau_0,1-\tau_0]}\max_{j\in \cG}\sqrt{n}\dfrac{\floor{nt}}{n}\Big(1-\dfrac{\floor{nt} }{n}\Big) \Big| \dfrac{\breve{\beta}^{b,(0,t)}_j-\breve{\beta}^{b,(t,1)}_j-\hat{\delta}_j (t)}{\sqrt{\hat{\sigma}_{\epsilon}^2\hat{\omega}_{j,j }}}\Big|.
			\end{equation}
			\textbf{Step~5:} Repeat the above procedures (\ref{equation: bootstrap observations})-(\ref{equation: bootstrap test statistic}) for $B$ times and obtain the bootstrap samples $\{T^1_{\cG},\ldots,T^B_{\cG}\}$. Let $c_{\cG,\alpha}:=\inf\{t:\P(T_{\cG}\leq t)\geq 1-\alpha\}$ be the theoretical critical value of $T_{\cG}$. Using the bootstrap samples $\{T^1_{\cG},\ldots,T^B_{\cG}\}$, we estimate $c_{\cG,\alpha}$ by
			\begin{equation}\label{equation: estimated critical values}
				\hat{c}_{\cG,\alpha}=\inf\Big\{t:{(B+1)^{-1}}\sum_{b=1}^B\mathbf{1}\{T^b_{\cG}\leq t |\Xb,\bY\}\geq 1-\alpha   \Big\}.
			\end{equation}
			\textbf{Step~6:} Define the new test for the hypothesis (\ref{hypothesis: H0}) as follows:
			\begin{equation}\label{equation: new test}
				\small
				\Phi_{\cG,\alpha}=\mathbf{1}\{T_{\cG}\geq \hat{c}_{\cG,\alpha} \}.
			\end{equation}
			Given a significance level $\alpha\in(0,1)$ and a prespecified subgroup $\cG$, for the hypothesis (\ref{hypothesis: H0}), we reject $\Hb_{0,\cG}$ if  $\Phi_{\cG,\alpha}=1$.
			
			It is shown in theory that the Bootsrap-II-based test statistic $T_{\cG}^b$ approximates the limiting null distribution of $T_{\cG}$. More importantly, by bootstrapping the whole de-biased lasso-based process, Bootstrap-II  enjoys better test size performance than Bootstrap-I under various candidate subgroups. This is supported by our extensive numerical studies in Section \ref{section: simulation studies}.

			\subsection{Extensions to multiple change points}\label{section: multiple change points}
			So far, we have proposed new methods for detecting a single change point as well as identifying its location using the argmax based estimator.  In this section, we aim to extend our new testing procedure for detecting and identifying multiple change points for high dimensional linear models. In particular, suppose there are $m$ change points $k_1,\ldots,k_m$ that divide the linear structures into $m+1$ segments with different regression coefficients:
			\begin{equation}\label{equation: linear model with multiple change point}
				\left\{\begin{array}{ll}
					Y_i=\bX_i^\top\bbeta^{(1)}+\epsilon_i, & \text{for}~i=1,\ldots, k_1 ,\\
					Y_i=\bX_i^\top\bbeta^{(2)}+\epsilon_i,& \text{for}~i=k_1+1,\ldots, k_2,\\
					\quad \quad\quad \vdots\\
					Y_i=\bX_i^\top\bbeta^{(m)}+\epsilon_i,& \text{for}~i=k_{m-1}+1,\ldots, k_m,\\
					Y_i=\bX_i^\top\bbeta^{(m+1)}+\epsilon_i,& \text{for}~i=k_m+1,\ldots, n.
				\end{array}\right.
			\end{equation}
			Based on Model (\ref{equation: linear model with multiple change point}), for any given subgroup $\cG\subset \{1,\ldots,p\}$, in the case of multiple change points, we consider the following hypothesis:
			\begin{equation}\label{hypothesis: H0 multiple}
				\begin{array}{ll}
					\Hb_{0,\cG}':\beta_{s}^{(1)}=\beta_{s}^{(2)}=\cdots=\beta_s^{(m)}=\beta_s^{(m+1)}~~\text{for~all}~ s\in\cG~~\text{v.s.}~~\\
					\Hb_{1,\cG}': \text{There exist}~ s\in\cG ~\text{and~at~least ~one}~j^*\in\{1,\ldots,m\}~\text{s.t.}~ \beta_{s}^{(j^*)}\neq \beta_{s}^{(j^*+1)}.
				\end{array}
			\end{equation}
			
			To solve Problem (\ref{hypothesis: H0 multiple}), we combine our bootstrap-based new testing procedure with the well-known binary segmentation technique \citep{Vostrikovadetecting} to simultaneously detect and identify multiple change points. 
			More specifically, for each candidate search interval $(s,e)$, we  detect the existence of a change point. If $\Hb_{0,\cG}$ is rejected, we identify the new change point $b$ by taking  the argmax in (\ref{equation: t-hat}). Then the  interval $(s,e)$ is  split into two subintervals  $(s, b)$ and $(b, e)$ and we conduct the above procedure on $(s, b)$ and $(b, e)$ separately. This algorithm is stopped until no subinterval can detect a change point.  Algorithm \ref{alg:third} describes our bootstrap-based multiple change point testing procedure. It is demonstrated by our numerical studies that  Algorithm \ref{alg:third} can automatically account for the data generating mechanism and 
			simultaneously detect and identify multiple change points, which enjoys better performance than  existing techniques.

			\begin{algorithm}[!h]
				\caption{: Bootstrap-based binary segmentation
					procedure for multiple change point detection in high dimensional  linear regression models. }\label{alg:third}
				\begin{description}
					\item[Input:] Given the data set $\{\Xb,\bY\}$, set the value for $\tau_0$, the number of bootstrap replications $B$, and the subset $\cG$. 
					\item[Step~1:] {Initialize the set of change point pairs $\cT=\{0,1\}$.}
					\item[Step~2:] {For each pair  $\{s,e\}$ in $\cT$},  detect the existence of a change point.  If $\Hb_{0,\cG}$ is rejected, identify the new change point $b$ by taking the argmax in (\ref{equation: t-hat}). {Then add new pairs of nodes $\{s,b\}$ and $\{b,e\}$ to $\cT$ and update $\cT$ as $\cT=\cT\cup\{s,b\}\cup\{b,e\}$. }
					
					\item[Step~3:] {Repeat Step 2 until no more new pair of nodes can be added. Denote the terminal set of change point pairs by $\cT_{\rm final}=\cup_{i=1}^{\hat{m}+1}\{\hat{t}_{i-1},\hat{t}_{i}\}$. }		
					\item[Output:] {Algorithm \ref{alg:third} provides the change point estimator ${\hat{\bt}=(\hat{t}_0,...,\hat{t}_{\hat{m}+1})^\top}$, where $\hat{m}=\#{\cT}_{\text{final}}-1$ and $0=\hat{t}_0<\hat{t}_1<...<\hat{t}_{\hat{m}}<\hat{t}_{\hat{m}+1}=1$, including  the number and locations. }
				\end{description}
			\end{algorithm}

			\section{Theoretical properties}\label{section: theoretical properties}
			In this section, we examine some theoretical properties of our proposed  method including the size, power and the change point estimation results.
			

			\subsection{Basic assumptions}\label{section: basic assumptions}
		We introduce some basic assumptions for making change point inference on high dimensional linear models. Assumption $\mathbf{(A.1)}$ is a basic requirement for the change point location. Assumptions $\mathbf{(A.1)}$ -- $\mathbf{(A.3)}$ impose some regular conditions on the design matrix as well as the error terms. Assumption $\mathbf{(A.4)}$ contains  basic requirements on model parameters. Assumption $\mathbf{(A.5)}$ is a technical condition on the regularity parameters in (\ref{equation: node-wise regression}) and (\ref{equation: lasso estimation before and after cpt}). 
		
		{Before giving the assumptions, we introduce the  concept of the restricted eigenvalue (RE) and uniform restricted eigenvalue (URE) conditions.}
		
		\begin{definition}(Restricted eigenvalue $\text{RE}(s_j,3)$).\label{def: RE}
			For integers $s_j$ such that $1\leq s_j\leq p-1$,  a set of indices $J_0\subset\{1,\ldots,p-1\}$ with $|J_0|\leq s_j$, define
			
			\begin{equation}\label{equation:  RE}
				\cR^{(j)}(s_j,3)=\min_{J_0\subset\{1,\ldots,p-1\}\atop|J_0|\leq s_j}\min_{\bdelta\neq 0\atop \|\bdelta_{J_0^c}\|_1\leq 3\|\bdelta_{J_0}\|_1}\dfrac{\|\Xb^{-j} \bdelta\|_2}{\sqrt{n}\|\bdelta_{J_{0}}\|_2},~\text{with}~1\leq j\leq p,
			\end{equation}
			{where $\Xb^{-j}\in \RR^{n\times (p-1)}$ is a submatrix of $\Xb$ with the $j$-th column being removed, and $\bdelta_{J_{0}}$ is the vector  that has the same coordinates as $\bdelta$ on $J_0$ and zero coordinates on the complement $J_0^c$ of $J_0$.}
		\end{definition}

		\begin{definition}(Uniform restricted eigenvalue $\text{URE}(s,3,\mathbb{T}))$.\label{def: URE}
			For integers $s$ such that $1\leq s\leq p$, a set of indices $J_0\subset\{1,\ldots,p\}$ with $|J_0|\leq s$, and $\mathbb{T}=[\tau_0,1-\tau_0]$, define
			
			\begin{equation}\label{equation:  URE before cpt}
				\cR_1(s,3,\mathbb{T})=\min_{t\in\mathbb{T}}\min_{J_0\subset\{1,\ldots,p\}\atop|J_0|\leq s}\min_{\bdelta\neq 0\atop \|\bdelta_{J_0^c}\|_1\leq 3\|\bdelta_{J_0}\|_1}\dfrac{\|\Xb_{(0,t)} \bdelta\|_2}{\sqrt{\lfloor nt \rfloor}\|\bdelta_{J_{0}}\|_2}.
			\end{equation}
			and 
			\begin{equation}\label{equation:  URE after cpt}
				\cR_2(s,3,\mathbb{T})=\min_{t\in\mathbb{T}}\min_{J_0\subset\{1,\ldots,p\}\atop|J_0|\leq s}\min_{\bdelta\neq 0\atop \|\bdelta_{J_0^c}\|_1\leq 3\|\bdelta_{J_0}\|_1}\dfrac{\|\Xb_{(t,1)} \bdelta\|_2}{\sqrt{\lfloor nt \rfloor^*}\|\bdelta_{J_{0}}\|_2}.
			\end{equation}
		\end{definition}
		
		{Note that Definition \ref{def: RE} is similar to the RE conditions introduced in \cite{bickel2009simultaneous} and is mainly used for the node-wise lasso estimators. It is well-known that the RE conditions are among the weakest assumptions on the design matrix and  are important for deriving the estimation error bounds of the  lasso solutions. See \cite{raskutti2010restricted,van2009conditions}. Moreover,  our testing procedure needs to calculate $\hat{\bbeta}^{(0,t)}$ and $\hat{\bbeta}^{(t,1)}$ as in (\ref{equation: lasso estimation before and after cpt}). For each search location $t\in[\tau_0,1-\tau_0]$, to guarantee $\hat{\bbeta}^{(0,t)}$ and $\hat{\bbeta}^{(t,1)}$ enjoy desirable properties toward their population counterpart ${\bbeta}^{(0,t)}$ and ${\bbeta}^{(t,1)}$, we introduce the  uniform restricted eigenvalue condition as in Definition 3, which is an extension of the RE condition.} 
		
		{With the above two definitions, we are ready to introduce  the assumptions, which are summarized as follows:}\\
		$\mathbf{Assumption ~(A.1)}$ The design matrix $\Xb$ has $i.i.d.$ rows following sub-Gaussian distributions. In other words, there exists a positive constant $K$ such that $\sup_{i,j}\E( \exp( |X_{i,j}|^2/K))\leq 1$ holds. \\
		$\mathbf{Assumption ~(A.2)}$ The error terms $\{\epsilon_i\}_{i=1}^n$ are $i.i.d.$ sub-Gaussian with finite variance $\sigma_{\epsilon}^2$. In other words, there exist positive constants 
		$K'$, $c_{\epsilon}$ and $C_{\epsilon}$ such that
		$\E(\exp(|\epsilon_i|^2/K'))\leq 1$ and $c_\epsilon\leq \text{Var}(\epsilon_i)\leq C_{\epsilon}$ hold. Furthermore, $\epsilon_i$ is independent with $\bX_i$ for $i=1,\ldots,n$.\\
		$\mathbf{Assumption~ (A.3)}$ {Assume that there are positive constants $\kappa_1$ and $\kappa_2$ such that $\max_{j}\Sigma_{j,j}<\kappa_1<\infty$ and $\max_{j}\|\btheta_{j}\|_2< \kappa_2<\infty$ hold, where $\btheta_{j}$ is the $j$-th row of $\bTheta=(\theta_{j,k}):=\bSigma^{-1}$. Moreover, for the RE and URE conditions, we require} 
		\begin{equation}\label{RE and URE condition}
			\min_{1\leq j\leq p}  	\cR^{(j)}(s_j,3)>\kappa_3,~~  \min\big(\cR_1(s,3,\mathbb{T}),\cR_2(s,3,\mathbb{T})\big)>\kappa_4
		\end{equation}
		for some $\kappa_3, \kappa_4>0$, where $s_j:=|\{1\leq k\leq p: \theta_{j,k}\neq 0, k\neq j\}|$.\\
		$\mathbf{Assumption ~(A.4)}$ For the change point model  in (\ref{equation: linear model2}), we assume the following:
		\begin{equation*}
			\begin{array}{l}
				(a)~ \text{Assume that}~ \log(pn)=O(\floor{n\tau_0}^{\zeta}) ~\text{holds for some}~ 0 <\zeta<1/7;\\
				(b) \text{We assume} \lfloor n\tau_0 \rfloor\rightarrow \infty,~
				\max\limits_{1\leq j\leq p}s_j\dfrac{\log(pn)}{\sqrt{ n }}\rightarrow0~ \text{and}
				~	s\sqrt{n}\dfrac{\log(pn)}{{\lfloor n\tau_0 \rfloor}}\rightarrow0~
				\text{as}~n,p\rightarrow\infty, \\\text{where } s:=s^{(1)}\vee s^{(2)};\\
				(c)~~ \text{There exists some constant}~ \gamma\in(0,1]~ \text{such that}~ |\cG|=p^\gamma.
			\end{array}
		\end{equation*}
		$\mathbf{Assumption ~(A.5)}$ For the node-wise regression in (\ref{equation: node-wise regression}), we require the regularization parameter $\lambda_{(j)}\asymp\sqrt{\log(p)/n}$ uniformly in $j$. For the lasso-based estimators in (\ref{equation: lasso estimation before and after cpt}), we require
		\begin{equation}\label{equation: regularization parameter}
			\begin{array}{l}
				\lambda_1(t)\asymp \sqrt{\dfrac{\log(p)}{\floor{nt}}},~~\lambda_2(t)\asymp \sqrt{\dfrac{\log(p)}{\floor{nt}^*}},~\text{for}~t\in[\tau_0,1-\tau_0].
			\end{array}
		\end{equation}

		Assumptions  $\mathbf{(A.1)}$ -- $\mathbf{(A.3)}$ are relatively weak conditions on 
		the covariates  and  error terms. In particular, they require that $\{\bX_i\}_{i=1}^n$ and  $\{\epsilon_i\}_{i=1}^n$
		are sub-Gaussian distributed with {``well-behaved" sample} covariance matrix  and {non-degenerate} variances $\sigma^2_{\epsilon}$, which covers a wide broad of distributional patterns and has been commonly adopted in high dimensional data analysis.
		Assumption $\mathbf{(A.4)}$ specifies the scaling relationships among parameters ($\{s,s_j,n,p,|\cG|\}$) in Model (\ref{equation: linear model2}). More specifically, (a) allows the number of variables ($p$) can grow exponentially with the number of data observations ($n$) as long as $\log(pn)=O(\floor{n\tau_0}^{\zeta})$ holds; (b) allows the number of active variables ($s$ and $s_j$) can go to infinity if $\max\limits_{1\leq j\leq p}s_j\dfrac{\log(pn)}{\sqrt{ n }}\rightarrow0$~ \text{and}
		~	$s\sqrt{n}\dfrac{\log(pn)}{{\lfloor n\tau_0 \rfloor}}\rightarrow0$  holds; (c) demonstrates that we can make change point inference on any large scale subgroup $\cG$ with $|\cG|=p^\gamma$. Lastly, Assumption $\mathbf{(A.5)}$ imposes some technical conditions on the regularity parameters of lasso and node-wise lasso, which is important for deriving desirable estimation error bounds of the corresponding estimators. It is worth mentioning that (\ref{equation: regularization parameter}) automatically accounts for the heterogeneity of the $\ell_1$ regularization problem (\ref{equation: lasso estimation before and after cpt}) and is consistent with the classical conditions as in  \cite{bickel2009simultaneous} when the data are homogenous (e.g. $\bbeta^{(1)}=\bbeta^{(2)}$).
		
		{Lastly, the following Proposition \ref{proposition: RE conditions} shows that the RE and URE conditions in (\ref{RE and URE condition}) of Assumption (A.2) hold with high probabilities.}
		\begin{proposition}\label{proposition: RE conditions}
			(i)	For integers $s_j$ such that $1\leq s_j\leq p-1$, a set of indices $J_0\subset\{1,\ldots,p-1\}$ with $|J_0|\leq s_j$ and $s_j\sqrt{\log(p)/n}=o(1)$. Under Assumption  (A.1), if $\bSigma$ satisfies
			\begin{equation}\label{equation: populational covariance node-wise}
				\min_{1\leq j\leq p}\min_{J_0\subset\{1,\ldots,p-1\}\atop|J_0|\leq s_j}\min_{\bdelta\neq 0\atop \|\bdelta_{J_0^c}\|_1\leq 3\|\bdelta_{J_0}\|_1} \dfrac{\|\bSigma^{-j,-j}\bdelta\|_2}{\|\bdelta_{J_0}\|}\geq 2\kappa_3,
			\end{equation}
			for some $\kappa_3>0$, then we have:
			\begin{equation*}
				\P(\min_{1\leq j\leq p}  	\cR^{(j)}(s_j,3)>\kappa_3)\geq 1-C_1(np)^{-C_2},
			\end{equation*}
			where $\bSigma^{-j,-j}:=\E[\Xb^{-j}(\Xb^{-j})^\top/n]$ and $C_1, C_2$ are universal positive constants not depending on $n$ or $p$.
			(ii) Similarly, for integers $s$ such that $1\leq s\leq p$, a set of indices $J_0\subset\{1,\ldots,p\}$ with $|J_0|\leq s$ and $s\sqrt{\log(p)/\floor{n\tau_0}}=o(1)$. Under Assumption  (A.1), if $\bSigma$ satisfies
			\begin{equation}\label{equation: populational covariance lasso}
				\min_{J_0\subset\{1,\ldots,p\}\atop|J_0|\leq s}\min_{\bdelta\neq 0\atop \|\bdelta_{J_0^c}\|_1\leq 3\|\bdelta_{J_0}\|_1}\dfrac{\|\bSigma \bdelta\|_2}{\|\bdelta_{J_{0}}\|_2}\geq 2\kappa_4,
			\end{equation}
			for some $\kappa_4>0$, then we have
			\begin{equation*}
				\P(\min\big(\cR_1(s,3,\mathbb{T}),\cR_2(s,3,\mathbb{T})\big)>\kappa_4)\geq 1-C_3(np)^{-C_4},
			\end{equation*}
			where $C_3, C_4>0$ are some universal constants not depending on $n$ or $p$.
		\end{proposition}
	
		\begin{remark}
			{The proof of Proposition \ref{proposition: RE conditions} is given in the Appendix. A sufficient condition for both (\ref{equation: populational covariance node-wise}) and (\ref{equation: populational covariance lasso}) hold is $\lambda_{\rm min}(\bSigma)>c$ for some $c>0$, where $\lambda_{\rm min}(\bSigma)$ is the smallest eigenvalue of $\bSigma$. Note that  the smallest eigenvalue condition is easy to verify and has been widely used in the literature such as \cite{kaul2019efficient,Wangdaren2021} for change point analysis of high dimensional linear models.} For example, many commonly used covariance matrices such as Toeplitz matrices, blocked diagonal matrices have positive smallest eigenvalue values. 
		\end{remark}

			\subsection{Main results}\label{section: main results}
			We derive some theoretical results of our proposed new test. In Section \ref{section: the validity of size}, we consider the control of Type I error.  In Section \ref{section: analysis under H1}, we examine the power performance as well as the accuracy of change point estimation.
			\subsubsection{The validity of test size}\label{section: the validity of size}
			Before giving the  test size results, we first consider the variance estimation. Theorem \ref{theorem: variance estimator under H0} shows that the pooled weighted variance estimator is uniformly consistent under the null hypothesis. It is crucial for deriving the Gaussian approximation results as in Theorem \ref{theorem: gaussian approximation}.
			\begin{theorem}\label{theorem: variance estimator under H0}
				Suppose Assumptions $\mathbf{(A.1)}$ -- $\mathbf{(A.5)}$ hold. Under $\Hb_{0,\cG}$, for the variance estimator, with probability at least $1-C_1(np)^{-C_2}$, we have
				\begin{equation*}\label{equation: variance estimation  under H0}
					\max_{1\leq j, k\leq p}|\hat{\sigma}_{\epsilon}^2\hat{\omega}_{j,k}-\sigma_{\epsilon}^2\omega_{j,k}|\leq C_3\big(\sqrt{\dfrac{\log(n)}{n}}+ \max_j\lambda_{(j)}\sqrt{s_j}\big),
				\end{equation*}
				where $C_1,\ldots,C_3$ are universal positive constants not depending on $n$ or $p$.
			\end{theorem}
			Based on Theorem \ref{theorem: variance estimator under H0} as well as other regularity conditions, the following Theorem \ref{theorem: gaussian approximation} justifies the validity of our bootstrap procedure.
			\begin{theorem}\label{theorem: bootstrap validity}\label{theorem: gaussian approximation}
				Suppose Assumptions $\mathbf{(A.1)}$ -- $\mathbf{(A.5)}$ hold.  Under $\Hb_{0,\cG}$, for any given subgroup $\cG\subset\{1,\ldots,p\}$, we have
				\begin{equation*}
					\sup_{z\in (0,\infty)}\big|\P(T_{\cG}\leq z)-\P(T_{\cG}^{b}\leq z|\{\Xb,\bY\})\big|=o_p(1), ~\text{as}~  n,p\rightarrow\infty.
				\end{equation*}
			\end{theorem}
			Theorem \ref{theorem: gaussian approximation} shows that we can uniformly approximate the distribution of $T_{\cG}$ using that of $T_{\cG}^b$. As a corollary, the following Corollary \ref{corollary: size} shows that our proposed new test can control the Type I error asymptotically  for any given pre-specified significance level $\alpha$.
			\begin{corollary}\label{corollary: size}
				Assume Assumptions $\mathbf{(A.1)}$--$\mathbf{(A.5)}$ hold. 	Under $\Hb_{0,\cG}$,  for any given subgroup $\cG\subset\{1,\ldots,p\}$, we have
					$				\P(\Phi_{\cG,\alpha}=1)\rightarrow\alpha, ~ \text{as} ~n,p,B\rightarrow\infty.$
			\end{corollary}

			\subsubsection{Analysis under $\Hb_{1,\cG}$}\label{section: analysis under H1}
			After analyzing the theoretical results under the null hypothesis, we next consider the performance under $\Hb_{1,\cG}$. To this end, some additional  assumptions are needed. \\
			$\mathbf{Assumption ~(A.6)}$. Let $\bdelta=\bbeta^{(1)}-\bbeta^{(2)}$. {For the signal jump, we require}
			there exists a constant $c^*\in[0,\infty)$ such that $\lim_{n,p\rightarrow\infty}s\|\bdelta\|_{\infty}\rightarrow c^*$.
			
			Note that  Assumption $\mathbf{(A.6)}$ is a signal strength  requirement for identifying the change point location $t_0$ with high accuracy. {It allows weak signals that can scale to zero as $(n,p)\rightarrow \infty$.}
			With the additional assumption as well as those of $\mathbf{(A.1)}$ -- $\mathbf{(A.5)}$, the following Theorem \ref{theorem: cpt estimation results} provides a non-asymptotic estimation error bound of  $\hat{t}_{0,\cG}$ for $t_0$.
			
			\begin{theorem}\label{theorem: cpt estimation results}
				Suppose Assumptions $\mathbf{(A.1)}$ - $\mathbf{(A.6)}$ hold. {Assume additionally 
					$\|\bdelta\|_{\cG,\infty}\gg \sqrt{\log(|\cG|n)/n}$ holds.} For any given subgroup $\cG\subset\{1,\ldots,p\}$, under $\Hb_{1,\cG}$,  with probability at least $1-C_1(np)^{-C_2}$, we have
				\begin{equation}\label{inequality: estimation error bound of cpt}
					\big|\hat{t}_{0,\cG}-t_0\big|\leq C^*\dfrac{\log(|\cG|n)}{n\|\bdelta\|_{\cG,\infty}^2},
				\end{equation}
				where $C^*$ is a  universal positive constant not depending on $n$ or $p$.
			\end{theorem}
			
			{Theorem \ref{theorem: cpt estimation results} shows that our subgroup-based change point estimator is asymptotically consistent, which allows the group size $|\cG|$ to grow with the sample size $n$ as long as $\|\bdelta\|_{\cG,\infty}\gg \sqrt{\log(|\cG|n)/n}$ holds.}
			
			\begin{remark}
				Note that \cite{Jirak2015Uniform,yu2021finite} considered the change point estimation for high dimensional mean vectors. They obtained the change point estimators by taking ``argmax'' of the corresponding partial sum processes  with an estimation error rate of $O_p\big(\log(p)/(n\|\bDelta\|_{\rm min}^2)\big)$, {where $\bDelta=(\Delta_1,\ldots,\Delta_p)^\top$ is the signal jump of mean vectors before and after the change point and $\|\bDelta\|_{\rm min}$ is the minimum signal jump for the coordinates with a change point.} {Different from \cite{Jirak2015Uniform,yu2021finite}, we adopt a different proof technique and derive an estimation error bound of $O_p\big(\log(p)/(n\|\bDelta\|_{\infty}^2)\big)$. Considering $\|\bDelta\|_{\infty}$ can be much larger than $\|\bDelta\|_{\rm min}$, our result is sharper than \cite{Jirak2015Uniform,yu2021finite}. More proof details can be found in the Appendix.}
			\end{remark}
			
			After analyzing the change point identification, we next consider the change point detection. Note that for the change point problem, variance estimation under the alternative is a difficult but important task. As pointed out in \cite{shao2010testing}, due to the unknown change point, any improper estimation may lead to non-monotonic power performance. This distinguishes the change point problem substantially from one-sample or two-sample 
			tests where homogenous data are used to construct  consistent variance estimation.
			
			Theorem \ref{theorem: variance estimation under H1} shows that the  pooled weighted variance estimation is uniformly consistent under $\Hb_{1,\cG}$. This guarantees that our new testing method has reasonable power performance. 
			\begin{theorem}\label{theorem: variance estimation under H1}
				Suppose Assumptions $\mathbf{(A.1)}$ - $\mathbf{(A.6)}$ hold.  Then,
				for the weighted variance estimation, under $\Hb_{1,\cG}$, we have
				\begin{equation}\label{equation: variance estimator under H11}
					\max_{1\leq j,k\leq p}|\hat{\sigma}_{\epsilon}^2\hat{\omega}_{j,k }-\sigma_{\epsilon}^2\omega_{j,k}|=o_p(1),~\text{as}~~n,p\rightarrow\infty.
				\end{equation}
			\end{theorem}
			
			From the proof of Theorem \ref{theorem: variance estimation under H1}, some interesting observations can be found. On one hand, if the signal strength is too weak such that  $\|\bdelta\|_{\cG,\infty}=O(\sqrt{\log(pn)/n})$ holds, then the pooled weighted variance estimator $\hat{\sigma}^2_{\epsilon}$ is a consistent estimator for $\sigma_{\epsilon}^2$ {even though we can not guarantee a consistent change point estimator in this case.} On the other hand, if  the signal strength is big enough such that  $\|\bdelta\|_{\cG,\infty}\gg \sqrt{\log(pn)/n}$ holds, then a consistent change point estimator $\hat{t}_{0,\cG}$ is needed to guarantee (\ref{equation: variance estimator under H11}) holds. {These are insightful findings for variance estimation in change point analysis, which is different from the $i.i.d.$ case.}
			
			Lastly, we discuss the power properties. To this end, we need some additional notations. Recall $\Pi=\{j: \beta^{(1)}_j\neq \beta^{(2)}_j\}$ as the set of coordinates having a change point.  Define the oracle signal to noise ratio vector $\bD=(D_1,\ldots,D_{p})^\top$ with
			\begin{equation}\label{equation: signal to noise ration}
				D_j:=\left\{\begin{array}{ll}
					0,& \text{for}~~ j\in \Pi^c\\
					\Big|\dfrac{t_0(1-t_0)(\beta_j^{(2)}-\beta_j^{(1)})}{(\sigma^2_{\epsilon}\omega_{j,j })^{1/2}}\Big|,& \text{for}~~j\in \Pi.
				\end{array}\right.
			\end{equation}  
			With the above notations and some regularity conditions, the following Theorem \ref{theorem: power results} shows that we can reject the null hypothesis of no change points with overwhelming probability.
			\begin{theorem}\label{theorem: power results}
				Suppose Assumptions $\mathbf{(A.1)}$ -- $\mathbf{(A.6)}$ hold. {Let $\epsilon_n=o(1)$}. For any given subgroup $\cG\subset\{1,\ldots,p\}$, if $\bD$ satisfies
				\begin{equation}\label{inequality: theoretical signal strengh}
					\sqrt{n}\big\|\bD\big\|_{\cG,\infty}\geq \dfrac{C_0}{1-\epsilon_n}\Big(\sqrt{2\log(|\cG|n)}+\sqrt{2\log(\alpha^{-1})}\Big),
				\end{equation}
				under $\Hb_{1,\cG}$, we have
					$		\P(\Phi_{\cG,\alpha}=1)\rightarrow 1, \text{as}~n,p,B\rightarrow\infty,$
				where $C_0$ is a large enough universal positive constant not depending on $n$ or $p$.
			\end{theorem}
			
			Theorem \ref{theorem: power results} demonstrates that with probability tending to one, our proposed new test can detect the existence of a change point for any given subgroup as long as the corresponding signal to noise ratio satisfies (\ref{inequality: theoretical signal strengh}). Combining (\ref{equation: signal to noise ration}) and (\ref{inequality: theoretical signal strengh}), we note that with a larger signal jump, a smaller noise level, and a closer change point location to the middle of data observations, it is more likely to trigger a rejection of  the null hypothesis.

			Lastly, we would like to point out that the requirements for identifying and detecting a change point are different. More specifically, from Theorem \ref{theorem: cpt estimation results}, to correctly identify the location of a change point with desirable accuracy, {the signal strength should at least satisfy $\|\bdelta\|_{\cG,\infty}\gg \sqrt{\log(|\cG|n)/n}$.} In contrast,  Theorem \ref{theorem: power results} shows that  it is sufficient to detect a change point if $\|\bD\big\|_{\cG,\infty}\geq C\sqrt{\log(|\cG|n)/n}$ holds. Hence, we need more stringent conditions for locating a change point than detecting its existence.

			\section{Numerical studies}\label{section: simulation studies}
			We examine the numerical performance of our proposed method and compare it with several existing state-of-art techniques. 

			We first consider single change point detection. For the design matrix $\Xb$, we generate $\bX_i$  ($i.i.d.$)  from $N(\mathbf{0},\bSigma)$, where the following two types of covariance structures are investigated: 
				\begin{description}
					\item[Model 1:] $\bSigma=\Ib_{p\times p}$;
						\item[Model 2:] $\bSigma=\bSigma^*$ with $\bSigma^*=(\sigma^*)_{i,j=1}^p$, where $\sigma^{*}_{i,j}=0.5^{|i-j|}$ for $1\leq i,j\leq p$.
					\end{description}
			
			To show the bootstrap performance, for each model, the error terms $(\epsilon_i)_{i=1}^n$ are $i.i.d.$ generated from  standard normal distributions, standardized $\rm Gamma(4,1)$ distributions as well as Student's $t_5$ distributions.
			
			For the regression coefficient $\bbeta^{(1)}$, for each replication, we  generate $s$ non-zero covariates randomly selected from  $\cS=\{1,\ldots,50\}$. The corresponding  selected coefficients are $i.i.d.$ from $\text{U}(0,2)$, and the remaining $p-s$ covariates are 0's. Note that we generate regression coefficients out of $\cS$, which is denoted as the active set.	Under $\Hb_{0,\cG}$, we set $\bbeta^{(2)}=\bbeta^{(1)}$. Throughout the simulations, we consider various combinations of the sample sizes $n$, data dimensions $p$, and overall sparsities $s$ by setting  $n\in\{200,300\}$,  $p\in\{100,200,300,400\}$ and  $s\in\{5,10\}$. The number of bootstrap replications is $B=100$. Without additional specifications, all numerical results are based on 2000 replications.
			\subsection{Empirical sizes}\label{section: detection}
			We investigate  the empirical sizes. We set the significance levels $\alpha=1\%, 5\%$. Furthermore, three different types of subgroups are investigated: $\cG=\cS$, $\cG=\cS^c$, and $\cG=\cS\cup\cS^c=\{1,\ldots,p\}$. To evaluate the numerical performance, in addition to our proposed methods, we consider four existing well-known techniques for  change point detection of high dimensional linear models: 
			the high dimensional lasso-based method in \cite{lee2016lasso} (Lee2016), 
			the sparse group lasso-based method in \cite{zhang2015change-point} (SGL),  the binary segmentation-based method in \cite{leonardi2016computationally} (L\&B), and the Variance Projected Wild Binary Segmentation in \cite{Wangdaren2021} (VPWBS).

			It is worth  noting that under $\Hb_{0,\cG}$ with $\bbeta^{(1)}=\bbeta^{(2)}$, SGL and L\&B can potentially select the true homogeneous model by  identifying the change points at $\{1,n\}$. Hence, we record their rates of false selections as their ``empirical sizes''. As for Lee2016, their main purpose is to simultaneously estimate the potential single change point as well as the regression coefficients. Therefore, we do not report their empirical sizes and powers here.

			\begin{table}[!hpt]
				\caption{Empirical sizes for Models 1 and 2. The errors are generated from $N(0,1)$. The  results are based on 2000 replications.}
				\label{table: empirical sizes}
				\vspace{-0.5cm}
				\addtolength{\tabcolsep}{-4pt}
				\begin{center}
					\scriptsize
					\begin{tabular}{ccccc cccc}
						\toprule[2pt]
						\multicolumn{9}{c}{\bf{Empirical sizes (\%)  with $(n,s)=(200,5)$}}\\
						Model	&$\cG$&$p$&Boot-I ($\alpha=1\%$)&Boot-II ($\alpha=1\%$)&Boot-I ($\alpha=5\%$)&Boot-II ($\alpha=5\%$)&SGL&L\&B\\		
						\hline
						$\bSigma=\Ib$	&$\cS$&200&7.61	&	1.70	&	18.52	&	3.86	&	NA	&	NA\\	
						&&400&10.70	&	1.80	&	23.05	&	5.30	&	NA	&	NA\\		
						&$\cS^c$&	200	&8.23	&	1.44	&	15.43	&	4.06	&	NA	&	NA
						\\
						&&	400	&11.93	&	0.93	&	21.60	&	3.40	&	NA	&	NA\\
						&$\cS\cup \cS^c$&	200	&	7.41	&	1.03	&	14.20	&	2.93	&	38.89	&	0.00
						\\
						&&	400	&	12.55	&	1.39	&	27.37	&	3.86	&	46.67	&	0.00	\\
						\cline{3-9}		
						$\bSigma=\bSigma^*$	&$\cS$&	200	&	7.61	&	1.49	&	14.40	&	4.73	&	NA	&	NA
						\\
						&&	400	&	8.64	&	1.65	&	16.26	&	4.68	&	NA	&	NA	\\	
						&$\cS^c$&	200	&	3.50	&	0.82	&	12.14	&	3.09	&	NA	&	NA\\
						&&	400	&	5.76	&	0.67	&	12.76	&	3.03	&	NA	&	NA	\\
						&$\cS\cup \cS^c$&	200	&	4.73	&	0.82	&	13.37	&	3.29	&	77.78	&	0.00
						\\
						&&	400	&	7.82	&	1.23	&	17.08	&	3.19	&	80.00	&	0.00\\
						
						\hline
						\multicolumn{9}{c}{\bf{Empirical sizes (\%) with $(n,s)=(300,10)$}}\\
						Model	&$\cG$&$p$&Boot-I ($\alpha=1\%$)&Boot-II ($\alpha=1\%$)&Boot-I ($\alpha=5\%$)&Boot-II ($\alpha=5\%$)&SGL&L\&B\\		
						\hline
						$\bSigma=\Ib$	&$\cS$&200	&	12.76	&	1.83	&	23.66	&	3.25	&	NA	&	NA
						\\
						&&	400	&	19.55	&	1.88	&	33.74	&	7.35	&	NA	&	NA\\	
						&$\cS^c$&	200	&	8.33	&	1.02	&	16.67	&	3.25	&	NA	&	NA
						\\
						&&	400	&	13.79	&	1.63	&	26.95	&	3.06	&	NA	&	NA\\
						&$\cS\cup \cS^c$&	200	&	11.52	&	0.82	&	22.43	&	3.27	&	56.67	&	0.00
						\\
						&&	400	&	17.49	&	2.45	&	32.30	&	5.71	&	62.30	&	0.00\\	
						\cline{3-9}		
						$\bSigma=\bSigma^*$	&$\cS$&	200	&	10.91	&	0.62	&	22.63	&	2.67	&	NA	&	NA
						\\
						&&	400	&	17.07	&	2.26	&	28.86	&	5.56	&	NA	&	NA\\	
						&$\cS^c$&	200	&	4.32	&	0.41	&	11.32	&	1.65	&	NA	&	NA
						\\
						&&	400	&	3.66	&	0.81	&	10.77	&	2.44	&	NA	&	NA\\
						&$\cS\cup \cS^c$&	200	&	6.50	&	1.85	&	16.06	&	4.32	&	56.67	&	0.00
						\\
						&&	400	&	7.06	&	0.61	&	17.57	&	3.25	&	55.30	&	0.00\\
						\bottomrule[2pt]
					\end{tabular}		
				\end{center}
			\end{table}

			Table \ref{table: empirical sizes}  summarizes the empirical sizes for Models 1 and 2 with different combinations of $(n,p,s)$ under $N(0,1)$ distributions.  We can see that both SGL and L\&B are only applicable for the case of the overall subset with $\cG=\{1,\ldots,p\}$. In those cases, SGL suffers from serious size distortions with too many false selections. One reasonable explanation is that SGL builds their algorithms on the sparse  group lasso which tends to overestimate the number of change points. 
			Moreover, 
			we observe that L\&B seems to be conservative although it
			can select the homogenous model with no false selections.  As for our proposed  methods, the empirical sizes of Boot-I are out of control (especially for the active set $\cS$). This suggests that for change point detection of high dimensional linear models, the residual term of the de-biased lasso-based process can not be ignored, even though it is asymptotically negligible in theory. As compared to Boot-I,  Boot-II benefits from bootstrapping the whole de-biased lasso-based process. In most cases, the empirical sizes for Boot-II are close to the nominal level across various dimensions and subgroups. Interestingly, it shows that the empirical performance of Boot-II is affected by  the candidate subgroups. More specifically, empirical sizes for the active set $\cS$ are sometimes larger than the nominal level and the size performance of the non-active set $\cS^c$ performs the best among all candidate subgroups. Note that similar findings are also observed in constructing simultaneous confidence intervals  in \cite{zhang2017simultaneous} for the given subgroup $\cG$. In addition, we can see that Boot-II can still have satisfactory size performance as the non-zero elements increase slowly from $s=5$ to $s=10$.
			
			In the supplemental materials, we report  the size performance under standardized $\rm Gamma(4,1)$ and Student's $t_5$ distributions in Tables \ref{table: empirical sizes gamma} and \ref{table: empirical sizes t5}. In both cases,  our proposed method can control the size under the nominal level. This suggests that the  bootstrap null distribution is correctly calibrated even for non-normal underlying errors.

			\subsection{Empirical powers}\label{sec: powers}
			\begin{table}[!hpt]
				\caption{Empirical powers (\%)  under Model 1.  The numerical results are  based on 2000 replications.}
				\label{table: empirical powers model 1}
				\vspace{-0.5cm}
				\addtolength{\tabcolsep}{1.8pt}
				\begin{center}
					\scriptsize
					\begin{tabular}{ccccc ccc  }
						\toprule[2pt]
						\multicolumn{8}{c}{\bf{Empirical powers (\%) with $\bdelta=0.5\sqrt{{\log(p)}/{n}}\times(2^3,2^2,2^1,2^0,2^{-1})$.}}\\
						&&&	\multicolumn{2}{c}{Change point at $k^*=0.5n$}&&\multicolumn{2}{c}{Change point at $k^*=0.3n$}\\
						Model	&$\cG$&$p$&Boot-II&L\&B&&Boot-II&L\&B\\
						\hline
						$\bSigma=\Ib$	&$\cS$&	200	&	58.33	&	NA	&&	36.46	&	NA
						\\
						&&	400	&	64.93	&	NA	&&	42.71	&	NA\\
						&$\cS^c$&	200	&	2.08	&	NA	&&	4.17	&	NA
						\\
						&&	400	&	3.47	&	NA	&&	3.82	&	NA\\
						&$\cS\cup \cS^c $&	200	&	43.75	&	0.00	&&	29.17	&	0.00
						\\
						&&	400	&	40.97	&	0.00	&&	27.17	&	0.00\\
						\hline
						\multicolumn{8}{c}{\bf{Empirical powers (\%) with $\bdelta=\sqrt{{\log(p)}/{n}}\times(2^3,2^2,2^1,2^0,2^{-1})$.}}\\
						&&&	\multicolumn{2}{c}{Change point at $k^*=0.5n$}&&\multicolumn{2}{c}{Change point at $k^*=0.3n$}\\
						Model	&$\cG$&$p$&Boot-II&L\&B&&Boot-II&L\&B\\
						\hline
						$\bSigma=\Ib$	&$\cS$&	200	&	100.00	&	NA	&&	99.38	&	NA
						\\
						&&	400	&	99.59	&	NA	&&	99.38	&	NA\\
						&$\cS^c$&1	200	&	3.50	&	NA	&&	3.91	&	NA
						\\
						&&	400	&	3.09	&	NA	&&	2.06	&	NA\\
						&$\cS\cup \cS^c $&	200	&	100.00	&	36.87	&&	99.18	&	29.29
						\\
						&&	400	&	99.38	&	38.38	&&	99.38	&	28.28\\
						\hline
						\multicolumn{8}{c}{\bf{Empirical powers (\%) with $\bdelta=2\sqrt{{\log(p)}/{n}}\times(2^3,2^2,2^1,2^0,2^{-1})$.}}\\
						&&&	\multicolumn{2}{c}{Change point at $k^*=0.5n$}&&\multicolumn{2}{c}{Change point at $k^*=0.3n$}\\
						Model	&$\cG$&$p$&Boot-II&L\&B&&Boot-II&L\&B\\
						\hline
						$\bSigma=\Ib$	&$\cS$&	200	&	100.00	&	NA	&&	100.00	&	NA
						\\
						&&	400	&	100.00	&	NA	&&	100.00	&	NA\\
						&$\cS^c$&	200	&	2.47	&	NA	&&	1.65	&	NA
						\\
						&&	400	&	3.50	&	NA	&&	2.88	&	NA\\
						&$\cS\cup \cS^c $&	200	&	100.00	&	99.49	&&	100.00	&	98.48
						\\
						&&	400	&	100.00	&	100.00	&&	100.00	&	97.98\\
						\bottomrule[2pt]
					\end{tabular}		
				\end{center}
			\end{table}

			We next analyze the empirical powers. Denote the signal jump
			\begin{equation*}
				\bdelta=C\sqrt{\log(p)/n}\times\big(2^3,2^2,2^1,2^0,2^{-1}\big)^\top.
			\end{equation*}
			{We set $n=200$. We first generate $\bbeta^{(1)}$ with $s=5$ non-zero elements following $\text{U}(0,2)$ distributions out of $\cS=\{1,\ldots,50\}$. Then, we add $\bdelta$  with $C\in\{0.5,1,2\}$ on the  corresponding $5$ non-zero covariates of $\bbeta^{(1)}$ to generate $\bbeta^{(2)}$. Note that in this setting, $\bbeta^{(1)}$ and $\bbeta^{(2)}$ have a common support.}
			
			Table \ref{table: empirical powers model 1}  shows the power results with $n=200$, where various data dimensions, change point locations, candidate subgroups,  and signal strength are considered. Note that we do not report the results of SGL and Boot-I  because of their serious size distortions. According to Table \ref{table: empirical powers model 1}, we see that our proposed method can detect a change point with a very high probability {across various data dimensions} when the candidate subgroup has a change point ($\cG=\cS$ and $\cG=\cS\cup\cS^c$). Interestingly, it is shown that the powers in $\cS^c$ are close to the nominal level since the coefficients in $\cS^c$ are zeros before and after the change point. {As for L\&B, we see that it  can successfully detect a change point when the signal jump is relatively strong ($C=2$). However, L\&B is not very sensitive to weak signals with $C=0.5$ and $C=1$.} The above analysis suggests that our  proposed method is very powerful to sparse alternatives and is more efficient and flexible  than the existing methods for change point detection of high dimensional linear models. Moreover, Table \ref{table: empirical powers model 2} in the supplemental materials shows the power performance similar to Table \ref{table: empirical powers model 1} for Model 2 with banded covariance structures.
			


			\subsection{Multiple change point detection}\label{section: multiple cpt}
			

			So far, we have considered the numerical performance of single change point detection and identification. Next, we investigate multiple change points detection for Problem (\ref{hypothesis: H0 multiple}). In this numerical study, we consider two model settings:\\
			{\bf{Case~2: Alternatives with three change points.}} In this case, we set $n=600$ and $p=200$ with three change points at $k_1=180$, $k_2=300$, and $k_3=420$, respectively. The above three change points divide the data into four segments with different regression coefficients: $\bbeta^{(1)}$, $\bbeta^{(2)}$, $\bbeta^{(3)}$, and $\bbeta^{(4)}$. We first generate $\bbeta^{(1)}$ and $\bbeta^{(2)}$.
			The generating mechanism  for $\bbeta^{(1)}$ and $\bbeta^{(2)}$ is the same as  Case 1  in the single change point setting except that we use a signal jump 
			\begin{equation*}
				\bdelta'=C\sqrt{\dfrac{\log(p)}{n}}\Big(2^4,2^3,2^2,2^1,2^{0}\Big)^\top.
			\end{equation*}
			Then, we set $\bbeta^{(3)}=\bbeta^{(1)}$ and $\bbeta^{(4)}=\bbeta^{(2)}$. In this case, we set $C\in\{1.5,3\}$.\\
			{\bf{Case~3: Alternatives with four change points.}} In this case, we set $n=1000$ and $p=200$ with four change points at $k_1=300$, $k_2=450$, $k_3=550$, and $k_4=700$, respectively. The above four change points divide the data into five segments with different regression coefficients: $\bbeta^{(1)}, \ldots, \bbeta^{(5)}$. We first generate $\bbeta^{(1)}$ and $\bbeta^{(2)}$ as introduced in Case 2. Then, we set $\bbeta^{(3)}=\bbeta^{(1)}$, $\bbeta^{(4)}=\bbeta^{(2)}$ and $\bbeta^{(5)}=\bbeta^{(1)}$. In this case, we set $C\in\{2,4\}$.

			We use Algorithm \ref{alg:third} to detect and identify multiple change points and compare our methods with SGL, L\&B, and VPWBS. Note that Lee2016 is not applicable here because they only considered single change point detection. Moreover, to evaluate their performance,  we report the mean for the number of identified change points (Mean) and the mean {adjusted} Rand index  between the identified change points and the true change points (Adj.Rand) as well as its standard deviations (Sd.Adj.Rand). Note that the {adjusted} Rand index with a value belonging to [$-1$, 1] is well adopted for measuring the similarity between two data clusterings. The {adjusted} Rand index with a value being one means that the data clusterings are exactly the same.  The results are reported in Table \ref{table: multiple cpt identification}. For detecting the number of multiple change points, SGL tends to   overestimate the numbers across all model settings. This is consistent with our numerical studies in the size control in Section \ref{section: detection}. For L\&B, it has satisfactory performance when the signal jump is strong with $C=3$ or $C=4$. However, L\&B fails to detect all relevant three or four  change points when the signal-to-noise ratio is low by setting $C=1.5$ or $C=2$. This suggests that L\&B is not very sensitive to weak signals and  this observation is consistent with our previous power analysis in Section \ref{sec: powers}. As for  our proposed method, it can correctly detect the three (or four) change points on average even for a small signal jump. For identifying the change point locations, VPWBS has better performance than L\&B when the signal is weak and L\&B becomes very competitive as the signal becomes stronger. In most cases, the Arg-max based methods can estimate the locations with high accuracy and have better performance than their competitors. This is supported by the high Adj.Rand. 
			Finally, we would like to point out that for all methods, their performance becomes better when the model has a stronger signal jump. 
			
			In summary, as compared to the existing works, our bootstrap-assistant method is more efficient and accurate for detecting and identifying multiple change points. Moreover, it is able to detect the structural changes for any given subgroup and is very flexible to use.

			\begin{table}[h]
				\caption{Multiple change point detection results for {\bf{Models 1 and 2}} under {\bf{Case 2}}. The significance level is $\alpha=5\%$. The numerical results are   based on 100 replications.}
				\label{table: multiple cpt identification}
				\vspace{-0.5cm}
				\addtolength{\tabcolsep}{2pt}
				\begin{center}	
					\scriptsize
					\begin{tabular}{ccccc cccc}
						\toprule[2pt]
						\multicolumn{9}{c}{\bf{Multiple change points with $(n,p)=(600,200)$ and three change points at $(180,300,420)$}}\\
						&\multicolumn{4}{c}{$\bSigma=\Ib$}&&\multicolumn{3}{c}{$\bSigma=\bSigma^*$}\\
						$C$&Method&Mean&Adj.Rand &Sd.Adj.Rand&&Mean&Adj.Rand &Sd.Adj.Rand\\
						\hline
						&$\mathbf{\cG=\cS}$&&	&&&&&		\\
						$C=1.5$	&Arg-max&3.265 	&	0.947 	&	0.056 	&&	3.133 	&	0.952 	&	0.043 	\\
						&	L\&B	&	NA	&	NA	&	NA	&&	NA	&	NA	&	NA
						\\
						&	SGL	&	NA	&	NA	&	NA	&&	NA	&	NA	&	NA
						\\
						&	VPWBS	&	NA	&	NA	&	NA	&&	NA	&	NA	&	NA\\
						
						&&&&&&&&\\
						
						&$\mathbf{\cG=\cS\cup \cS^c}$&&	&&&&&		\\
						&Arg-max&3.177 	&	0.950 	&	0.048 	&&	2.983 	&	0.940 	&	0.045 	\\
						&	L\&B	&	1.000 	&	0.398 	&	0.013 	&&	1.133 	&	0.439 	&	0.148 
						\\
						&	SGL	&	4.000 	&	0.722 	&	0.111 	&&	5.417 	&	0.753 	&	0.083 
						\\
						&	VPWBS	&	2.857 	&	0.899 	&	0.133 	&&	2.949 	&	0.918 	&	0.086\\ 	
						\cline{2-9}						
						&$\mathbf{\cG=\cS}$&&	&&&&&		\\
						$C=3$&Arg-max&3.112 	&	0.967 	&	0.034 	&&	3.200 	&	0.955 	&	0.049 	\\
						&	L\&B	&	NA	&	NA	&	NA	&&	NA	&	NA	&	NA
						\\
						&	SGL	&	NA	&	NA	&	NA	&&	NA	&	NA	&	NA
						\\
						&	VPWBS	&	NA	&	NA	&	NA	&&	NA	&	NA	&	NA\\
						&&&&&&&&\\
						
						&$\mathbf{\cG=\cS\cup \cS^c}$&&	&&&&&		\\
						&Arg-max&3.104 	&	0.968 	&	0.032 	&&	3.250 	&	0.951 	&	0.035 	\\
						&	L\&B	&	3.000 	&	0.991 	&	0.006 	&&	3.000 	&	0.992 	&	0.007 
						\\
						&	SGL	&	7.000 	&	0.767 	&	0.093 	&&	8.000 	&	0.873 	&	0.118 
						\\
						&	VPWBS	&	2.878 	&	0.945 	&	0.066 	&&	2.898 	&	0.944 	&	0.060\\ 
						\bottomrule[2pt]
					\end{tabular}		
				\end{center}
			\end{table}

			\begin{table}[h]
				\caption{Multiple change point detection results for {\bf{Models 1 and 2}} under {\bf{Case 3}}. The significance level is $\alpha=5\%$. The numerical results are   based on 100 replications.}
				\label{table: multiple cpt identification four cpts}
				\vspace{-0.1cm}
				\addtolength{\tabcolsep}{1pt}
				\begin{center}	
					\scriptsize
					\begin{tabular}{ccccc cccc}
						\toprule[2pt]			
						\multicolumn{9}{c}{\bf{Multiple change points with $(n,p)=(1000,200)$ and four change points at $(300,450,550,700)$}}\\
						&\multicolumn{4}{c}{$\bSigma=\Ib$}&&\multicolumn{3}{c}{$\bSigma=\bSigma^*$}\\
						$C$&Method&Mean&Adj.Rand &Sd.Adj.Rand&&Mean&Adj.Rand &Sd.Adj.Rand\\
						\hline
						&$\mathbf{\cG=\cS}$&&	&&&&&		\\
						$C=2$	&Arg-max&4.100 	&	0.967 	&	0.047 	&&	4.183 	&	0.968 	&	0.036 	\\
						&	L\&B	&	NA	&	NA	&	NA	&&	NA	&	NA	&	NA
						\\
						&	SGL	&	NA	&	NA	&	NA	&&	NA	&	NA	&	NA
						\\
						&	VPWBS	&	NA	&	NA	&	NA	&&	NA	&	NA	&	NA\\
						
						&&&&&&&&\\
						
						&$\mathbf{\cG=\cS\cup\cS^c}$&&	&&&&&		\\
						&Arg-max&4.067 	&	0.949 	&	0.052 	&&	4.200 	&	0.961 	&	0.044 	\\
						&	L\&B	&	1.600 	&	0.589 	&	0.296 	&&	1.867 	&	0.688 	&	0.185 
						\\
						&	SGL	&	6.167 	&	0.664 	&	0.054 	&&	6.500 	&	0.708 	&	0.104 
						\\
						&	VPWBS	&	3.296 	&	0.882 	&	0.093 	&&	3.276 	&	0.882 	&	0.106 \\
						\cline{2-9}
						&$\mathbf{\cG=\cS}$&&	&&&&&		\\
						$C=4$&Arg-max&4.150 	&	0.971 	&	0.031 	&&	4.067 	&	0.968 	&	0.029 	\\
						&	L\&B	&	NA	&	NA	&	NA	&&	NA	&	NA	&	NA
						\\
						&	SGL	&	NA	&	NA	&	NA	&&	NA	&	NA	&	NA
						\\
						&	VPWBS	&	NA	&	NA	&	NA	&&	NA	&	NA	&	NA\\
						&&&&&&&&\\
						&$\mathbf{\cG=\cS\cup \cS^c}$&&	&&&&&		\\
						&Arg-max&4.050 	&	0.979 	&	0.026 	&&	4.183 	&	0.967 	&	0.040 \\
						&	L\&B	&	3.956 	&	0.988 	&	0.038 	&&	4.000 	&	0.994 	&	0.004 
						\\
						&	SGL	&	8.833 	&	0.799 	&	0.111 	&&	8.583 	&	0.807 	&	0.112 
						\\
						&	VPWBS	&	3.520 	&	0.932 	&	0.052 	&&	3.592 	&	0.939 	&	0.046 	\\				
						\bottomrule[2pt]
					\end{tabular}		
				\end{center}
			\end{table}

	\section{Application to Alzheimer’s Disease Data Analysis }\label{sec: real data application}
In this section, we apply our proposed method to analyze data from the Alzheimer’s Disease Neuroimaging Initiative (\url{http://adni.loni.usc.edu/}). It is known that AD accounts for most forms of dementia characterized by progressive cognitive and memory deficits. This makes  it a very important health issue which attracts a lot of scientific attentions in recent years. To study AD, Mini-Mental State Examination (MMSE) \citep{Folstein1975} is a 30-point questionnaire that is commonly used to measure cognitive impairment. According to MMSE, any score of 24 or more (out of 30) indicates a normal cognition. Below this, scores can indicate severe ($\leq $9 points), moderate (10–18 points) or mild (19–23 points) cognitive impairment. Because of  the strong relationship between the MMSE score and AD, it can be interesting and useful to predict the MMSE score  using some biomarkers for  diagnosing the current disease status of AD as well as to identify important predictive biomarkers. According to previous studies \citep{2016Sparse,yu2019}, structural magnetic resonance imaging (MRI) data are very useful  for the prediction of the MMSE score. However, these studies typically ignored the effect of other covariates such as ages, education years, or genders on the linear models. Hence, an interesting question is whether there is a change point in the linear structure between the MMSE score and MRI data due to some other covariates. If a change point exists, we would like to identify the location of the change point.  To answer these questions, we use our proposed change point detection method to address these issues. We focus on the covariate age which is of particular interest in AD studies.
We obtain the dataset for our analysis from the ADNI database. After proper image preprocessing steps such as anterior commissure posterior commissure correction and intensity inhomogeneity correction, we obtain the final dataset with 410 subjects  with 225  normal controls and 185 AD patients. For each subject with known age, there is one MMSE score and 93 MRI features corresponding to 93 manually labeled regions of interest (ROI) \citep{Zhang2012}. We treat  the MMSE score as the response variable and MRI features as predictors in our model. The dataset is first scaled to have mean 0 and variance 1 for the MMSE score and each MRI feature. We are interested in detecting a change point in the linear structure due to the change of ages.  Considering potential effect variations of different samples, 
we randomly select 370 subjects from the whole 410 subjects according to the empirical distribution of ages shown in Figure \ref{figure: age} (left)  as the training data and use the remaining 40 subjects as the testing data. Then, we sort the training subjects by the value of ages and use our proposed method to detect and identify a change point in the covariate age. 
We repeat the above process for 50 times. As a comparison, for each  random split, we also use lasso to select variables on the training data via 10-fold cross-validation. For this study, we set the significance level at 5\%. The number of bootstraps  is 200.

\begin{figure}[!h]
	\begin{center}
		\includegraphics[width=11cm]{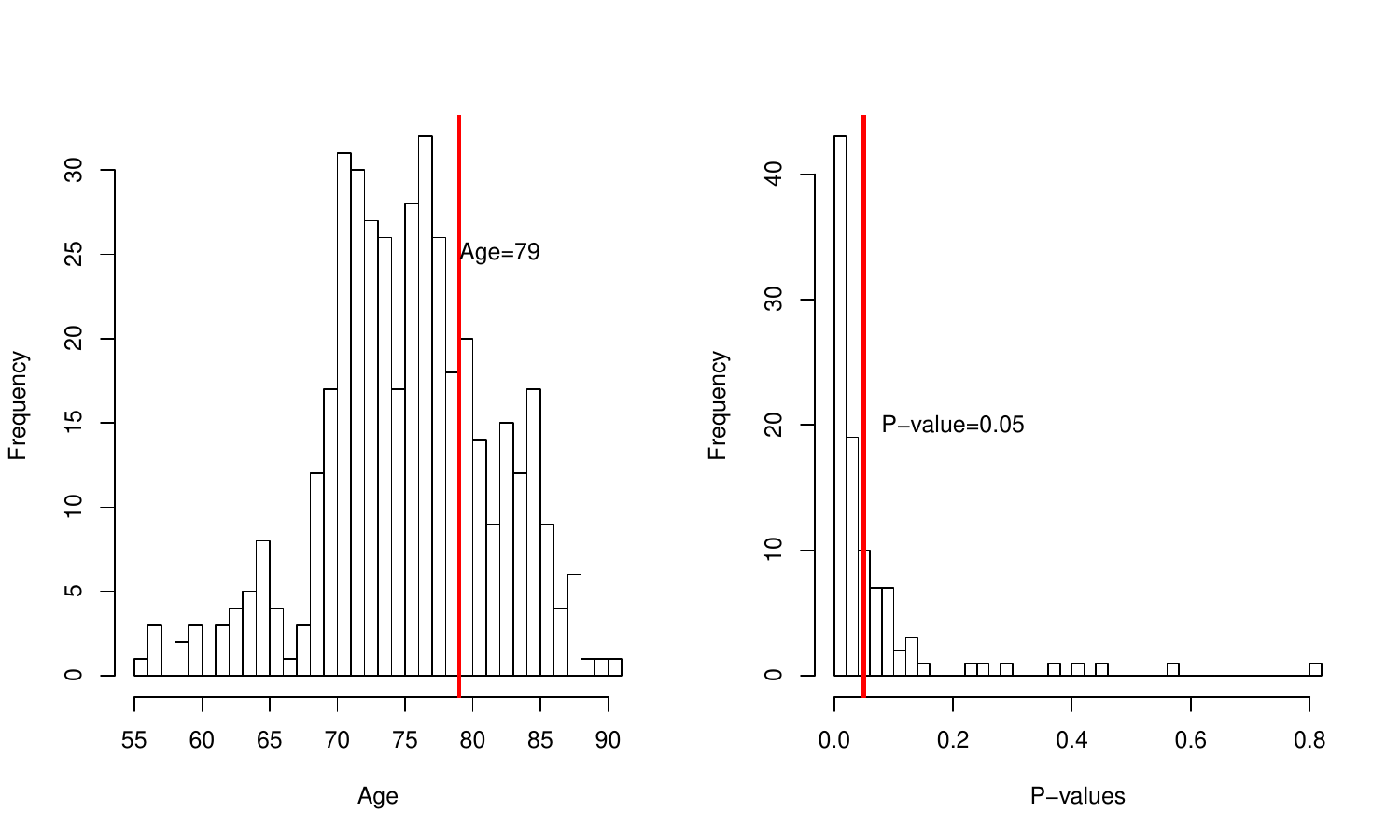}
		\vspace{-0.5cm}
		\caption{Left: Distribution of ages among the 410 subjects. Right:  Empirical $p$-values for change point detection out of 50 random data splits.}
		\label{figure: age}
	\end{center}
\end{figure}

Figure \ref{figure: age} (right) shows the  empirical $p$-values for the 50 random data splits.  Based on our results, 82\% of the random splits with an estimated $p$-value lower than 0.05 have detected a change point. This strongly suggests that there is a change point in the linear structure due to the covariate age. Moreover, for the above 82\% random splits, we record the estimated change points in  Figure \ref{figure: cpt adni}. We can see that in most cases,  the argmax-based  estimator identify the change point at the age of 79. The above analysis indicates that the linear structure between the MMSE score and MRI may be different before  and after the age of 79. To see this more clearly, among the random splits with a change point, Figure \ref{figure: varible selection adni} reports the features (with estimated coefficients bigger than 0.01) which are  selected for more than 80\% times before and after the change point, respectively. There are 16 features selected before the change point and 6 features  selected after the change point. In other words, those 16 features shown in Figure \ref{figure: varible selection adni} (left) are very predictive for the MMSE score for people with an age smaller or equal to 79. Once 
the age exceeds 79, it is better to predict the MMSE score using the other 6 features in Figure \ref{figure: varible selection adni} (right). To verify this, for those random splits with a change point, we calculate the mean squared error for the corresponding testing data, based on the selected models using the training data.
Figure \ref{figure: rmse adni} shows the results of our proposed method and lasso.  We can see that our proposed method has better prediction performance by segmenting the model by the covariate age, with about 5.34\% lower averaged MSE than that of  lasso.

Lastly, as for the selected variables, some interesting observations  can be made. For example,  ROI 83 is predictive for the MMSE score across all ages. ROIs 30 and 69 are only very predictive for the MMSE score under the age of 79 and above 79, respectively. 
It is known that the  83th ROI corresponds to the amygdala region, and the 30th and 69th features correspond to the hippocampal regions. According to many previous studies \citep{Zhang2012}, 
those regions are known to be related to AD based on group comparison methods. For these and other selected features, it would be very interesting to investigate their relationship with AD 
by some group comparison studies according to the segmentation of ages.

\begin{figure}[!h]
	\begin{center}
		\includegraphics[width=6cm]{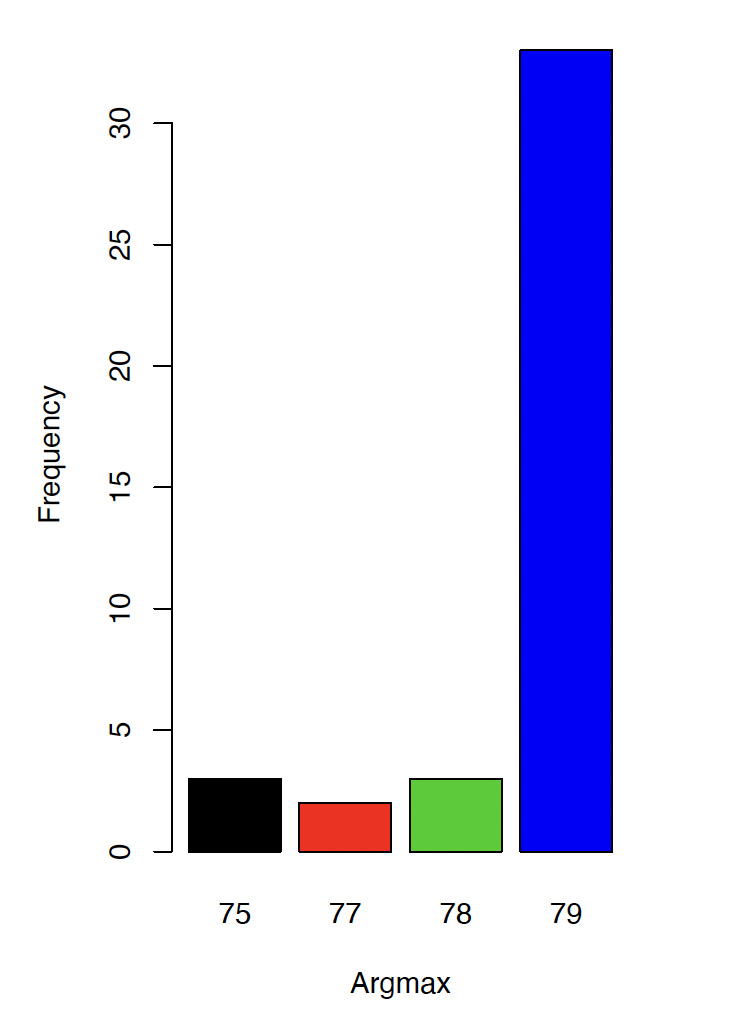}
		\vspace{-0.5cm}
		\caption{Estimated change points for the 82\% random splits with change points among the 50 replications. }
		\label{figure: cpt adni}
	\end{center}
\end{figure}

\begin{figure}[!hpt]
	\begin{center}
		\includegraphics[width=12.8cm]{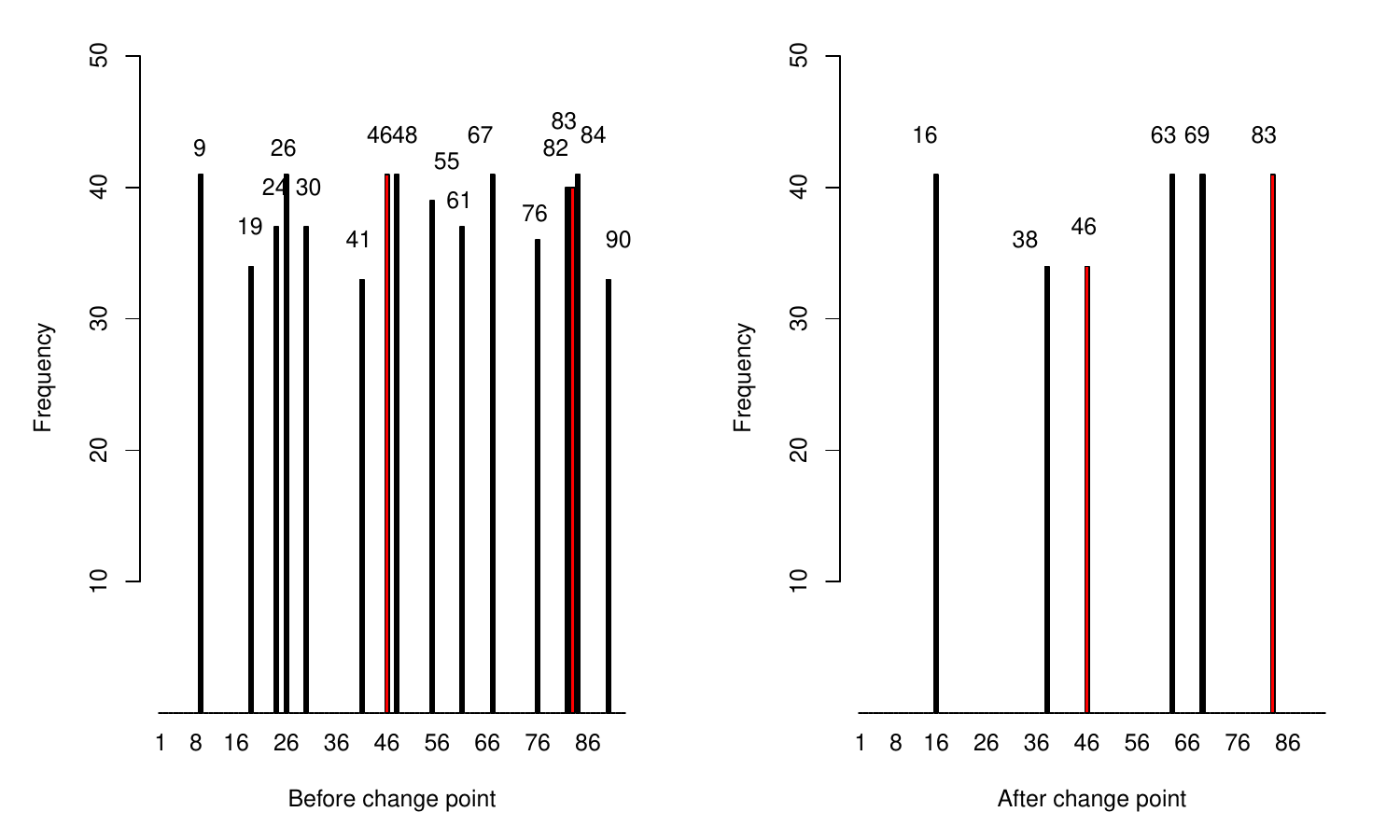}
		\vspace{-0.4cm}
		\caption{Frequency of features selected before the change point (left) and after the change point (right) for the ADNI data out of  50 random splits. Red corresponds to the features   that are selected both before and after the change point.}
		\label{figure: varible selection adni}
	\end{center}
\end{figure}

\begin{figure}[!hpt]
	\begin{center}
		\includegraphics[width=7cm]{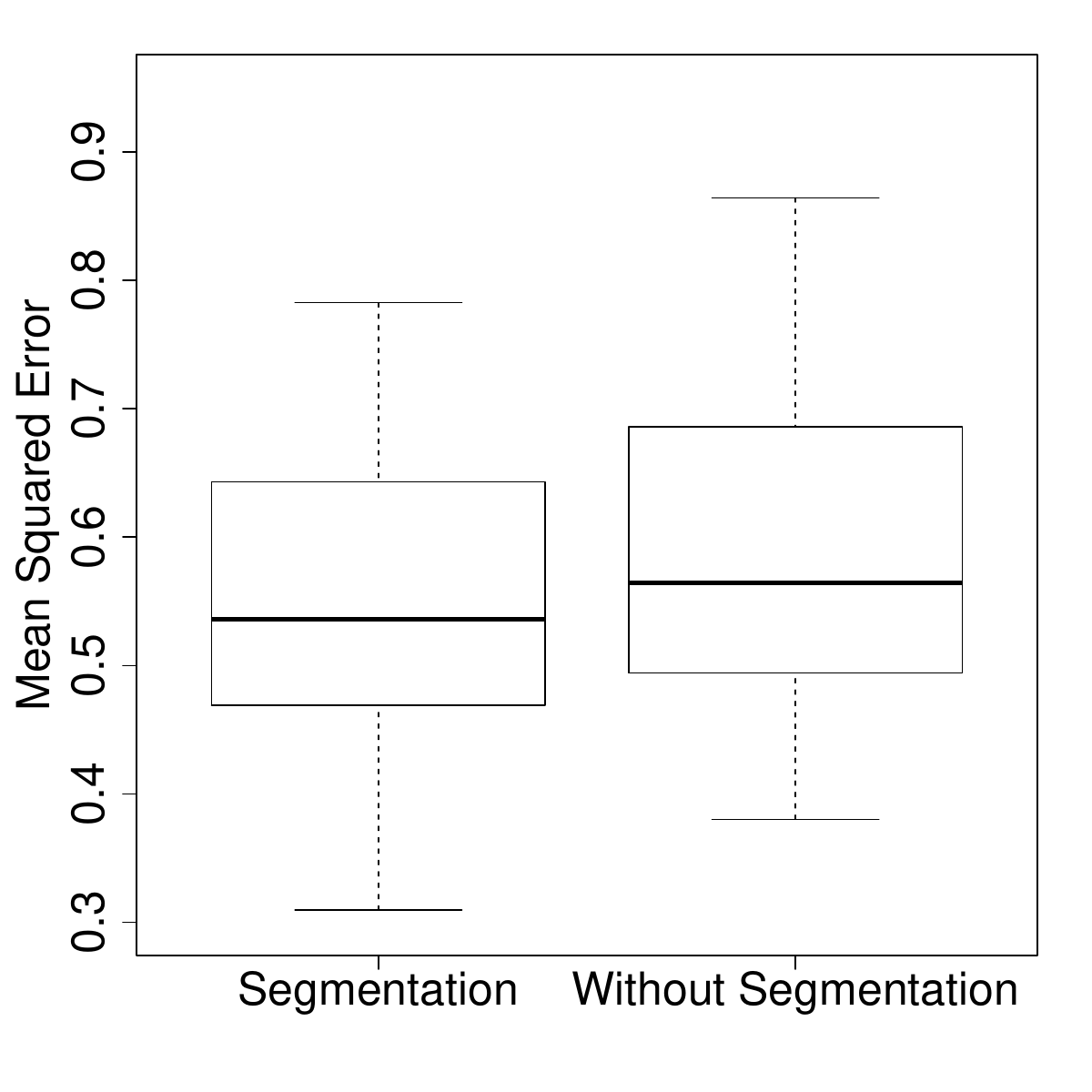}
		\vspace{-0.7cm}
		\caption{Mean squared errors for the prediction of the MMSE score with and without change point models.
		}
		\label{figure: rmse adni}
	\end{center}
\end{figure}

			\section{Conclusions}\label{section: summary}
			In this paper, we propose a new method for change point inference in the context of high dimensional linear models. For any given subgroup $\cG\subset\{1,\ldots,p\}$,  a $L_{\infty}$-norm-based test statistic $T_{\cG}$ is constructed for testing the homogeneity of regression coefficients across the observations. To approximate its  limiting null distribution, a novel multiplier bootstrap procedure is introduced. Our new method is powerful against sparse alternatives with only a few entries in $\cG$ having a change point, and allows the group size $|\cG|$ {to} grow exponentially with the sample size $n$.
			As for the change point identification, 
			a new change point estimator is obtained by taking ``argmax" of the $L_{\infty}$-aggregated process $H_{\cG}(t)$. Theoretically,  the change point estimator is shown to be consistent,  allowing the overall sparsity $s$ of regression coefficients and the group size $|\cG|$ {to} grow simultaneously with the sample size $n$.  In addition to single change point detection, we further combine our proposed method with the binary segmentation-based technique for detecting and identifying multiple change points. Our new testing method is relatively easy to implement and is justified via extensive numerical studies. 

		\vskip 0.2in
		{
				\bibliographystyle{ims}
			\bibliography{RefDatBas_Reg}
		}

\renewcommand{\baselinestretch}{1}
\setcounter{footnote}{0}
\clearpage
\setcounter{page}{1}
\title{
	\begin{center}
		\Large Supplementary Materials to
		``Simultaneous Change Point Detection and Identification  for High Dimensional Linear Models''
	\end{center}
}
\date{}
\begin{center}
	\author{
		Bin Liu
		\footnotemark[1],
		Xinsheng Zhang
		\footnotemark[1]
		Yufeng Liu
		\footnotemark[3]
	}
	\renewcommand{\thefootnote}{\fnsymbol{footnote}}
		\footnotetext[1]{Department of Statistics and Data Science, School of Management at Fudan University, China; e-mail:{\tt liubin0145@gmail.com; xszhang@fudan.edu.cn}}
		\footnotetext[2]{Department of Statistics and Operations Research, Department of Genetics, and Department of Biostatistics, University of North Carolina at Chapel Hill, U.S.A; e-mail:{\tt yfliu@email.unc.edu }}
\end{center}
\maketitle
\appendix

The Appendix provides detailed proofs and additional results of the main paper. In Section \ref{section: some notations}, we introduce some additional notations.
In Section \ref{sec: Additional numerical results}, some additional numerical results, including size, power as well as detecting multiple change points, are provided. 
In Section \ref{section: useful lemmas}, some useful lemmas are provided. In Section \ref{section: proofs of main results}, we give the detailed proofs of theoretical results in the main paper. In Sections \ref{section: proofs of lemmas in the main proof} and \ref{section: proof of useful lemmas}, we prove the useful lemmas in Section \ref{section: useful lemmas} as well as the  lemmas used in Section \ref{section: proofs of main results}.

\section{Some notations}\label{section: some notations}
\indent Under $\Hb_0$, we set $\bbeta^{(0)}:=\bbeta^{(1)}=\bbeta^{(2)}$ and $s^{(0)}:=s^{(1)}=s^{(2)}$. We set $s:=s^{(1)}\vee s^{(2)}$. For a given subgroup $\cG$, set $\Pi_{\cG}=\{j\in\cG: \beta^{(1)}_j-\beta^{(2)}_j\neq 0\}$ as the subset of coordinates with a change point. 
For a vector $\bv\in \RR^p$, we set $\cM(\bv)$ as the number of non-zero elements of $\bv$, i.e. $\cM(\bv)=\sum_{j=1}^{p}\mathbf{1}\{v_j\neq 0\}$. We denote $J(\bv)=\{1\leq j\leq p: v_j\neq 0\}$ as the set of non-zero elements of $\bv$.  For a set $J$ and $\bv\in \RR^p$, denote $\bv_J$ as the vector in $\RR^p$ that has the same coordinates as $\bv$ on $J$ and zero coordinates on the complement $J^c$ of $J$. 
Denote $\cX=\{\Xb,\bY\}$. 
We use $C_1, C_2,\ldots$ to denote constants that may vary from line to line.

\section{Additional numerical results}\label{sec: Additional numerical results}

\subsection{Implementations of the existing techniques}
\indent {Before reporting additional numerical results, we first demonstrate how to implement the mentioned techniques in this paper. }

{\bf{Implementation of the existing methods:}} 
For Lee2016, we use the {\bf{package-glmnet}} to implement their proposed algorithm. Note that Lee2016 involves a selection of the tuning parameter $\lambda$. For each replication, we generate a sequence from $2^{-5}$ to $2^5$ and select the ``best" $\lambda$ by 10-fold cross-validation. 
{For L\&B, we use the binary segmentation-based method with parameters suggested by the authors using the {\bf{package-glmnet}}. Moreover, we use a three folded cross-validation procedure to select the tuning parameters in L\&B. For VPWBS, we implement the algorithm using the codes provided by the authors at GitHub (https://github.com/darenwang/VPBS).} {For SGL, we use the {\bf{package-SGL}} with parameters in favor of their method and use three folded cross-validation to select the  tuning parameters. Note that SGL solves the following optimization problem:}
\begin{equation*}
	\begin{array}{ll}
		\{\hat{\bbeta}_1,\ldots,\hat{\bbeta}_n\}\\
		=	\argmin\limits_{\bbeta_1,\ldots,\bbeta_n\in \RR^p}\sum\limits_{i=1}^n(Y_i-\bX_i^\top\bbeta_i)^2+\lambda_n\alpha\sum\limits_{i=1}^n\|\bbeta_i-\bbeta_{i-1}\|_2+\lambda_n(1-\alpha)\sum\limits_{i=1}^n\|\bbeta_i-\bbeta_{i-1}\|_1.
	\end{array}
\end{equation*}
{Based on the above optimization, SGL finds a change point at $i^*$ if $\hat{\bbeta}_{i^*}-\hat{\bbeta}_{i^*-1}\neq \mathbf{0}$. 
	It is well-known that lasso tends to over select the variables. In addition, SGL essentially solves a group lasso problem by calculating $n\times p$ parameters using only $n$ observations. As a result, SGL may yield false alarms by identifying some $\{i:\bbeta_{i}-\bbeta_{i-1}=\mathbf{0}_p\}$ as a change point.
	This can be seen by our following empirical size performance in Section \ref{section: detection} as well as the multiple change point detection results in Section \ref{section: multiple cpt}. Moreover, we note that this phenomenon was also observed by \cite{Wangdaren2021}.}

{\bf{Implementation of our method:}} As for our proposed method, we use the {\bf{package-hdi}} to obtain the node-wise lasso estimator $\hat{\bTheta}$. {Note that the calculation of the  lasso processes $\hat{\bbeta}^{(0,t)}$ and $\hat{\bbeta}^{(t,1)}$  with $t\in[\tau_0,1-\tau_0]$ involves the selection of tuning parameters  $\lambda_1(t)$ and $\lambda_2(t)$ defined in (\ref{equation: lasso estimation before and after cpt}). We select the tuning parameters via three folded cross-validation. Specifically, for each search location $t\in[\tau_0,1-\tau_0]$, we set
	\begin{equation*}
		\lambda_1(t)=C\sqrt{\dfrac{\log(p)}{\floor{nt}}},~~\text{and}~~\lambda_2(t)=C\sqrt{\dfrac{\log(p)}{\floor{nt}^*}},~~\text{with}~C\in\{1,2,\ldots,8\}.
	\end{equation*}
	Then, we  use the {\bf{package-glmnet}} to select the best ``C" via  three folded cross-validation, which enjoys satisfactory performance in change point detection and identification.}

\subsection{Additional size performance}
In addition to	$N(0,1)$, we also report the size performance under standardized $\rm Gamma(4,1)$ (Table \ref{table: empirical sizes gamma}) and Student's $t_5$ (Table \ref{table: empirical sizes t5}) distributions which have very similar performance to Table \ref{table: empirical sizes} of the main paper. In this case, our proposed method can control the size under the nominal level. This suggests that the  bootstrap null distribution is correctly calibrated even for non-normal underlying errors.

\subsection{Additional power performance}

Table \ref{table: empirical powers model 2} shows the power performance for Model 2 with banded covariance structures of $\bX$, which is similar to Table \ref{table: empirical powers model 1} in the main paper.



	\subsection{{Computational cost}}
In this section, we compare the computational cost of the existing methods. In theory, for detecting a single change point, the computational costs for the existing methods are
$O(n\text{Lasso}(n,p))$ (Lee2016), $O(n\text{Lasso}(n,p))$ (L\&B), $O(Mn\text{GroupLasso}(n,p))$ (VPWBS),  $O(\text{GroupLasso}(n,np))$ (SGL), and $(B+1)O(n\text{Lasso}(n,p))$ (our proposed method), where $\text{Lasso}(n,p)$ and $\text{GroupLasso}(n,p)$ denote the computational cost for solving lasso and group lasso problems  with the sample size $n$ and the data dimension $p$, $M$ is the number of random intervals in \cite{Wangdaren2021}, and $B$ is the number of bootstrap replications. Empirically, we implement the  corresponding program  independently on a {CPU (Linux) with 2.50GHz and 256G RAM and report the average computational time (seconds) based on 5 replications. Note that the computational cost for our proposed method 
	mainly relies on the bootstrap procedure which can be time-consuming. Since the $B$ bootstrap replications can be done separately, we can use parallel computation in modern computer techniques  to  further reduce the computational time via implementing the $B$ bootstrap replications in a parallel fashion on different cores of the Linux server. Specifically, for our method, we report the computational cost by using 8, 16, and 32 logical cores, respectively. Figure \ref{figure: time} reports the computational time  for the existing methods with various $n\in\{200,400,600,800,1000\}$ (upper)  and $p\in\{100,200,300,400\}$ (bottom). In general, Lee2016 and L\&B are the most efficient and have very close performance. The computational time for SGL is the most expensive among all methods. For our proposed algorithm, we can see that it has a tolerable computational cost and can even be comparable to  its competitors using more cores.  Lastly, Figure \ref{figure: time} shows that for all methods, the computational 
	time grows linearly  with $n$ and $p$, and it appears that the computational cost is more sensitive to the growth of the sample size $n$ than the data dimension $p$. 
	
	\begin{figure}[!h]
		\begin{center}
			\includegraphics[width=17cm]{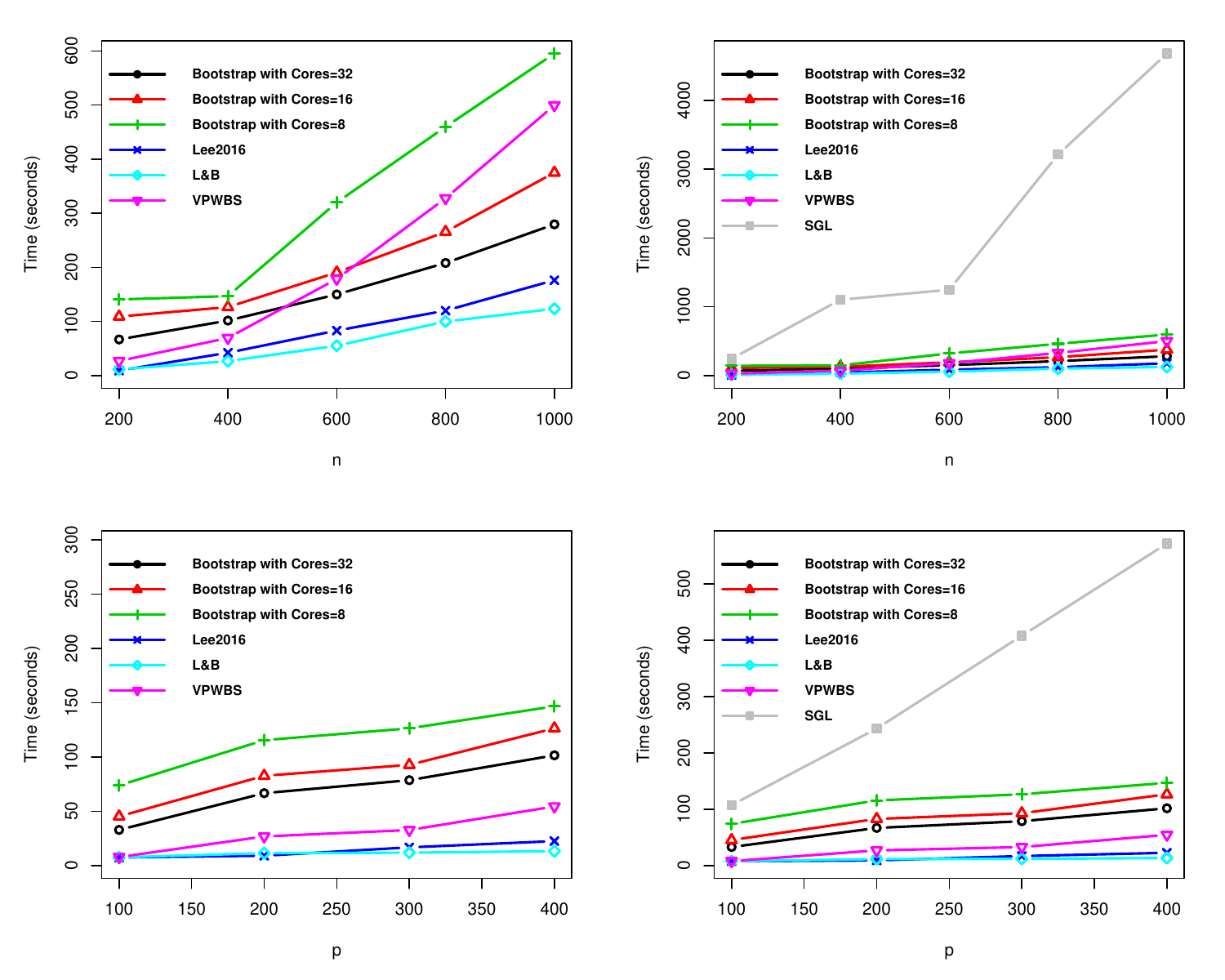}
			\vspace{-0.45cm}
			\caption{{Computational time (seconds) for the existing methods based on an average of 5 replications. Upper: Computational time for $p=200$ and   $n\in\{200,400,600,800,1000\}$  without  the plot of SGL (left)  and  with SGL (right), respectively. Bottom: Computational time for $n=200$ and   $p\in\{100,200,300,400\}$  without  the plot of SGL (left)  and  with SGL (right), respectively.}}
			\label{figure: time}
		\end{center}
	\end{figure}

	\clearpage
	\begin{table}[h]
		\caption{Empirical sizes for {\bf{Models 1 and 2}}  under various combinations of $(n,p,s)$. The errors are generated from standardized {\bf{Gamma(4,1)}}  distributions. The results are based on 2000 replications.}
		\label{table: empirical sizes gamma}
		\vspace{-0.1cm}
		\addtolength{\tabcolsep}{-4pt}
		\begin{center}
			\begin{tabular}{ccccc cccc}
				\toprule[2pt]
				\multicolumn{9}{c}{\bf{Empirical sizes (\%) for Gamma(4,1) with $(n,s)=(200,5)$}}\\
				Model	&$\cG$&$p$&Boot-I ($\alpha=1\%$)&Boot-II ($\alpha=1\%$)&Boot-I ($\alpha=5\%$)&Boot-II ($\alpha=5\%$)&SGL&L\&B\\		
				\hline
				&&&&&&&&\\
				$\bSigma=\Ib$	&$\cS$&100&7.00	&	1.70&		14.81	&	4.63&		NA	&	NA\\	
				&&	200	&	8.64	&	1.29	&	17.70	&	4.32	&	NA	&	NA
				\\
				&&	300	&	9.67	&	2.11	&	16.67	&	5.14	&	NA	&	NA
				\\
				&&	400	&	13.99	&	1.80	&	23.66	&	5.14	&	NA	&	NA\\	
				&$\cS^c$&100&4.32	&	0.98	&	9.67	&	3.60	&	NA	&	NA\\	
				&&	200	&	6.38	&	1.23	&	15.02	&	3.81	&	NA	&	NA
				\\
				&&	300	&	11.11	&	1.08	&	20.99	&	3.86	&	NA	&	NA
				\\
				&&	400	&	13.58	&	1.80	&	24.90	&	4.27	&	NA	&	NA\\
				&$\cS\cup \cS^c$&100&6.17	&	1.49	&	15.43	&	4.73	&	56.67	&	0.00\\	
				&&	200	&	9.05	&	1.54	&	17.28	&	3.96	&	43.33	&	0.00
				\\
				&&	300	&	10.91	&	1.44	&	23.25	&	4.22	&	40.00	&	0.00
				\\
				&&	400	&	18.31	&	2.11	&	30.66	&	4.94	&	40.00	&	0.00	\\
				\cline{3-9}		
				&&&&&&&&\\
				$\bSigma=\bSigma^*$	&$\cS$&100&4.94	&	1.92	&	11.73	&	4.87	&	NA	&	NA\\	
				&&	200	&	6.79	&	1.58	&	15.64	&	4.46	&	NA	&	NA
				\\
				&&	300	&	8.23	&	2.12	&	17.90	&	5.81	&	NA	&	NA
				\\
				&&	400	&	12.55	&	2.06	&	24.07	&	4.65	&	NA	&	NA\\
				&$\cS^c$&100&3.91	&	1.44	&	10.08	&	4.03	&	NA	&	NA\\	
				&&	200	&	3.70	&	1.57	&	10.29	&	3.82	&	NA	&	NA
				\\
				&&	300	&	7.61	&	1.30	&	14.61	&	3.69	&	NA	&	NA
				\\
				&&	400	&	4.73	&	0.89	&	15.84	&	2.73	&	NA	&	NA\\
				&$\cS\cup \cS^c$&100&8.64	&	1.36	&	16.87	&	3.35	&	51.11	&	0.00\\	
				&&	200	&	7.00	&	1.37	&	12.96	&	3.14	&	40.00	&	0.00
				\\
				&&	300	&	8.02	&	1.36	&	19.55	&	3.14	&	50.00	&	0.00
				\\
				&&	400	&	7.20	&	1.16	&	15.02	&	3.76	&	37.78	&	0.00\\
				\bottomrule[2pt]
			\end{tabular}		
		\end{center}
	\end{table}

	\begin{table}[h]
		\caption{Empirical sizes for {\bf{Models 1 and 2}}  under various combinations of $(n,p,s)$. The errors are generated from standardized {\bf{Student's $t_5$}} distributions. The results are based on 2000 replications.}
		\label{table: empirical sizes t5}
		\vspace{-0.1cm}
		\addtolength{\tabcolsep}{-4pt}
		\begin{center}
			\begin{tabular}{ccccc cccc}
				\toprule[2pt]
				\multicolumn{9}{c}{\bf{Empirical sizes (\%) for Student's $t_5$ with $(n,s)=(200,5)$}}\\
				Model	&$\cG$&$p$&Boot-I ($\alpha=1\%$)&Boot-II ($\alpha=1\%$)&Boot-I ($\alpha=5\%$)&Boot-II ($\alpha=5\%$)&SGL&L\&B\\		
				\hline
				&&&&&&&&\\
				$\bSigma=\Ib$	&$\cS$&100&5.35	&	1.29	&	15.23	&	4.17	&	NA	&	NA\\	
				&&	200	&	9.26	&	1.95	&	21.40	&	5.61	&	NA	&	NA
				\\
				&&	300	&	9.05	&	1.95	&	20.16	&	5.30	&	NA	&	NA
				\\
				&&	400	&	14.40	&	2.37	&	22.84	&	6.43	&	NA	&	NA\\	
				&$\cS^c$&100&5.97	&	1.18	&	10.29	&	4.22	&	NA	&	NA\\	
				&&	200	&	9.67	&	1.59	&	20.99	&	4.42	&	NA	&	NA\\
				&&	300	&	10.70	&	2.16	&	22.22	&	4.78	&	NA	&	NA\\
				&&	400	&	11.93	&	1.85	&	21.60	&	4.48	&	NA	&	NA\\
				&$\cS\cup \cS^c$&100&7.20	&	1.65	&	16.05	&	4.63	&	61.11	&	0.00\\	
				&&	200	&	10.29	&	1.80	&	20.78	&	4.68	&	45.56	&	0.00\\
				&&	300	&	12.76	&	1.75	&	26.13	&	5.20	&	50.00	&	0.00\\
				&&	400	&	16.46	&	2.42	&	30.45	&	5.04	&	54.44	&	0.00\\
				\cline{3-9}		
				&&&&&&&&\\
				$\bSigma=\bSigma^*$	&$\cS$&100&6.17	&	1.33	&	13.58	&	3.90	&	NA	&	NA\\	
				&&	200	&	9.05	&	1.89	&	18.31	&	5.38	&	NA	&	NA
				\\
				&&	300	&	9.05	&	2.72	&	18.31	&	5.78	&	NA	&	NA
				\\
				&&	400	&	10.91	&	2.04	&	21.19	&	5.32	&	NA	&	NA	\\
				&$\cS^c$&100&4.53	&	1.48	&	10.29	&	4.35	&	NA	&	NA\\	
				&&	200	&	3.91	&	1.64	&	10.08	&	4.26	&	NA	&	NA
				\\
				&&	300	&	6.79	&	1.44	&	14.40	&	3.65	&	NA	&	NA
				\\
				&&	400	&	7.61	&	1.80	&	16.46	&	4.41	&	NA	&	NA\\
				&$\cS\cup \cS^c$&100&6.79	&	1.64	&	13.58	&	5.19	&	51.11	&	0.00\\	
				&&	200	&	5.14	&	1.59	&	12.96	&	4.41	&	44.44	&	0.00
				\\
				&&	300	&	9.26	&	2.10	&	18.11	&	4.87	&	31.11	&	0.00
				\\
				&&	400	&	9.47	&	2.05	&	18.93	&	4.87	&	36.67	&	0.00\\	
				\bottomrule[2pt]
			\end{tabular}		
		\end{center}
	\end{table}

	\begin{table}[h]
		\caption{Empirical powers (\%) for {\bf{Case 1 under Model 2}} with various dimensions, candidate subgroups, and change point locations. The sample size is $n=200$. The significance level is $\alpha=5\%$. The numerical results are  based on 2000 replications.}
		\label{table: empirical powers model 2}
		\vspace{-0.2cm}
		\addtolength{\tabcolsep}{4pt}
		\begin{center}
			\begin{tabular}{ccccc ccc  }
				\toprule[2pt]
				\multicolumn{8}{c}{\bf{Empirical powers (\%) with $\bdelta=0.5\sqrt{{\log(p)}/{n}}\times(2^3,2^2,2^1,2^0,2^{-1})$.}}\\
				&&&	\multicolumn{2}{c}{Change point at $k^*=0.5n$}&&\multicolumn{2}{c}{Change point at $k^*=0.3n$}\\
				Model	&$\cG$&$p$&Boot-II&L\&B&&Boot-II&L\&B\\
				\hline
				$\bSigma=\bSigma^*$	&$\cS$&	200	&	49.33	&	NA	&&	30.27	&	NA
				\\
				&&	400	&	45.33	&	NA	&&	33.33	&	NA\\
				&$\cS^c$&	200	&	1.67	&	NA	&&	3.00	&	NA
				\\
				&&	400	&	2.67	&	NA	&&	1.83	&	NA\\
				&$\cS\cup \cS^c $&	200	&	34.00	&	0.00	&&	21.43	&	0.00
				\\
				&&	400	&	28.00	&	0.00	&&	18.67	&	0.00\\
				\hline
				\multicolumn{8}{c}{\bf{Empirical powers (\%) with $\bdelta=\sqrt{{\log(p)}/{n}}\times(2^3,2^2,2^1,2^0,2^{-1})$.}}\\
				&&&	\multicolumn{2}{c}{Change point at $k^*=0.5n$}&&\multicolumn{2}{c}{Change point at $k^*=0.3n$}\\
				Model	&$\cG$&$p$&Boot-II&L\&B&&Boot-II&L\&B\\
				\hline
				$\bSigma=\bSigma^*$		&$\cS$&	200	&	100.00	&	NA	&&	99.18	&	NA
				\\
				&&	400	&	100.00	&	NA	&&	99.18	&	NA\\
				&$\cS^c$&	200	&	2.06	&	NA	&&	2.67	&	NA
				\\
				&&	400	&	2.06	&	NA	&&	1.65	&	NA\\
				&$\cS\cup \cS^c $&	200	&	99.59	&	60.42	&&	97.53	&	40.63
				\\
				&&	400	&	99.18	&	57.29	&&	95.68	&	47.92\\
				\hline
				\multicolumn{8}{c}{\bf{Empirical powers (\%) with $\bdelta=2\sqrt{{\log(p)}/{n}}\times(2^3,2^2,2^1,2^0,2^{-1})$.}}\\
				&&&	\multicolumn{2}{c}{Change point at $k^*=0.5n$}&&\multicolumn{2}{c}{Change point at $k^*=0.3n$}\\
				Model	&$\cG$&$p$&Boot-II&L\&B&&Boot-II&L\&B\\
				\hline
				$\bSigma=\bSigma^*$		&$\cS$	&200	&	100.00	&	NA	&&	100.00	&	NA
				\\
				&&	400	&	100.00	&	NA	&&	100.00	&	NA\\
				&$\cS^c$&	200	&	2.67	&	NA	&&	1.82	&	NA
				\\
				&&	400	&	2.26	&	NA	&&	1.65	&	NA\\
				&$\cS\cup \cS^c $&	200	&	100.00	&	100.00	&&	100.00	&	99.49
				\\
				&&	400	&	100.00	&	100.00	&&	100.00	&	99.49\\
				\bottomrule[2pt]
			\end{tabular}		
		\end{center}
	\end{table}	

\clearpage
\section{Useful lemmas}\label{section: useful lemmas}
Let $\bZ_1,\ldots,\bZ_n$ be independent centered random vectors in $\mathbb{R}^{p}$ with $\bZ_i=(Z_{i,1},\ldots, Z_{i,p})^\top$ for $i=1,\ldots,n$. Let $\bG_1,\ldots,\bG_n$ be independent centered Gaussian random vectors in $\mathbb{R}^{p}$ such that each $\bG_i$ has the same covariance matrix as $\bZ_i$. We then require the following conditions:
\begin{description}
	\item[\bf (M1)] There is a constant $b>0$ such that $\inf_{1\leq j\leq p}\E(Z_{i,j})^{2}\geq b$  for  $i=1,\ldots,n$. 
	\item[\bf (M2)] There exists a constant $K>0$ such that $\max\limits_{1\leq j
		\leq p} \dfrac{1}{n}\sum\limits_{i=1}^{n}\E|Z_{i,j}|^{2+\ell}\leq K^{\ell}$ for $\ell=1,2$.
	\item[\bf (M3)] There exists a constant $K'>0$ such that $\E\big(\exp(|Z_{i,j}|/K')\big)\leq 2$ for $j=1,\ldots,d$ and $i=1,\ldots,n$.
\end{description}
\begin{lemma}(\cite{Liu2019Unified})\label{lemma: key lemma for gaussian approximation}
	Assume that $\log(pn)=O(\floor{n\tau_0}^{\zeta})$ holds for some $0 <\zeta<1/7$. Let
	\begin{equation*}
		\bS^{\bZ}(\floor{nt})=\dfrac{1}{\sqrt{n}}\sum_{i=1}^{n}\bZ_i\big(\mathbf{1}(i\leq\floor{nt})-\floor{nt}/n\big),~~\bS^{\bG}(\floor{nt})=\dfrac{1}{\sqrt{n}}\sum_{i=1}^{n}\bG_i\big(\mathbf{1}(i\leq\floor{nt})-\floor{nt}/n\big)
	\end{equation*}
	be the partial sum processes for $(\bZ_i)_{i\geq 1}$ and $(\bG_i)_{i\geq 1}$, respectively.
	If $\bZ_1,\ldots,\bZ_n$ satisfy $\bf (M1)$, $\bf (M2)$ and $\bf (M3)$, then there is a constant $\zeta_0>0$ such that
	\begin{equation}\label{equation: key lemma for gaussian approximation1}
		\sup_{z\in(0,\infty)} \big|\P(\max_{\tau_0\leq t\leq 1-\tau_0}\|\bS^{\bZ}(\floor{nt})\|_{\infty}\leq z\big)-\P(\sup_{\tau_0\leq t\leq 1-\tau_0}\|\bS^{\bG}(\floor{nt})\|_{\infty}\leq z\big)\big|\leq Cn^{-\zeta_0},
	\end{equation}
	where $C$ is a constant only depending on $b$, $K$, and $K'$.
\end{lemma}

\begin{lemma}[Nazarovs inequality in \cite{nazarov2003maximal}] \label{lemma:anti consentration inequality}
	Let $\bW=(W_1,W_2,\cdots,W_p)^\top \in \mathbb{R}^{p}$ be centered Gaussian random vector with  $\inf_{1\leq k\leq p}\E(W_k)^2\geq b>0$. Then for
	any $\bx\in \mathbb{R}^p$ and $a>0$, we have
	\begin{equation*}
		\P(\bW\leq \bx+a)-\P(\bW\leq \bx)\leq Ca\sqrt{\log p},
	\end{equation*}
	where $C$ is a constant only depending on $b$.
\end{lemma}

\begin{lemma}(\cite{Zhou2017An})\label{lemma:maximum inequality}
	Let $\bW=(W_1,\ldots,W_p)^\top$ be a random vector with a marginal distribution  $N(0,\sigma_i^2)$ ($1\leq i\leq p$).
	Suppose $\exists A_0>0$ such that $\max_i\sigma_i^2\leq A^2_0$. Then, for any $t>0$, we have
	\begin{equation*}
		\E \big(\max_{1\leq i\leq p}|W_i|\big)\leq \dfrac{\log(2p)}{t}+\dfrac{tA_0^2}{2}.
	\end{equation*}
\end{lemma}

\begin{lemma}[\cite{van2014asymptotically}]\label{lemma: upper bounds for precision matrix}
	Suppose Assumptions $\mathbf{(A.1)}$ -- $\mathbf{(A.3)}$ hold. Assume additionally $\max_{j}\sqrt{s_j\log(p)/n}=o(1)$ holds. For the node-wise regression in (\ref{equation: node-wise regression}), choosing the tuning parameters $\lambda_{(j)}\approx \sqrt{\log(p)/n}$ uniformly over $j$, we have 
	\begin{equation}
		\begin{array}{l}
			\|\hat{\bTheta}_j-\bTheta_j\|_q=O_p\Big(s_j^{1/q}\sqrt{\dfrac{\log(p)}{n}}\Big),~\text{for}~q=1,2.\\
		\end{array}
	\end{equation}
\end{lemma}	

\begin{lemma}\label{lemma: exponential inequality for partial sum process}
	Let $\bZ_1,\ldots,\bZ_n$ be independent centered random vectors in $\mathbb{R}^{p}$ with $\bZ_i=(Z_{i,1},\ldots$
	$, Z_{i,p})^\top$ for $i=1,\ldots,n$. Assume that $\Zb_i$ follows the sub-exponential distribution. Then,  for any given subgroup $\cG\subset\{1,\ldots,p\}$, with probability at least $1-C_1(pn)^{-C_2}$, we have
	\begin{equation}
		\max_{t\in[\tau_0,1-\tau_0]}\max\limits_{j\in\cG}\Big|\dfrac{1}{\sqrt{n}}\Big( \sum_{i=1}^{\lfloor nt \rfloor}Z_{i,j}-\dfrac{\lfloor nt \rfloor}{n} \sum_{i=1}^{n}Z_{i,j}\Big)\Big|\leq C_3  \sqrt{\log(|\cG|n)},
	\end{equation}
	where $C_1$, $C_2$, and $C_3$ are universal positive constants not depending on $p$ or $n$.
\end{lemma}	

We next provide some useful results for the lasso estimators from heterogeneous data observations. To this end, for each $t\in[\tau_0,1-\tau_0]$, define
\begin{equation}\label{equation: four basic sets}
	\begin{array}{ll}
		\cA(t)=\left\{\Big\|\dfrac{1}{\lfloor nt \rfloor}\big(\Xb_{(0,t)})^\top(\bY_{(0,t)}-\Xb_{(0,t)}\bbeta^{(0,t)}\Big\|_{\infty}\leq \lambda^{(1)}\right\}, \\\\
		\cB(t)=\left\{\Big\|\dfrac{1}{\lfloor nt \rfloor^*}\big(\Xb_{(t,1)})^\top(\bY_{(t,1)}-\Xb_{(t,1)}\bbeta^{(t,1)}\Big\|_{\infty}\leq \lambda^{(2)}\right\}, \\\\
	\end{array}
\end{equation}
where $\lambda^{(1)}:=K_1\sqrt{\dfrac{\log(p)}{\lfloor nt \rfloor}}$ and $\lambda^{(2)}:=K_2\sqrt{\dfrac{\log(p)}{\lfloor nt \rfloor^*}}$, and $K_1,\ldots,K_2$ are some universal positive constants not depending on $n$ or $p$. 

The following Lemma \ref{lemma: basic inequality} provides a basic inequality for the lasso estimators, which is important for deriving the precise estimation error bound as well as prediction error bound (see Lemma \ref{lemma: esimation error bound} below). The proof of Lemma \ref{lemma: basic inequality} is given in Section \ref{section: proof of basic inequality}.
\begin{lemma}\label{lemma: basic inequality}
	Suppose Assumptions $\mathbf{(A.1)}$ -- $\mathbf{(A.3)}$ hold. Assume $\|\bbeta^{(2)}-\bbeta^{(1)}\|_2\leq C_{\bDelta}$ for some $C_{\bDelta}>0$. Recall $\bbeta^{(0,t)}$ and $\bbeta^{(t,1)}$ defined in (\ref{equation: true parameters before and after cpt}).	Let $\hat{\bbeta}^{(0,t)}$ and $\hat{\bbeta}^{(t,1)}$ be the lasso estimators as defined in (\ref{equation: lasso estimation before and after cpt}). Then, for each $t\in[\tau_0,1-\tau_0]$, under the event $\cA(t)\cap\cB(t)$,  we have
	\begin{equation}\label{inequality: basic inequality1}
		\dfrac{\big\|\Xb_{(0,t)}\big(\hat{\bbeta}^{(0,t)}-\bbeta^{(0,t)}\big)\big\|_2^2}{\lfloor nt \rfloor}+\lambda_1(t) \big\|(\hat{\bbeta}^{(0,t)}-\bbeta^{(0,t)})_{J^c(\bbeta^{(0,t)})}\big\|_1\leq 3\lambda_1(t)\big\|(\hat{\bbeta}^{(0,t)}-\bbeta^{(0,t)})_{J(\bbeta^{(0,t)})}\big\|_1,
	\end{equation}
	and
	\begin{equation}\label{inequality: basic inequality2}
		\dfrac{\big\|\Xb_{(t,1)}\big(\hat{\bbeta}^{(t,1)}-\bbeta^{(t,1)}\big)\big\|_2^2}{\lfloor nt \rfloor^*}+\lambda_2(t) \big\|(\hat{\bbeta}^{(t,1)}-\bbeta^{(t,1)})_{J^c(\bbeta^{(t,1)})}\big\|_1\leq 3\lambda_2(t)\big\|(\hat{\bbeta}^{(t,1)}-\bbeta^{(t,1)})_{J(\bbeta^{(t,1)})}\big\|_1,
	\end{equation}
	where $\lambda_1(t):=2\lambda^{(1)}$, $\lambda_2(t):=2\lambda^{(2)}$.
\end{lemma}	

The following Lemma \ref{lemma: key estimation error bound for lasso} provides the estimation error bounds for the lasso estimators $\hat{\bbeta}^{(0,t)}$ and $\hat{\bbeta}^{(t,1)}$ in terms of $\ell_q$-norm. The proof of Lemma \ref{lemma: key estimation error bound for lasso} is given in Section \ref{section: proof of key  estimation error bound}. 
\begin{lemma}\label{lemma: key estimation error bound for lasso}
	Suppose Assumptions $\mathbf{(A.1)}$ -- $\mathbf{(A.3)}$ hold. Assume $\|\bbeta^{(2)}-\bbeta^{(1)}\|_2\leq C_{\bDelta}$ for some $C_{\bDelta}>0$. Recall $\bbeta^{(0,t)}$ and $\bbeta^{(t,1)}$ defined in (\ref{equation: true parameters before and after cpt}).	Let $\hat{\bbeta}^{(0,t)}$ and $\hat{\bbeta}^{(t,1)}$ be the lasso estimators as defined in (\ref{equation: lasso estimation before and after cpt}). For each $t\in[\tau_0,1-\tau_0]$, let $s_1(t):=\cM(\bbeta^{(0,t)})$ and $s_2(t):=\cM(\bbeta^{(t,1)})$. Then, under the event $\cA(t)\cap\cB(t)$, we have
	\begin{equation}\label{inequality: key estimation error bound of lasso for pooled samples}
		\begin{array}{l}
			\big\|\hat{\bbeta}^{(0,t)}-\bbeta^{(0,t)}\big\|_{q}\leq C_1(s_1(t))^{\frac{1}{q}}\sqrt{\dfrac{\log p}{\lfloor nt \rfloor}},~~\big\|\hat{\bbeta}^{(t,1)}-\bbeta^{(t,1)}\big\|_{q}\leq C_2(s_2(t))^{\frac{1}{q}}\sqrt{\dfrac{\log p}{\lfloor nt \rfloor^*}},~q=1,2,\\\\
			\dfrac{\big\|\Xb_{(0,t)}\big(\hat{\bbeta}^{(0,t)}-\bbeta^{(0,t)}\big)\big\|_2^2}{\lfloor nt \rfloor}\leq 
			C_3s_1(t)\dfrac{\log p}{\lfloor nt \rfloor},
			~~\dfrac{\big\|\Xb_{(t,1)}\big(\hat{\bbeta}^{(t,1)}-\bbeta^{(t,1)}\big)\big\|_2^2}{\lfloor nt \rfloor^*}\leq 
			C_4s_2(t)\dfrac{\log p}{\lfloor nt \rfloor^*},\\\\
			\cM(\hat{\bbeta}^{(0,t)})\leq C_5s_1(t), ~~~~\cM(\hat{\bbeta}^{(t,1)})\leq C_6s_2(t),
		\end{array}
	\end{equation}	
	where $C_1,\ldots,C_6$ are some universal positive constants not depending on $n$ or $p$.
\end{lemma}

Lastly, as a by product of  Lemma \ref{lemma: key estimation error bound for lasso}, the following Lemma \ref{lemma: esimation error bound} provides the estimation error bounds for  $\hat{\bbeta}^{(0,t)}-\bbeta^{(1)}$ and $\hat{\bbeta}^{(t,1)}-\bbeta^{(2)}$ in terms of the $\ell_q$-norm, which is frequently used in the proofs. 
\begin{lemma}\label{lemma: esimation error bound}
	Suppose Assumptions $\mathbf{(A.1)}$ -- $\mathbf{(A.3)}$ hold. Assume $\|\bbeta^{(2)}-\bbeta^{(1)}\|_2\leq C_{\bDelta}$ for some $C_{\bDelta}>0$. Recall $s:=s^{(1)}\vee s^{(2)}$. Let $\hat{\bbeta}^{(0,t)}$ and $\hat{\bbeta}^{(t,1)}$ be the lasso estimators as defined in (\ref{equation: lasso estimation before and after cpt}). Then, under the event $\cA(t)\cap\cB(t)$, for each $t\in[\tau_0,1-\tau_0]$, we have
	\begin{equation}\label{inequality: estimation error bound}
		\begin{array}{l}
			\big\|\hat{\bbeta}^{(0,t)}-\bbeta^{(1)}\big\|_{q}\leq C_1\max\Big\{s^{\frac{1}{q}}\sqrt{\dfrac{\log p}{\lfloor nt \rfloor}},\dfrac{\lfloor nt\rfloor-\lfloor nt_0\rfloor}{\lfloor nt \rfloor}\big\|\bbeta^{(2)}-\bbeta^{(1)}\big\|_q\mathbf{1}\{t\geq t_0\}\Big\},q=1,2,\\\\
			\big\|\hat{\bbeta}^{(t,1)}-\bbeta^{(2)}\big\|_{q}\leq C_2\max\Big\{s^{\frac{1}{q}}\sqrt{\dfrac{\log p}{\lfloor nt \rfloor^*}},\dfrac{\lfloor nt_0\rfloor-\lfloor nt\rfloor}{\lfloor nt \rfloor^*}\big\|\bbeta^{(2)}-\bbeta^{(1)}\big\|_q\mathbf{1}\{t\leq t_0\}\Big\},q=1,2,\\\\
			\dfrac{\big\|\Xb_{(0,t)}\big(\hat{\bbeta}^{(0,t)}-\bbeta^{(1)}\big)\big\|_2^2}{\lfloor nt \rfloor}\leq 
			C_3\max\Big\{s\dfrac{\log p}{\lfloor nt \rfloor},\big(\dfrac{\lfloor nt\rfloor-\lfloor nt_0\rfloor}{\lfloor nt \rfloor}\big)^2\big\|\bbeta^{(2)}-\bbeta^{(1)}\big\|_2^2\mathbf{1}\{t\geq t_0\}\Big\},
			\\\\
			\dfrac{\big\|\Xb_{(t,1)}\big(\hat{\bbeta}^{(t,1)}-\bbeta^{(2)}\big)\big\|_2^2}{\lfloor nt \rfloor^*}\leq 
			C_4\max\Big\{s\dfrac{\log p}{\lfloor nt \rfloor^*},\big(\dfrac{\lfloor nt_0\rfloor-\lfloor nt\rfloor}{\lfloor nt \rfloor^*})^2\big\|\bbeta^{(2)}-\bbeta^{(1)}\big\|_2^2\mathbf{1}\{t\leq t_0\}\Big\},
			\\\\	
			\cM(\hat{\bbeta}^{(0,t)})\leq C_5s, ~~~~\cM(\hat{\bbeta}^{(t,1)})\leq C_6s,
		\end{array}
	\end{equation}
	where $C_1,\ldots,C_6$ are some universal positive constants not depending on $n$ or $p$.
\end{lemma}	

The following Lemma \ref{lemma: large probabilty for intersection} shows that the results in Lemmas \ref{lemma: basic inequality} -- \ref{lemma: esimation error bound} occur uniformly over $t\in[\tau_0,1-\tau_0]$ with high probability. The proof of Lemma \ref{lemma: large probabilty for intersection} is given in Section \ref{section: proof of intersection}.
\begin{lemma}\label{lemma: large probabilty for intersection}
	Suppose Assumptions $\mathbf{(A.1)}$ -- $\mathbf{(A.3)}$ hold. Assume $\|\bbeta^{(2)}-\bbeta^{(1)}\|_2\leq C_{\bDelta}$ for some $C_{\bDelta}>0$. Then we have
	\begin{equation}\label{inequality: large probability for intersection}
		\P\big(\bigcap_{t\in[\tau_0,1-\tau_0]}\big\{\cA(t)\cap\cB(t)\big\}        \big)\geq 1-C_1(np)^{-C_2},
	\end{equation}
	where $C_1, C_2$ are some big enough universal positive constants not depending on $n$ or $p$.
\end{lemma}

\section{Proof of main results}\label{section: proofs of main results}

\subsection{Proof of Proposition \ref{proposition: RE conditions}}
\begin{proof}
	Note that the proof of Part (i) is  easier than Part (ii). To save space, we  give the proof of Part (ii). Firstly, we consider $\cR_1(s,3,\mathbb{T})$. The proof proceeds in two steps. \\
	\textbf{Step~1:} we prove  $\sup_{t\in[\tau_0,1-\tau_0]}\|\hat{\bSigma}_{(0,t)}-\bSigma\|_{\infty}=O_p(\sqrt{\log(p)/\floor{n\tau_0}})$. For any fixed $t\in[\tau_0,1-\tau_0]$ and $j,k\in\{1,\ldots,p\}$, by Assumption (A.1), using exponential inequality, we have 
	\begin{equation*}
		\P\big(|\dfrac{1}{\floor{nt}}\sum_{i=1}^{\floor{nt}}(X_{ij}X_{ik}-\E[X_{ij}X_{ik}])|\geq x\big)\leq C_1\exp(-C_2\floor{nt}x^2)\leq C_1\exp(-C_2\floor{n\tau_0}x^2).
	\end{equation*}
	Hence, taking $x=C_3\sqrt{\log(pn)/\floor{n\tau_0}}$ for some big constant $C_3>0$, we have:
	\begin{equation*}
		\P\big(|\dfrac{1}{\floor{nt}}\sum_{i=1}^{\floor{nt}}(X_{ij}X_{ik}-\E[X_{ij}X_{ik}])|\geq x\big)\leq C_1(np)^{-C_3}.
	\end{equation*}
	As a result, we have:
	\begin{equation*}
		\begin{array}{ll}
			\P(\sup\limits_{t\in[\tau_0,1-\tau_0]}\|\hat{\bSigma}_{(0,t)}-\bSigma\|_{\infty}\geq x)\\
			=\P\Big(\bigcup\limits_{t}\bigcup\limits_{j,k}\big\{ |\dfrac{1}{\floor{nt}}\sum\limits_{i=1}^{\floor{nt}}(X_{ij}X_{ik}-\E[X_{ij}X_{ik}])|\geq x\big\} \Big  )\\
			\leq np^2\max_{t,j,k}\P\big(|\dfrac{1}{\floor{nt}}\sum\limits_{i=1}^{\floor{nt}}(X_{ij}X_{ik}-\E[X_{ij}X_{ik}])|\geq x\big)\\
			\leq C_1(np)^{-C_4},
		\end{array}
	\end{equation*}
	where $C_1-C_4$ are some big enough universal constants. This yields $\sup_{t\in[\tau_0,1-\tau_0]}\|\hat{\bSigma}_{(0,t)}-\bSigma\|_{\infty}=O_p(\sqrt{\log(p)/\floor{n\tau_0}})$. \\
	\textbf{Step~2:} For integers $s$ such that $1\leq s\leq p$, a set of indices $J_0\subset\{1,\ldots,p\}$ with $|J_0|\leq s$, and any vector $\bdelta$ satisfying  $\|\bdelta_{J_0^c}\|_1\leq 3\|\bdelta_{J_0}\|_1$, we have:
	\begin{equation}\label{inequality: step 3 for URE}
		\begin{array}{rl}
			\dfrac{\bdelta^\top\hat{\bSigma}_{(0,t)}\bdelta }{|\bdelta_{J_0}|_2^2}&=_{(1)}\dfrac{\bdelta^\top\bSigma\bdelta }{|\bdelta_{J_0}|_2^2}+\dfrac{\bdelta^\top(\bSigma-\hat{\bSigma}_{(0,t)})\bdelta }{|\bdelta_{J_0}|_2^2},\\
			&\geq_{(2)} \dfrac{\bdelta^\top\bSigma\bdelta }{|\bdelta_{J_0}|_2^2}-\dfrac{\sup_{t\in[\tau_0,1-\tau_0]}\|\hat{\bSigma}_{(0,t)}-\bSigma\|_{\infty}}{|\bdelta_{J_0}|_2^2}|\bdelta|_1^2,\\
			&\geq_{(3)} \dfrac{\bdelta^\top\bSigma\bdelta }{|\bdelta_{J_0}|_2^2}-\dfrac{\sup_{t\in[\tau_0,1-\tau_0]}\|\hat{\bSigma}_{(0,t)}-\bSigma\|_{\infty}}{|\bdelta_{J_0}|_2^2}(1+c_0)^2|\bdelta_{J_0}|_1^2,\\
			& \geq_{(4)} \dfrac{\bdelta^\top\bSigma\bdelta }{|\bdelta_{J_0}|_2^2}-\sup\limits_{{t\in[\tau_0,1-\tau_0]}}\|\hat{\bSigma}_{(0,t)}-\bSigma\|_{\infty}(1+c_0)^2s.\\
			&\geq_{(5)} 4\kappa_4^4-sO_p(\sqrt{\log(p)/\floor{n\tau_0}})\geq_{(6)} \kappa_4^2,
		\end{array}
	\end{equation}
	where $(5)$ comes from Condition  (\ref{equation: populational covariance lasso}) and the result in Step 1, $(6)$ comes from the assumption  $s\sqrt{\log(p)/\floor{n\tau_0}}=o(1)$. Lastly, combining Steps 1 and 2, we finish the proof.

\end{proof}
\subsection{Proof of Theorem \ref{theorem: variance estimator under H0}}
\begin{proof}
	Under $\Hb_0$, the change point $t_0$ is not identifiable. Hence, to prove Theorem \ref{theorem: variance estimator under H0}, we need to prove the  convergence of $\hat{\sigma}_{\epsilon}^2(t)\hat{\omega}_{j,k}\}$ to $\{\sigma_{\epsilon}^2\omega_{j,k }\}$ uniformly over $ \tau_0\leq t\leq 1-\tau_0$ and $1\leq j,k\leq p$, where  $\hat{\sigma}_{\epsilon}^2(t)$ is defined in (\ref{equation: pooled variance estimation}). Note that for each $t$, $j$ and $k$,
	\begin{equation}\label{inequality: proof of large deviation}
		\begin{array}{ll}
			|\hat{\sigma}_{\epsilon}^2(t)\hat{\omega}_{j,k}-\sigma_{\epsilon}^2\omega_{j,k}|\\
			\leq|\hat{\sigma}_{\epsilon}^2(t)\hat{\omega}_{j,k}-\sigma_{\epsilon}^2\hat{\omega}_{j,k}|+\sigma_{\epsilon}^2|\hat{\omega}_{j,k}-\omega_{j,k}|\\
			\leq|\hat{\sigma}_{\epsilon}^2(t)-\sigma_{\epsilon}^2||\hat{\omega}_{j,k}-\omega_{j,k}|+|\hat{\sigma}_{\epsilon}^2(t)-\sigma_{\epsilon}^2|\omega_{j,k}+\sigma_{\epsilon}^2|\hat{\omega}_{j,k}-\omega_{j,k}|\\
			\leq C(|\hat{\sigma}_{\epsilon}^2(t)-\sigma_{\epsilon}^2|+|\hat{\omega}_{j,k}-\omega_{j,k}|),
		\end{array}
	\end{equation}
	where the last inequality comes from Assumptions $\mathbf{(A.2)}$ and $\mathbf{(A.3)}$ and $C$ is a universal positive constant not depending on $n$ or $p$. Hence, by (\ref{inequality: proof of large deviation}), to prove Theorem \ref{theorem: variance estimator under H0}, we need to bound $\max_{t\in[\tau_0,1-\tau_0]}|\hat{\sigma}_{\epsilon}^2(t)-\sigma_{\epsilon}^2|$ and $\max_{1\leq j,k\leq p}|\hat{\omega}_{j,k}-\omega_{j,k}|$, respectively.
	
	For bounding $\max\limits_{t\in[\tau_0,1-\tau_0]}|\hat{\sigma}_{\epsilon}^2(t)-\sigma_{\epsilon}^2|$, by the definition of $\hat{\sigma}_{\epsilon}^2(t)$ in (\ref{equation: pooled variance estimation}), under $\Hb_0$, using some straightforward calculations, we have 
	\begin{equation}\label{equation: decomposition for pooled variance}
		\begin{array}{ll}
			&\hat{\sigma}_{\epsilon}^2(t)-\sigma_{\epsilon}^2\\
			&=n^{-1}\Big( \big\|\bepsilon_{(0,t)}+\Xb_{(0,t)  }\big(\hat{\bbeta}^{(0,t) }-\bbeta^{(0)}\big)\big\|_2^2\Big)+ \big\|\bepsilon_{(t,1)}+\Xb_{(t,1)}\big(\hat{\bbeta}^{(t,1) }-\bbeta^{(0)}\big)\big\|_2^2\Big)-\sigma_{\epsilon}^2,\\
			&=n^{-1}\big\|\Xb_{(0,t)  }\big(\hat{\bbeta}^{(0,t) }-\bbeta^{(0)}\big)\big\|_2^2+2\dfrac{\lfloor nt \rfloor}{n}\dfrac{(\bepsilon_{(0,t)})^\top\Xb_{(0,t)  }}{\lfloor nt \rfloor }\big(\hat{\bbeta}^{(0,t) }-\bbeta^{(0)}\big)\\
			&\quad+n^{-1}\big\|\Xb_{(t,1)}\big(\hat{\bbeta}^{(t,1) }-\bbeta^{(0)}\big)\big\|_2^2+2\dfrac{\lfloor nt \rfloor^*}{n}\dfrac{(\bepsilon_{(t,1)})^\top\Xb_{(t,1)}}{\lfloor nt \rfloor^* }\big(\hat{\bbeta}^{(t,1) }-\bbeta^{(0)}\big)\\
			&\quad\quad\quad\quad+n^{-1}\sum\limits_{i=1}^{n}(\epsilon_i^2-\sigma_{\epsilon}^2).
		\end{array}
	\end{equation}
	By (\ref{equation: decomposition for pooled variance}), to bound $\max_{t\in[\tau_0,1-\tau_0]}|\hat{\sigma}_{\epsilon}^2(t)-\sigma_{\epsilon}^2|$, we need to consider the 
	five parts on the RHS of (\ref{equation: decomposition for pooled variance}), respectively. For the first four parts, by Lemma \ref{lemma: esimation error bound}, we have
	\begin{equation}\label{inequality: upper bounds for the first four terms}
		\begin{array}{ll}
			\dfrac{1}{n}\big\|\Xb_{(0,t)  }\big(\hat{\bbeta}^{(0,t) }-\bbeta^{(0)}\big)\big\|_2^2\leq\dfrac{\lfloor nt \rfloor }{n}O_p\Big(s^{(0)}\dfrac{\log(p)}{\lfloor nt \rfloor }\Big)=O_p\Big(s^{(0)}\dfrac{\log(p)}{n }\Big),\\\\
			\dfrac{1}{n}\big\|\Xb_{(t,1)}\big(\hat{\bbeta}^{(t,1) }-\bbeta^{(0)}\big)\big\|_2^2\leq\dfrac{\lfloor nt \rfloor^* }{n}O_p\Big(s^{(0)}\dfrac{\log(p)}{\lfloor nt \rfloor^* }\Big)=O_p\Big(s^{(0)}\dfrac{\log(p)}{n }\Big),\\\\
			\Big|2\dfrac{\lfloor nt \rfloor}{n}\dfrac{(\bepsilon_{(0,t)})^\top\Xb_{(0,t)  }}{\lfloor nt \rfloor }\big(\hat{\bbeta}^{(0,t) }-\bbeta^{(0)}\big)\Big|\leq O_p\Big(\lambda^{(1)} \big\|\hat{\bbeta}^{(0,t) }-\bbeta^{(0)}\big\|_1\Big)\leq O_p\Big(s^{(0)}\dfrac{\log(p)}{\lfloor nt \rfloor}\Big),\\\\
			\Big|2\dfrac{\lfloor nt \rfloor^*}{n}\dfrac{(\bepsilon_{(t,1)})^\top\Xb_{(t,1)}}{\lfloor nt \rfloor^* }\big(\hat{\bbeta}^{(t,1) }-\bbeta^{(0)}\big)\Big|\leq O_p\Big(\lambda^{(3)} \big\|\hat{\bbeta}^{(t,1) }-\bbeta^{(0)}\big\|_1\Big)\leq O_p\Big(s^{(0)}\dfrac{\log(p)}{\lfloor nt \rfloor^*}\Big).
		\end{array}
	\end{equation}
	Note that $\epsilon_i^2-\sigma^2_{\epsilon}$ follows the sub-exponential distribution.	For $\sum\limits_{i=1}^{n}(\epsilon_i^2-\sigma_{\epsilon}^2)/n$, under Assumption $\mathbf{(A.2)}$, using Bernstein's  inequalities, we can prove
	\begin{equation}\label{inequality: upper bounds for residual term}
		\sum\limits_{i=1}^{n}(\epsilon_i^2-\sigma_{\epsilon}^2)/n\leq O_p\Big(\sqrt{\dfrac{\log(n)}{n}}\Big).
	\end{equation}
	Hence, combining (\ref{equation: decomposition for pooled variance}), (\ref{inequality: upper bounds for the first four terms}), and (\ref{inequality: upper bounds for residual term}), and using Assumptions  $\mathbf{(A.1)} - \mathbf{(A.3)}$, we have 
	\begin{equation}\label{inequality: final upper bounds for pooled variance}
		\max_{t\in[\tau_0,1-\tau_0]}|\hat{\sigma}_{\epsilon}^2(t)-\sigma_{\epsilon}^2|\leq O_p\Big(\sqrt{\dfrac{\log(n)}{n}}\Big).
	\end{equation}
	Next, we bound $\max_{1\leq j,k\leq p}|\hat{\omega}_{j,k}-\omega_{j,k}|$. By Lemmas 5.3 and 5.4 in \cite{van2014asymptotically}, we have 
	\begin{equation}\label{inequality: upper bounds for omega matrix}
		\max_{1\leq j,k\leq p}|\hat{\omega}_{j,k}-\omega_{j,k}|=\max_{1\leq j,k\leq p}|\hat{\bTheta}_j^\top\hat{\bSigma}\hat{\bTheta}_k-\bTheta_j^\top\bSigma\bTheta_k|=O_p(\max_j\lambda_{(j)}\sqrt{s_j}).
	\end{equation}
	Finally, combining (\ref{inequality: final upper bounds for pooled variance}) and (\ref{inequality: upper bounds for omega matrix}), we have
	\begin{equation}
		\max_{t\in[\tau_0,1-\tau_0]}\max_{1\leq j,k\leq p}|\hat{\sigma}_{\epsilon}^2(t)\hat{\omega}_{j,k}-\sigma_{\epsilon}^2\omega_{j,k}|\leq O_p\Big(\sqrt{\dfrac{\log(n)}{n}}+ \max_j\lambda_{(j)}\sqrt{s_j}\Big),
	\end{equation}
	which completes the proof of Theorem \ref{theorem: variance estimator under H0}.
\end{proof}
\subsection{Proof of Theorem \ref{theorem: gaussian approximation}}\label{section: proof of gaussian approximation}
\begin{proof}
	In this section, we aim to prove 
	\begin{equation}\label{equation: proof of gaussian approximation}
		\sup_{z\in (0,\infty)}\big|\P(T_{\cG}\leq z)-\P(T_{\cG}^{b}\leq z|\mathcal{X})\big|=o_p(1), ~\text{as}~  n,p\rightarrow\infty.
	\end{equation}
	The proof proceeds in four steps. In Steps 1 and 2, we decompose $T_{\cG}$ and $T_{\cG}^b$ into a leading term and a residual term and show that the corresponding residual terms can be asymptotically negligible. In Step 3, we prove that it is possible to approximate the leading term of $T_{\cG}$ by that of $T^b_{\cG}$. In Step 4, we combine the previous results to complete the proof.\\ 
	\textbf{Step~1} (Decomposition of $T_{\cG}$). Note that under the null hypothesis  of no change point, we have $\beta^{(1)}_s=\beta^{(2)}_s=\beta^{(0)}_s$ for $1\leq s\leq p$. By the definition of the de-biased lasso estimators $\breve{\bbeta}^{(0,t)}$ and $\breve{\bbeta}^{(t,1)}$  in (\ref{equation: de-biased before and after cpt}),  we can write them as follows:
	\begin{equation}
		\begin{array}{l}
			\breve{\bbeta}^{(0,t)}=\bbeta^{(0)}+\hat{\bTheta}(\Xb_{(0,t)})^\top\bepsilon_{(0,t)}/\lfloor nt \rfloor +\bDelta^{(0,t)},\\\\
			\breve{\bbeta}^{(t,1)}=\bbeta^{(0)}+\hat{\bTheta}(\Xb_{(t,1)})^\top\bepsilon_{(t,1)}/\lfloor nt \rfloor^* +\bDelta^{(t,1)},
		\end{array}
	\end{equation}
	where $\bDelta^{(0,t)}=\big(\Delta_1^{(0,t)},\ldots, \Delta_p^{(0,t)}\big)^\top$ and $\bDelta^{(t,1)}=\big(\Delta_1^{(t,1)},\ldots, \Delta_p^{(t,1)}\big)^\top$ are defined as
	\begin{equation}\label{equation: residual term before and after cpt}
		\begin{array}{l}
			\bDelta^{(0,t)}:=-\big(\hat{\bTheta}\hat{\bSigma}_{(0,t)}-\Ib\big)\big(\hat{\bbeta}^{(0,t)}-\bbeta^{(0)} \big),\\\\
			\bDelta^{(t,1)}:=-\big(\hat{\bTheta}\hat{\bSigma}_{(t,1)}-\Ib\big)\big(\hat{\bbeta}^{(t,1)}-\bbeta^{(0)} \big),
		\end{array}
	\end{equation}
	with 
	$\hat{\bSigma}_{(0,t)}:=(\Xb_{(0,t)})^\top\Xb_{(0,t)}/\lfloor nt \rfloor$ and
	$\hat{\bSigma}_{(t,1)}:=(\Xb_{(t,1)})^\top\Xb_{(t,1)}/\lfloor nt \rfloor^*$. Denote $\hat{\bTheta}_i$, $\Xb_{(0,t),i}$, $\Xb_{(t,1),i}$ as the $i$-th row of 
	$\hat{\bTheta}$, $\Xb_{(0,t)}$, and $\Xb_{(t,1)}$, respectively. Then, for each coordinate $j$ at time point $\lfloor nt\rfloor$, we can write  each coordinate of the de-biased lasso estimator in the following form:
	\begin{equation}\label{equation: coordinate-wise de-biased estimators}
		\begin{array}{l}
			\breve{\beta}^{(0,t)}_j=\beta^{(0)}_j+\dfrac{1}{\lfloor nt \rfloor}\sum\limits_{i=1}^{\lfloor nt \rfloor}\hat{\bTheta}_j^\top\Xb_i\epsilon_i+\Delta^{(0,t)}_j,\\\\
			\breve{\beta}^{(t,1)}_j=\beta^{(0)}_j+\dfrac{1}{\lfloor nt \rfloor^*}\sum\limits_{i=\lfloor nt \rfloor+1}^{n}\hat{\bTheta}_j^\top\Xb_i\epsilon_i+\Delta^{(t,1)}_j.
		\end{array}
	\end{equation}
	For each $t\in[\tau_0,1-\tau_0]$ and $1 \leq j\leq p$, define the coordinate-wise process as
	\begin{equation}\label{equation: partial sum process for each coordinate }
		C_j(\lfloor nt \rfloor)=\sqrt{n}\dfrac{\lfloor nt \rfloor}{n}\dfrac{\lfloor nt \rfloor^* }{n}\dfrac{\big(\breve{\beta}^{(0,t)}_j-\breve{\beta}^{(t,1)}_j\big)}{\sqrt{\hat{\sigma}_{\epsilon}^2\hat{\omega}_{j,j }}}.
	\end{equation}
	By the definition of $T_{\cG}$ in (\ref{equation: test statistics}) , we have $T_{\cG}=\max\limits_{t\in[\tau_0,1-\tau_0]}\max\limits_{j\in\cG}|C_j(\lfloor nt \rfloor)|$. Furthermore, by (\ref{equation: coordinate-wise de-biased estimators}), we can decompose $C_j(\lfloor nt \rfloor)$ into two parts:
	\begin{equation}\label{equation: decomposition for partial sum process}
		C_j(\lfloor nt \rfloor)=C^{\rm I}_j(\lfloor nt \rfloor)+C^{\rm II}_j(\lfloor nt \rfloor), ~\text{for}~t\in[\tau_0,1-\tau_0], ~~1\leq j\leq p,
	\end{equation}
	with 
	\begin{equation}\label{equation: leading and residual term of partial sum process}
		\small
		\begin{array}{l}
			C^{\rm I}_j(\lfloor nt \rfloor):=\sqrt{n}\dfrac{\lfloor nt \rfloor}{n}\dfrac{\lfloor nt \rfloor^* }{n}\dfrac{\Big(\dfrac{1}{\lfloor nt \rfloor}\sum\limits_{i=1}^{\lfloor nt \rfloor}\hat{\bTheta}_j^\top\Xb_i\epsilon_i-\dfrac{1}{\lfloor nt \rfloor^*}\sum\limits_{i=\lfloor nt \rfloor+1}^{n}\hat{\bTheta}_j^\top\Xb_i\epsilon_i\Big)}{\sqrt{\hat{\sigma}_{\epsilon}^2\hat{\omega}_{j,j }}},\\\\
			C^{\rm II}_j(\lfloor nt \rfloor):=\sqrt{n}\dfrac{\lfloor nt \rfloor}{n}\dfrac{\lfloor nt \rfloor^* }{n}\dfrac{\Big(\Delta^{(0,t)}_j-\Delta^{(t,1)}_j\Big)}{\sqrt{\hat{\sigma}_{\epsilon}^2\hat{\omega}_{j,j }}},~\text{with}~1\leq j\leq p~\text{and}~t\in[\tau_0,1-
			\tau_0].
		\end{array}
	\end{equation}
	Note that we can regard $C^{\rm I}_j(\lfloor nt \rfloor)$ as the leading term and $C^{\rm II}_j(\lfloor nt \rfloor)$ as the residual term of $C_j(\lfloor nt \rfloor)$.
	Furthermore, by replacing $\hat{\sigma}_{\epsilon}^2$, $\hat{\omega}_{j,j }$, and $\hat{\bTheta}_j$ by their true values $\sigma_{\epsilon}^2$, $\omega_{j,j}$, and $\bTheta_j$, we can define the oracle leading term as follows:
	\begin{equation}\label{equation: oracle leading term}
		\tilde{C}^{\rm I}_j(\lfloor nt \rfloor):=\sqrt{n}\dfrac{\lfloor nt \rfloor}{n}\dfrac{\lfloor nt \rfloor^* }{n}\dfrac{\Big(\dfrac{1}{\lfloor nt \rfloor}\sum\limits_{i=1}^{\lfloor nt \rfloor}\bTheta_j^\top\Xb_i\epsilon_i-\dfrac{1}{\lfloor nt \rfloor^*}\sum\limits_{i=\lfloor nt \rfloor+1}^{n}\bTheta_j^\top\Xb_i\epsilon_i\Big)}{\sqrt{\sigma_{\epsilon}^2\omega_{j,j }}}.
	\end{equation}
	Based on (\ref{equation: partial sum process for each coordinate }), (\ref{equation: leading and residual term of partial sum process}), and (\ref{equation: oracle leading term}), define the following four vector-valued processes:
	\begin{equation}\label{equation: 4 vectorized process}
		\small
		\begin{array}{ll}
			\bC(\floor{nt})=\big(C_1(\floor{nt}),\ldots,C_p(\floor{nt})\big)^\top,& \bC^{\rm I}(\floor{nt})=\big(C^{\rm I}_1(\floor{nt}),\ldots,C^{\rm I}_p(\floor{nt})\big)^\top,\\\\ \bC^{\rm II}(\floor{nt})=\big(C^{\rm II}_1(\floor{nt}),\ldots,C^{\rm II}_p(\floor{nt})\big)^\top,& \tilde{\bC}^{\rm I}(\floor{nt})=\big(\tilde{C}_{1}^{\text{\rm I}}(\floor{nt}),\ldots,\tilde{C}^{\text{I}}_{p}(\floor{nt})\big)^{\top}.
		\end{array}
	\end{equation}
	The following Lemma  \ref{lemma: negligible} shows that the residual term $|C_j^{\rm II}|$ can be uniformly negligible over $t\in[\tau_0,1-\tau_0]$ and $1\leq j\leq p$. The proof of Lemma  \ref{lemma: negligible} is provided in Section \ref{section: proof of negligible}.
	\begin{lemma}\label{lemma: negligible}
		Assume Assumptions $\mathbf{(A.1)}$ -- $\mathbf{(A.5)}$ hold. Under $\Hb_0$, we have
		\begin{equation}\label{equation: Lemma B.1}
			\P\Big(\max_{\tau_0\leq t\leq 1-\tau_0} \big\|\bC(\floor{nt})-\tilde{\bC}^{\text{I}}(\floor{nt})\big\|_{\cG,\infty}\geq \epsilon \Big)=o(1),
		\end{equation}
		where $\epsilon=C\max(\max\limits_{1\leq j\leq p}s_j\dfrac{\log(pn)}{\sqrt{ n }},s\sqrt{n}\dfrac{\log(pn)}{{\lfloor n\tau_0 \rfloor}})$, and $C$ is a universal constant not depending on $n$ or $p$.
	\end{lemma}
	\textbf{Step~2} (Decomposition of $T^b_{\cG}$).   In this step, we analyze the bootstrap version of the test statistic and decompose $T_{\cG}^b$ into a leading term and a residual term.  To this end, we need some additional notations. For $0\leq t_1\leq t_2\leq 1$, define 
	
	\begin{equation*}
		\begin{array}{ll}
			\bY_{(t_1,t_2)}=(Y_{\floor{nt_1}+1},\ldots,Y_{\floor{nt_2}})^\top,~~\bepsilon_{(t_1,t_2)}=(\epsilon_{\floor{nt_1}+1},\ldots,\epsilon_{\floor{nt_2}})^\top,\\
			\Xb_{(t_1,t_2)}=(\bX_{\floor{nt_1}+1},\ldots,\bX_{\floor{nt_2}})^\top,~~~\hat{\bSigma}_{(t_1,t_2)}=\dfrac{(\Xb_{(t_1,t_2)})^\top\Xb_{(t_1,t_2)}}{\lfloor nt_2 \rfloor-\lfloor nt_1 \rfloor+1}.
		\end{array}
	\end{equation*}
	Note that the decomposition for  $T_{\cG}^b$ is different from that of $T_{\cG}$. The main difficulty is that the bootstrap based samples involve a change point estimator $\hat{t}_{0,\cG}$ and the data are split into two sub-samples (before and after $\hat{t}_{0,\cG}$ ), which requires a careful discussion about the location. 	To analyze $\breve{\bbeta}^{b,(0,t)}$ and $\breve{\bbeta}^{b,(t,1)}$  in (\ref{equation: boot de-biased before and after cpt}), we need to consider the following cases:\\
	$\mathbf{Case~1:}$ The search location $t$ at $t\in[\tau_0,\hat{t}_{0,\cG}]$. In this case, since $\breve{\bbeta}^{b,(0,t)}$ is constructed using homogeneous bootstrap samples, similar to Step 1, we can decompose $\breve{\bbeta}^{b,(0,t)}$ as:
	\begin{equation}\label{equation: bootstraped de-biased before cpt1}
		\breve{\bbeta}^{b,(0,t)}=\hat{\bbeta}^{(0,\hat{t}_{0,\cG})}+\dfrac{\hat{\bTheta}(\Xb_{(0,t)})^\top\bepsilon^{b,(0,t)}}{\lfloor nt \rfloor} +\bDelta^{b,(0,t),\rm I},
	\end{equation}
	where $\bDelta^{b,(0,t),\rm I}=(\Delta^{b,(0,t),\rm I}_1,\ldots,\Delta^{b,(0,t),\rm I}_p)^\top$ are defined as
	\begin{equation}\label{equation: bootstraped residual before cpt1}
		\bDelta^{b,(0,t),\rm I}:=-\big(\hat{\bTheta}\hat{\bSigma}_{(0,t)}-\Ib\big)\big(\hat{\bbeta}^{b,(0,t)}-\hat{\bbeta}^{(0,\hat{t}_{0,\cG})} \big).
	\end{equation}
	For $\breve{\bbeta}^{b,(t,1)}$, since it is constructed using data both before $\lfloor n\hat{t}_{0,\cG} \rfloor$ and after $\lfloor n\hat{t}_{0,\cG} \rfloor$, 
	using tedious calculations, we can decompose $\breve{\bbeta}^{b,(t,1)}$ into
	\begin{equation}\label{equation: bootstraped de-biased after cpt11}
		\begin{array}{ll}
			\breve{\bbeta}^{b,(t,1)}=\dfrac{\lfloor n\hat{t}_{0,\cG}\rfloor -\lfloor nt \rfloor}{\lfloor nt \rfloor^* }\hat{\bbeta}^{(0,\hat{t}_{0,\cG})}+\dfrac{n-\lfloor n\hat{t}_{0,\cG}\rfloor }{\lfloor nt \rfloor^* }\hat{\bbeta}^{(\hat{t}_{0,\cG},1)}+\dfrac{\hat{\bTheta}(\Xb_{(t,1)})^\top\bepsilon^{b}_{(t,1)}}{\lfloor nt \rfloor^*}+\bDelta^{b,(t,1),\rm I},\\
		\end{array}
	\end{equation}
	where $\bDelta^{b,(t,1),\rm I}=(\Delta^{b,(t,1),\rm I}_1,\ldots,\Delta^{b,(t,1),\rm I}_p)^\top$ are defined as
	\begin{equation}\label{equation: bootstraped residual after cpt1}
		\begin{array}{ll}
			\bDelta^{b,(t,1),\rm I}&:=-\dfrac{\lfloor n\hat{t}_{0,\cG}\rfloor -\lfloor nt \rfloor}{\lfloor nt \rfloor^* }\big(\hat{\bTheta}\hat{\bSigma}_{(t,\hat{t}_{0,\cG})}-\Ib\big)\big(\hat{\bbeta}^{(\hat{t}_{0,\cG},1)}-\hat{\bbeta}^{(0,\hat{t}_{0,\cG})} \big)\\
			&-\big(\hat{\bTheta}\hat{\bSigma}_{(t,1)}-\Ib\big)\big(\hat{\bbeta}^{b,(t,1)}-\hat{\bbeta}^{(\hat{t}_{0,\cG},1)} \big).
		\end{array}
	\end{equation}
	$\mathbf{Case~2:}$ The search location $t$ at $t\in[\hat{t}_{0,\cG},1-\tau_0]$. Similar to the analysis of Case 1, using some basic calculations, we can decompose $\breve{\bbeta}^{b,(0,t)}$ and $\breve{\bbeta}^{b,(t,1)}$ into 
	\begin{equation}\label{equation: bootstraped de-biased before+after cpt2}
		\begin{array}{l}
			\breve{\bbeta}^{b,(0,t)}=\dfrac{ \lfloor n\hat{t}_{0,\cG}\rfloor }{\lfloor nt\rfloor}\hat{\bbeta}^{(0,\hat{t}_{0,\cG})}+\dfrac{\lfloor nt \rfloor- \lfloor n\hat{t}_{0,\cG}\rfloor}{\lfloor nt \rfloor}\hat{\bbeta}^{(\hat{t}_{0,\cG},1)}+\dfrac{\hat{\bTheta}(\Xb_{(0,t)})^\top\bepsilon^{b,(0,t)}}{\lfloor nt \rfloor}+\bDelta^{b,(0,t),\rm II},\\\\
			\breve{\bbeta}^{b,(t,1)}=\hat{\bbeta}^{(\hat{t}_{0,\cG},1)}+\dfrac{\hat{\bTheta}(\Xb_{(t,1)})^\top\bepsilon^{b}_{(t,1)}}{\lfloor nt \rfloor^*} +\bDelta^{b,(t,1),\rm II},
		\end{array}
	\end{equation}
	where $\bDelta^{b,(0,t), \rm II}=(\Delta^{b,(0,t), \rm II}_1,\ldots,\Delta^{b,(0,t), \rm II}_p)^\top$ and $\bDelta^{b,(t,1), \rm II}=(\Delta^{b,(t,1) \rm II}_1,\ldots,\Delta^{b,(t,1),\rm II}_p)^\top$ are defined as 
	\begin{equation}\label{equation: bootstraped residual before+after cpt2}
		\begin{array}{ll}
			\bDelta^{b,(0,t), \rm II}&:=-\dfrac{\lfloor nt \rfloor-\lfloor n\hat{t}_{0,\cG}\rfloor}{\lfloor nt \rfloor }\big(\hat{\bTheta}\hat{\bSigma}_{(\hat{t}_{0,\cG},t)}-\Ib\big)\big(\hat{\bbeta}^{(0,\hat{t}_{0,\cG})}-\hat{\bbeta}^{(\hat{t}_{0,\cG},1)} \big)\\
			&-\big(\hat{\bTheta}\hat{\bSigma}_{(0,t)}-\Ib\big)\big(\hat{\bbeta}^{b,(0,t)}-\hat{\bbeta}^{(0,\hat{t}_{0,\cG})} \big),\\\\
			\bDelta^{b,(t,1),\rm II}&:=-\big(\hat{\bTheta}\hat{\bSigma}_{(t,1)}-\Ib\big)\big(\hat{\bbeta}^{b,(t,1)}-\hat{\bbeta}^{(\hat{t}_{0,\cG},1)} \big).
		\end{array}
	\end{equation}
	Based on the above decompositions, we next give a unified form of the de-biased lasso estimator for the bootstrap-based samples. To this end,
	define $\hat{\bdelta}(t)=(\hat{\delta}_1(t),\ldots,\hat{\delta}_p(t))^\top$:
	\begin{equation}
		\hat{\bdelta}(t):=\left\{\begin{array}{ll}
			\dfrac{n-\lfloor n\hat{t}_{0,\cG}\rfloor }{n-\lfloor nt\rfloor }\Big(\hat{\bbeta}^{(0,\hat{t}_{0,\cG})}-\hat{\bbeta}^{(\hat{t}_{0,\cG},1)}\Big),& \text{for}~~ t\in [\tau_0,\hat{t}_{0,\cG}],\\\\
			\dfrac{\lfloor n\hat{t}_{0,\cG}\rfloor }{\lfloor nt\rfloor }\Big(\hat{\bbeta}^{(0,\hat{t}_{0,\cG})}-\hat{\bbeta}^{(\hat{t}_{0,\cG},1)}\Big),& \text{for}~~t\in[\hat{t}_{0,\cG},1-\tau_0].
		\end{array}\right.
	\end{equation}  
	Let $\bDelta^{b,(0,t)}=(\Delta^{b,(0,t)}_1,\ldots,\Delta^{b,(0,t)}_p)^\top$ and $\bDelta^{b,(t,1)}=(\Delta^{b,(t,1)}_1,\ldots,\Delta^{b,(t,1)}_p)^\top$ with
	\begin{equation}\label{equation: bootstraped residual term}
		\begin{array}{l}
			\bDelta^{b,(0,t)}:=\bDelta^{b,(0,t),\rm I}\mathbf{1}\big\{t\in[\tau_0,\hat{t}_{0,\cG}]\big\}+\bDelta^{b,(0,t),\rm II}\mathbf{1}\big\{t\in[\hat{t}_{0,\cG},1-\tau_0]\big\},\\\\
			\bDelta^{b,(t,1)}:=\bDelta^{b,(t,1),\rm I}\mathbf{1}\big\{t\in[\tau_0,\hat{t}_{0,\cG}]\big\}+\bDelta^{b,(t,1),\rm II}\mathbf{1}\big\{t\in[\hat{t}_{0,\cG},1-\tau_0]\big\}.
		\end{array}
	\end{equation}
	With above notations, we are ready to analyze $T_{\cG}^b$. Similar to the analysis of Step~1, for each coordinate $j$ at time point $\lfloor nt \rfloor $, we define the coordinate-wise process as
	\begin{equation}\label{equation: bootstraped partial sum process for each coordinate }
		C_j^b(\lfloor nt \rfloor)=\sqrt{n}\dfrac{\lfloor nt \rfloor}{n}\dfrac{\lfloor nt \rfloor^* }{n}\big(\hat{\sigma}_{\epsilon}^2\hat{\omega}_{j,j }\big)^{-1/2}\big(\breve{\beta}^{b,(0,t)}_j-\breve{\beta}^{b,(t,1)}_j-\hat{\delta}_j(t)\big).
	\end{equation}
	By the definition of $T^b_{\cG}$ in (\ref{equation: bootstrap test statistic}) , we have $T^b_{\cG}=\max\limits_{t\in[\tau_0,1-\tau_0]}\max\limits_{j\in\cG}|C_j^b(\lfloor nt \rfloor)|$. Furthermore, by (\ref{equation: bootstraped de-biased before cpt1}), (\ref{equation: bootstraped de-biased after cpt11}), (\ref{equation: bootstraped de-biased before+after cpt2}), and (\ref{equation: bootstraped residual term}), we can decompose $C^b_j(\lfloor nt \rfloor)$ into 
	\begin{equation}\label{equation: bootstraped decomposition for partial sum process}
		C_j^b(\lfloor nt \rfloor)=C^{b,\rm I}_j(\lfloor nt \rfloor)+C^{b,\rm II}_j(\lfloor nt \rfloor), 
	\end{equation}
	with 
	\begin{equation}\label{equation: bootstraped leading and residual term of partial sum process}
		\small
		\begin{array}{l}
			C^{b,\rm I}_j(\lfloor nt \rfloor):=\sqrt{n}\dfrac{\lfloor nt \rfloor}{n}\dfrac{\lfloor nt \rfloor^* }{n}\dfrac{\big(\dfrac{1}{\lfloor nt \rfloor}\sum\limits_{i=1}^{\lfloor nt \rfloor}\hat{\bTheta}_j^\top\Xb_i\epsilon_i^b-\dfrac{1}{\lfloor nt \rfloor^*}\sum\limits_{i=\lfloor nt \rfloor+1}^{n}\hat{\bTheta}_j^\top\Xb_i\epsilon_i^b\big)}{\sqrt{\hat{\sigma}_{\epsilon}^2\hat{\omega}_{j,j }}},\\\\
			C^{b,\rm II}_j(\lfloor nt \rfloor):=\sqrt{n}\dfrac{\lfloor nt \rfloor}{n}\dfrac{\lfloor nt \rfloor^* }{n}\dfrac{\big(\Delta^{b,(0,t) }_j-\Delta^{b,(t,1) }_j\big)}{\sqrt{\hat{\sigma}_{\epsilon}^2\hat{\omega}_{j,j }}},~\text{with}~1\leq j\leq p~\text{and}~t\in[\tau_0,1-
			\tau_0].
		\end{array}
	\end{equation}
	By replacing $\hat{\sigma}_{\epsilon}^2$, $\hat{\omega}_{j,j }$, and $\hat{\bTheta}_j$ by their true values $\sigma_{\epsilon}^2$, $\omega_{j,j}$, and $\bTheta_j$, for the bootstrap based process $C^{b,\rm I}_j(\lfloor nt \rfloor)$, we can define the oracle leading term as follows:
	\begin{equation}\label{equation: bootstraped oracle leading term}
		\tilde{C}^{b,\rm I}_j(\lfloor nt \rfloor):=\sqrt{n}\dfrac{\lfloor nt \rfloor}{n}\dfrac{\lfloor nt \rfloor^* }{n}\dfrac{\Big(\dfrac{1}{\lfloor nt \rfloor}\sum\limits_{i=1}^{\lfloor nt \rfloor}\bTheta_j^\top\Xb_i\epsilon_i^b-\dfrac{1}{\lfloor nt \rfloor^*}\sum\limits_{i=\lfloor nt \rfloor+1}^{n}\bTheta_j^\top\Xb_i\epsilon_i^b\Big)}{\sqrt{\sigma_{\epsilon}^2\omega_{j,j }}}.
	\end{equation}
	Based on (\ref{equation: bootstraped decomposition for partial sum process}),    (\ref{equation: bootstraped leading and residual term of partial sum process}), and (\ref{equation: bootstraped oracle leading term}), define the following four vector-valued processes:
	\begin{equation}\label{equation: bootstraped 4 vectorized process}
		\small
		\begin{array}{ll}
			\bC^b(\floor{nt})=\big(C^b_1(\floor{nt}),\ldots,C^b_p(\floor{nt})\big)^\top,& \bC^{b,\rm I}(\floor{nt})=\big(C^{b,\rm I}_1(\floor{nt}),\ldots,C^{b,\rm I}_p(\floor{nt})\big)^\top,\\\\ \bC^{b,\rm II}(\floor{nt})=\big(C^{b,\rm II}_1(\floor{nt}),\ldots,C^{b,\rm II}_p(\floor{nt})\big)^\top,& \tilde{\bC}^{b,\rm I}(\floor{nt})=\big(\tilde{C}_{1}^{b,\text{\rm I}}(\floor{nt}),\ldots,\tilde{C}^{b,\text{I}}_{p}(\floor{nt})\big)^{\top}.
		\end{array}
	\end{equation}
	The following Lemma \ref{lemma: bootstraped negligible} shows that the residual term $C^{b,\rm II}_j(\lfloor nt \rfloor)$ can be uniformly negligible over $t\in[\tau_0,1-\tau_0]$ and $1\leq j\leq p$. The proof of Lemma \ref{lemma: bootstraped negligible} is given in Section \ref{section: proof of bootstraped negligible}.
	\begin{lemma}\label{lemma: bootstraped negligible}
		Assume Assumptions $\mathbf{(A.1)}$ -- $\mathbf{(A.5)}$ hold. Under $\Hb_0$, we have
		\begin{equation}\label{equation: bootstraped negligible}
			\P\Big(\max_{\tau_0\leq t\leq 1-\tau_0} \big\|\bC^b(\floor{nt})-\tilde{\bC}^{b,\text{I}}(\floor{nt})\big\|_{\cG,\infty}\geq \epsilon|\cX \Big)=o(1),
		\end{equation}
		where
		$\epsilon=C\max(\max\limits_{1\leq j\leq p}s_j\dfrac{\log(pn)}{\sqrt{ n }},s\sqrt{n}\dfrac{\log(pn)}{{\lfloor n\tau_0 \rfloor}})$, and $C$ is a universal constant not depending on $n$ or $p$.
	\end{lemma}	
	\textbf{Step~3} (Gaussian approximation). In Step~1, we have defined the oracle leading term $\tilde{\bC}^{\rm I}(\lfloor nt \rfloor)$.
	Let
	\begin{equation}
		\Vb=\diag\big((\omega_{1,1}\sigma_{\epsilon}^2)^{-\frac{1}{2}},\ldots,(\omega_{p,p}\sigma_{\epsilon}^2)^{-\frac{1}{2}}\big).
	\end{equation}
	By the definition of $\tilde{\bC}^{\rm I}(\lfloor nt \rfloor)$ in (\ref{equation: 4 vectorized process}), we can rewrite it  in the  form of partial sum process:
	\begin{equation}
		\tilde{\bC}^{\rm I}(\lfloor nt \rfloor)=\dfrac{1}{\sqrt{n}}\sum_{i=1}^{n}\Vb\cdot \bTheta\Xb_i\epsilon_i\Big(\mathbf{1}(i\leq\floor{nt})-\dfrac{\lfloor nt \rfloor }{n}\Big),~\text{with}~ \tau_0\leq t\leq 1-\tau_0.
	\end{equation}
	In Step 2, we have introduced the oracle leading term $\tilde{\bC}^{b,\rm I}(\floor{nt})$ in (\ref{equation: bootstraped 4 vectorized process}) for the bootstrap based test statistic. Similar to $\tilde{\bC}^{\rm I}(\lfloor nt \rfloor)$, we can write $\tilde{\bC}^{b,\rm I}(\lfloor nt \rfloor)$
	in the following form:
	\begin{equation}
		\tilde{\bC}^{b,\rm I}(\lfloor nt \rfloor)=\dfrac{1}{\sqrt{n}}\sum_{i=1}^{n}\Vb\cdot \bTheta\Xb_i\epsilon^b_i\big(\mathbf{1}(i\leq\floor{nt})-\floor{nt}/n\big),~\text{with}~ \tau_0\leq t\leq 1-\tau_0.
	\end{equation}
	Let $\bZ_i=\Vb\cdot \bTheta\Xb_i\epsilon_i$ and $\bG_i=\Vb\cdot \bTheta\Xb_i\epsilon_i^b$ for $i=1,\ldots,n$. Note that $\bG_i$ follows multivariate Gaussian distributions with mean zero and the same covariance matrix as $\bZ_i$. We aim to use $\max_{\tau_0\leq t\leq 1-\tau_0}\|\tilde{\bC}^{b,\rm I}(\floor{nt})\|_{\cG,\infty}$ to approximate $\max_{\tau_0\leq t\leq 1-\tau_0}\|\tilde{\bC}^{\rm I}(\floor{nt})\|_{\cG,\infty}$. Hence, it remains to verify that the conditions of Lemma \ref{lemma: key lemma for gaussian approximation} hold. 
	In fact, by Assumptions $\mathbf{(A.1)}$ and $\mathbf{(A.2)}$, we can show that Assumptions $\mathbf{(M1)}$ - $\mathbf{(M3)}$ hold for $\Vb\cdot \bTheta\Xb_i\epsilon_i$ with $1\leq i\leq n$. Hence, by Lemma \ref{lemma: key lemma for gaussian approximation}, we have
	\begin{equation}
		\small
		\sup_{z\in(0,\infty)} \big|\P(\max_{\tau_0\leq t\leq 1-\tau_0}\|\tilde{\bC}^{\rm I}(\floor{nt})\|_{\cG,\infty}\leq z\big)-\P(\max_{\tau_0\leq t\leq 1-\tau_0}\|\tilde{\bC}^{b,\rm I}(\floor{nt})\|_{\cG,\infty}\leq z\big)\big|\leq Cn^{-\zeta_0}.
	\end{equation}
	\textbf{Step~4}. In this step, we aim to combine the previous results to prove
	\begin{equation}\label{equation: final proof of gaussian approximation}
		\sup_{z\in (0,\infty)}\big|\P(T_{\cG}\leq z)-\P(T_{\cG}^{b}\leq z|\mathcal{X})\big|=o_p(1), ~\text{as}~  n,p\rightarrow\infty.
	\end{equation}
	In particular, we need to obtain the upper and lower bounds of $\rho_0$, where 
	\begin{equation}
		\rho_0:=\P(T_{\cG}> z)-\P(T_{\cG}^{b}> z|\mathcal{X}).
	\end{equation}
	We first consider the upper bound. Note that $T_{\cG}=\max\limits_{t\in[\tau_0,1-\tau_0]}\|\bC(\lfloor nt \rfloor)\|_{\cG,\infty}$. By plugging $\tilde{\bC}^{\rm I}(\lfloor nt \rfloor)$ in $T_{\cG}$ and using 
	the triangle inequality of $\|\cdot\|_{\cG,\infty}$, we have
	\begin{equation}
		\P(T_{\cG}> z)\leq \P(\max_{t\in[\tau_0,1-\tau_0]}\|\tilde{\bC}^{\rm I}(\lfloor nt \rfloor)\|_{\cG,\infty}>z-\epsilon)+\rho_1,
	\end{equation}
	where $\rho_1:=\P(\max\limits_{t\in[\tau_0,1-\tau_0]}\|\bC(\lfloor nt \rfloor)-\tilde{\bC}^{\rm I}(\lfloor nt \rfloor)\|_{\cG,\infty}>\epsilon)$. By Lemma \ref{lemma: negligible}, we have $\rho_1=o(1)$. Recall $\tilde{\bC}^{b,\rm I}(\lfloor nt \rfloor)$ defined in (\ref{equation: bootstraped 4 vectorized process}). For $\P(\max_{t\in[\tau_0,1-\tau_0]}\|\tilde{\bC}^{\rm I}(\lfloor nt \rfloor)\|_{\cG,\infty}>z-\epsilon)$, we then have
	\begin{equation}
		\P(\max_{t\in[\tau_0,1-\tau_0]}\|\tilde{\bC}^{\rm I}(\lfloor nt \rfloor)\|_{\cG,\infty}>z-\epsilon)\leq \underbrace{\P(\max_{t\in[\tau_0,1-\tau_0]}\|\tilde{\bC}^{b, \rm I}(\lfloor nt \rfloor)\|_{\cG,\infty}>z-\epsilon|\cX)}_{\rho_3}+\rho_2,
	\end{equation}
	where $$\rho_2:=\sup_{x\in (0,\infty)}|\P(\max_{t\in[\tau_0,1-\tau_0]}\|\tilde{\bC}^{\rm I}(\lfloor nt \rfloor)\|_{\cG,\infty}>x)-\P(\max_{t\in[\tau_0,1-\tau_0]}\|\tilde{\bC}^{b, \rm I}(\lfloor nt \rfloor)\|_{\cG,\infty}>x|\cX) |.$$ By Step 3, we have proved $\rho_2\leq Cn^{-\zeta_0}$ holds. For $\rho_3$, we have $\rho_3=\rho_4+\rho_5$, where 
	\begin{equation}
		\begin{array}{l}
			\rho_4:=\P(z-\epsilon<\max\limits_{t\in[\tau_0,1-\tau_0]}\|\tilde{\bC}^{b, \rm I}(\lfloor nt \rfloor)\|_{\cG,\infty}<z+\epsilon|\cX),\\
			\rho_5:=\P(\max\limits_{t\in[\tau_0,1-\tau_0]}\|\tilde{\bC}^{b, \rm I}(\lfloor nt \rfloor)\|_{\cG,\infty}>z+\epsilon|\cX).
		\end{array}
	\end{equation}
	By Lemma \ref{lemma:anti consentration inequality}, we have proved $\rho_4=o_p(1)$. So far, we have proved that
	\begin{equation}\label{inequality: last2}
		\P(T_{\cG}> z)\leq \P(\max_{t\in[\tau_0,1-\tau_0]}\|\tilde{\bC}^{b,\rm I}(\floor{nt})\|_{\cG,\infty}>z+\epsilon|\cX)+o_p(1).
	\end{equation} 
	Note that $T^b_{\cG}:=\max_{t\in[\tau_0,1-\tau_0]}\|\bC^b(\lfloor nt \rfloor)\|_{\cG,\infty}$. By the triangle inequality, we have
	\begin{equation}\label{inequality: rho7}
		\P(\max_{t\in[\tau_0,1-\tau_0]}\|\tilde{\bC}^{b,\rm I}(\floor{nt})\|_{\cG,\infty}>z+\epsilon|\cX)\leq \P(\max_{t\in[\tau_0,1-\tau_0]}\|\bC^b(\lfloor nt \rfloor)\|_{\cG,\infty}>z|\cX)+\rho_6,
	\end{equation} 
	where $\rho_6:=\P(\max_{t\in[\tau_0,1-\tau_0]}\|\bC^b(\lfloor nt \rfloor)-\tilde{\bC}^{b,\rm I}(\floor{nt})\|_{\cG,\infty}>\epsilon|\cX)$. By Lemma \ref{lemma: bootstraped negligible}, we have  proved $\rho_6=o_p(1)$. Combining (\ref{inequality: last2}) and (\ref{inequality: rho7}), we have
	\begin{equation}\label{inequality: last1}
		\P(T_{\cG}> z)\leq \P(T^b_{\cG}>z|\cX)+o_p(1).
	\end{equation} 
	With a similar proof technique, we can also obtain the lower bound and prove
	\begin{equation}
		|\P(T_{\cG}> z)- \P(T^b_{\cG}>z|\cX)|=o_p(1)
	\end{equation}
	holds uniformly in $z\in(0,\infty)$, which finishes the proof of Theorem \ref{theorem: bootstrap validity}.
\end{proof}
\subsection{Proof of Theorem \ref{theorem: cpt estimation results}}
\begin{proof}
	Without loss of generality, we assume $\delta_j:=\beta^{(1)}_j-\beta^{(2)}_j\geq 0$. As a mild technical assumption, throughout this section, we assume  ${s\sqrt{\log(p)/n\tau_0}\|\bdelta\|_{\infty}/\|\bdelta\|_{\cG,\infty}=o(1)}$. For each $t\in[\tau_0,1-\tau_0]$, define
	$\bZ(\lfloor nt \rfloor)=\big(Z_1(\lfloor nt \rfloor),\ldots,Z_p(\lfloor nt \rfloor)\big)^\top$ with
	\begin{equation}\label{equation: zj(nt)}
		Z_j(\lfloor nt \rfloor):=\sqrt{n}\dfrac{\lfloor nt \rfloor}{n}\Big(1-\dfrac{\lfloor nt \rfloor }{n}\Big)(\breve{\beta}^{(0,t)}_j-\breve{\beta}^{(t,1)}_j),~\text{for}~1\leq j\leq p.
	\end{equation}
	Note that there is no variance estimator in $Z_j(\lfloor nt \rfloor)$.
	By definition, we have $$\hat{t}_{0,\cG}:=\argmax\limits_{t\in[\tau_0,1-\tau_0]}\|\bZ(\lfloor nt \rfloor)\|_{\cG,\infty}.$$ 
	For notational simplicity, we abbreviate $\hat{t}_{0,\cG}$ to $\hat{t}_0$. Moreover, we assume $\hat{t}_0\in[t_0,1-\tau_0]$. To give the proof, we need to make decompositions on $\bZ(\lfloor nt \rfloor)$. We first define $\bdelta(t)=(\delta_1(t),\ldots,\delta_p(t))^\top$:
	\begin{equation}\label{equation: delta (nt)}
		\begin{array}{ll}
			\bdelta(t):=\sqrt{n}\dfrac{\lfloor nt \rfloor}{n}
			\dfrac{\lfloor nt_0\rfloor^* }{n}\big(\bbeta^{(1)}-\bbeta^{(2)}\big)\mathbf{1}\{ t\in [\tau_0,t_0]\}\\
			\quad\quad\quad+
			\sqrt{n}\dfrac{\lfloor nt_0\rfloor }{n }\dfrac{\lfloor nt \rfloor ^*}{n}\big(\bbeta^{(1)}-\bbeta^{(2)}\big) \mathbf{1}\{t\in[t_0,1-\tau_0]\},
		\end{array}
	\end{equation}  
	and $\bR^{(0,t)}=(R^{(0,t)}_1,\ldots,R^{(0,t)}_p)^\top$ and $\bR^{(t,1)}=(R^{(t,1)}_1,\ldots,R^{(t,1)}_p)^\top$:
	\begin{equation}\label{inequality: R(nt) and R(nt2)}
		\begin{array}{l}
			\bR^{(0,t)}:=\bR^{(0,t),\rm I}\mathbf{1}\big\{t\in[\tau_0,t_0]\big\}+\bR^{(0,t),\rm II}\mathbf{1}\big\{t\in[t_0,1-\tau_0]\big\},\\\\
			\bR^{(t,1)}:=\bR^{(t,1),\rm I}\mathbf{1}\big\{t\in[\tau_0,t_0]\big\}+\bR^{(t,1),\rm II}\mathbf{1}\big\{t\in[t_0,1-\tau_0]\big\},
		\end{array}
	\end{equation}
	where $	\bR^{(0,t),\rm I} - \bR^{(t,1),\rm II}$ are defined as
	\begin{equation}\label{equation: residual before cpt1}
		\begin{array}{ll}
			\bR^{(0,t),\rm I}&:=-\big(\hat{\bTheta}\hat{\bSigma}_{(0,t)}-\Ib\big)\big(\hat{\bbeta}^{(0,t)}-\bbeta^{(1)} \big),\\\\
			\bR^{(0,t), \rm II}&:=-\dfrac{\lfloor nt\rfloor-\lfloor nt_0 \rfloor}{\lfloor nt \rfloor }\big(\hat{\bTheta}\hat{\bSigma}_{(t_0,t)}-\Ib\big)\big(\bbeta^{(1)}-\bbeta^{(2)} \big)-\big(\hat{\bTheta}\hat{\bSigma}_{(0,t)}-\Ib\big)\big(\hat{\bbeta}^{(0,t)}-\bbeta^{(1) } \big),\\\\
			\bR^{(t,1),\rm I}&:=-\dfrac{\lfloor nt_0\rfloor -\lfloor nt \rfloor}{\lfloor nt \rfloor^* }\big(\hat{\bTheta}\hat{\bSigma}_{(t,t_0)}-\Ib\big)\big(\bbeta^{(1)}-\bbeta^{(2)} \big)-\big(\hat{\bTheta}\hat{\bSigma}_{(t,1)}-\Ib\big)\big(\hat{\bbeta}^{(t,1)}-\bbeta^{(2)}\big),\\\\
			\bR^{(t,1),\rm II}&:=-\big(\hat{\bTheta}\hat{\bSigma}_{(t,1)}-\Ib\big)\big(\hat{\bbeta}^{(t,1)}-\bbeta^{(2)} \big).
		\end{array}
	\end{equation}
	Then, by the definitions of $\breve{\bbeta}^{(0,t)}$ and $\breve{\bbeta}^{(t,1)}$, similar to the analysis of Step 2 in Section \ref{section: proof of gaussian approximation},
	under $\Hb_1$, we can write $\bZ(\lfloor nt \rfloor)$ as follows:
	\begin{equation}\label{equation: decomposition under H1}
		\bZ(\lfloor nt \rfloor)=\bdelta(t)+\sqrt{n}\dfrac{\lfloor nt \rfloor}{n}\dfrac{\lfloor nt \rfloor^* }{n}\big(\dfrac{1}{\lfloor nt \rfloor}\sum\limits_{i=1}^{\lfloor nt \rfloor}\hat{\bxi}_{i}-\dfrac{1}{\lfloor nt \rfloor^*}\sum\limits_{i=\lfloor nt \rfloor+1}^{n}\hat{\bxi}_{i}+\bR^{(0,t)}-\bR^{(t,1)}\big),
	\end{equation}
	where
	$\hat{\bxi}_i:=(\hat{\xi}_{i,1},\ldots,\hat{\xi}_{i,p})^\top$ with $\hat{\xi}_{i,j}=\hat{\bTheta}_j^\top\Xb_i\epsilon_i$ for $i=1,\ldots,n$ and $j=1,\ldots,p$. 
	
	In addition to the decomposition, let $\bdelta=\bbeta^{(1)}-\bbeta^{(2)}$ and we assume 
	\begin{equation*}
		\|\bdelta\|_{\cG,\infty}\gg \sqrt{\dfrac{\log(|\cG|n)}{n}}.
	\end{equation*}
	Let $j^*\in\cG$ such that $Z_{j^*}(\lfloor n\hat{t}_0 \rfloor)=\max\limits_{j\in\cG}Z_j(\lfloor n\hat{t}_0 \rfloor)$. The following Lemma \ref{lemma: maximum signal at the identified cpt} shows that $\liminf_{n\rightarrow \infty}\delta_{j^*}/\|\bdelta\|_{\cG,\infty}\geq 1$. The proof of Lemma \ref{lemma: maximum signal at the identified cpt} is given in Section \ref{section: proof of maximum signal at the identified cpt}.
	\begin{lemma}\label{lemma: maximum signal at the identified cpt}
		Suppose Assumptions $\mathbf{(A.1)}$ -- $\mathbf{(A.5)}$ hold. Then, with probability tending to one,  we have
		$\liminf_{n\rightarrow \infty}\delta_{j^*}/\|\bdelta\|_{\cG,\infty}\geq 1$.
	\end{lemma}
	Furthermore, define the event 
	\begin{equation}
		\begin{array}{ll}
			\cH_1=\Big\{\max_{j\in\cG}Z_j(\lfloor n\hat{t}_0 \rfloor)=\max_{j\in\cG}|Z_j(\lfloor n\hat{t}_0 \rfloor)|:=\|\bZ(\lfloor n\hat{t}_0 \rfloor)\|_{\cG,\infty}\Big\},\\
			\cH_2=\Big\{Z_{j^*}(\lfloor n{t}_0 \rfloor)=|Z_{j^*}(\lfloor n{t}_0 \rfloor)|\Big\}.
		\end{array}	
	\end{equation}
	The following Lemma \ref{lemma: sign consistent} shows that $\cH_1\cap\cH_2$ occurs with high probability. The proof of Lemma \ref{lemma: sign consistent} is provided in Section \ref{section: proof of sign consistent}.
	\begin{lemma}\label{lemma: sign consistent}
		Suppose Assumptions $\mathbf{(A.1)}$ -- $\mathbf{(A.5)}$ hold. Then we have
		\begin{equation}
			\P(\cH_1\cap\cH_2)\geq 1-C_1(np)^{-C_2},
		\end{equation}
		where $C_1$ and $C_2$ are universal positive constants not depending on $n$ or $p$.
	\end{lemma}
	Using Lemmas \ref{lemma: maximum signal at the identified cpt} and \ref{lemma: sign consistent}, we are ready to give the proof. Specifically, by the above two lemmas, we have:
	\begin{equation*}
		\begin{array}{ll}
			\|\bZ(\lfloor nt_0 \rfloor)\|_{\cG,\infty}-	\|\bZ(\lfloor n\hat{t}_0 \rfloor)\|_{\cG,\infty}
			&=\|\bZ(\lfloor nt_0 \rfloor)\|_{\cG,\infty}-Z_{j^*}(\lfloor n\hat{t}_0 \rfloor)\\
			&\geq Z_{j^*}(\lfloor n{t}_0 \rfloor)-Z_{j^*}(\lfloor n\hat{t}_0 \rfloor).	
		\end{array}
	\end{equation*}
	Moreover, by the decomposition of $\bZ(\lfloor nt \rfloor)$ in (\ref{equation: decomposition under H1}), we have:
	\begin{equation}\label{equation: cpt est key1}
		Z_{j^*}(\lfloor n{t}_0 \rfloor)-Z_{j^*}(\lfloor n\hat{t}_0 \rfloor)\geq \sqrt{n}\dfrac{\lfloor nt_0\rfloor }{n }\dfrac{\floor{n\hat{t}_0}-\floor{nt_0}}{n}\delta_{j^*}+I-II,
	\end{equation}
	where
	\begin{equation}
		\begin{array}{ll}
			I=\dfrac{1}{\sqrt{n}}\big( \sum\limits_{i=\lfloor nt_0 \rfloor+1}^{\lfloor n\hat{t}_0 \rfloor}\hat{\xi}_{i,j}-\dfrac{\lfloor n\hat{t}_0 \rfloor-\lfloor nt_0 \rfloor}{n} \sum\limits_{i=1}^{n}\hat{\xi}_{i,j}\big),\\
			II=\sqrt{n}\dfrac{\lfloor n\hat{t}_0 \rfloor}{n}\dfrac{\lfloor n\hat{t}_0 \rfloor^* }{n}\big( \|\bR^{(0,\hat{t}_0),II}\|_{\infty}+\|\bR^{(\hat{t}_0,1),II}\|_{\infty}\big)\\
			\qquad\qquad\qquad+\sqrt{n}\dfrac{\lfloor n{t}_0 \rfloor}{n}\dfrac{\lfloor n{t}_0 \rfloor^* }{n}\big( \|\bR^{(0,{t}_0),II}\|_{\infty}+\|\bR^{({t}_0,1),II}\|_{\infty}\big).
		\end{array}
	\end{equation}
	Note that  by Assumptions $\mathbf{(A.1)}$ -- $\mathbf{(A.3)}$, $\hat{\xi}_{i,j}$ follows the sub-exponential distribution. Using Bernstein's inequalities, we can prove that:
	\begin{equation}\label{equation: cpt est key2}
		\max_{t\in[t_0,1-\tau_0]}\max_{j\in \cG}\dfrac{	|\dfrac{1}{\sqrt{n}}\big( \sum\limits_{i=\lfloor nt_0 \rfloor+1}^{\lfloor nt \rfloor}\hat{\xi}_{i,j}-\dfrac{\lfloor nt \rfloor-\lfloor nt_0 \rfloor}{n} \sum\limits_{i=1}^{n}\hat{\xi}_{i,j}\big)|}{(\floor{nt}-\floor{nt_0})^{1/2}}=O_p(\sqrt{\dfrac{\log(|\cG|)}{n}}).
	\end{equation}
	Moreover, the following Lemma \ref{lemma: three terms for the residual terms of cpt} shows that $II$ can be decomposed into three terms. The proof of Lemma \ref{lemma: three terms for the residual terms of cpt} is given in Section \ref{section: proof of three terms for the residual terms of cpt}.
	\begin{lemma}\label{lemma: three terms for the residual terms of cpt}
		Suppose Assumptions $\mathbf{(A.1)}$ -- $\mathbf{(A.5)}$ hold. For $II$ in (\ref{equation: cpt est key1}), with probability tending to 1, we have
		\begin{equation*}
			II\leq C_1\sqrt{{\log(|\cG|n)}\dfrac{\lfloor n\hat{t}_0 \rfloor-\lfloor nt_0\rfloor}{n}}s\|\bdelta\|_{\infty}+C_2\sqrt{n}s\dfrac{\log(|\cG|n)}{{n}}+o\big(\sqrt{n}\dfrac{\lfloor nt_0 \rfloor}{n}\dfrac{\lfloor n\hat{t}_0\rfloor -\lfloor nt_0\rfloor}{n}\|\bdelta\|_{\cG,\infty}\big).
		\end{equation*}	
		where $C_1,C_2>0$ are some constants not depending on $n$ or $p$.
	\end{lemma}
	Considering (\ref{equation: cpt est key1}) - (\ref{equation: cpt est key2}), by Lemmas \ref{lemma: maximum signal at the identified cpt} and \ref{lemma: three terms for the residual terms of cpt}, we have:
	\begin{equation}
		\begin{array}{ll}
			\|\bZ(\lfloor nt_0 \rfloor)\|_{\cG,\infty}-	\|\bZ(\lfloor n\hat{t}_0 \rfloor)\|_{\cG,\infty}\\
			\geq Z_{j^*}(\lfloor n{t}_0 \rfloor)-Z_{j^*}(\lfloor n\hat{t}_0 \rfloor)\\
			\geq \sqrt{n}\dfrac{\lfloor nt_0\rfloor }{n }\dfrac{\floor{n\hat{t}_0}-\floor{nt_0}}{n}\|\bdelta\|_{\cG,\infty}-C_1\sqrt{\dfrac{\floor{n\hat{t}_0}-\floor{nt_0}\log(|\cG|)}{n}}\\
			-C_2\sqrt{{\log(|\cG|n)}\dfrac{\lfloor n\hat{t}_0 \rfloor-\lfloor nt_0\rfloor}{n}}s\|\bdelta\|_{\infty}-C_3\sqrt{n}s\dfrac{\log(|\cG|n)}{{n}}-o\big(\sqrt{n}\dfrac{\lfloor nt_0 \rfloor}{n}\dfrac{\lfloor n\hat{t}_0\rfloor -\lfloor n\tau_0\rfloor}{n}\|\bdelta\|_{\cG,\infty}\big).
		\end{array}
	\end{equation}
	Note that $	\|\bZ(\lfloor nt_0 \rfloor)\|_{\cG,\infty}-	\|\bZ(\lfloor n\hat{t}_0 \rfloor)\|_{\cG,\infty}\leq 0$. Hence, by (\ref{equation: cpt est key2}), we have:
	\begin{equation*}
		\begin{array}{ll}
			\dfrac{1}{2}\sqrt{n}\dfrac{\lfloor nt_0\rfloor }{n }\dfrac{\floor{n\hat{t}_0}-\floor{nt_0}}{n}\|\bdelta\|_{\cG,\infty}\\
			\leq 3\max\Big(C_1\sqrt{\dfrac{\floor{n\hat{t}_0}-\floor{nt_0}\log(|\cG|)}{n}},  C_2\sqrt{{\log(|\cG|n)}\dfrac{\lfloor n\hat{t}_0 \rfloor-\lfloor nt_0\rfloor}{n}}s\|\bdelta\|_{\infty}, C_3\sqrt{n}s\dfrac{\log(|\cG|n)}{{n}}\Big).
		\end{array}
	\end{equation*}
	This implies that with probability tending to 1, we must have
	\begin{equation*}
		\begin{array}{ll}
			\dfrac{\floor{n\hat{t}_0}-\floor{nt_0}}{n}&\leq C^*\max\Big(   \dfrac{\log(|\cG|)}{n\|\bdelta\|^2_{\cG,\infty}},  \dfrac{\log(|\cG|)s^2\|\bdelta\|_\infty^2}{n\|\bdelta\|^2_{\cG,\infty}},\dfrac{\log(|\cG|)s\|\bdelta\|_\infty}{n\|\bdelta\|^2_{\cG,\infty}}   \Big)\leq C^*\dfrac{\log(|\cG|)}{n\|\bdelta\|^2_{\cG,\infty}},
		\end{array}
	\end{equation*}
	where the second inequality comes from Assumption {(\bf{A.6})}, and $C^*$ is some big enough constant not depending on $n$ or $p$, which completes the proof of Theorem \ref{theorem: cpt estimation results}.
\end{proof}

\subsection{Proof of Theorem \ref{theorem: variance estimation under H1}}
\begin{proof}
	Note that by  (\ref{inequality: upper bounds for omega matrix}) in Theorem \ref{theorem: variance estimator under H0} and by Assumption $\mathbf{(A.4)}$, we have shown that 	$\max_{1\leq j,k\leq p}|\hat{\omega}_{j,k}-\omega_{j,k}|=o_p(1)$. Hence, to prove Theorem \ref{theorem: variance estimation under H1}, it remains to prove that $|\hat{\sigma}^2_{\epsilon}-\sigma^2_{\epsilon}|=o_p(1)$, where $\hat{\sigma}^2_{\epsilon}$ is the weighted variance estimator as defined in (\ref{equation: variance estimator under H1}). Without loss of generality, we assume the change point estimator $\hat{t}_{0,\cG}\in[\tau_0,t_0]$, where $\hat{t}_{0,\cG}$ is obtained by (\ref{equation: t-hat}). To simplify notations, we denote $\hat{t}_{0,\cG}$ by $\hat{t}_{0}$. Thoughout this section, we denote 
	\begin{equation*}
		\epsilon_n:=\dfrac{\floor{nt_0}-\floor{n\hat{t}_0}}{n},~~\text{and}~\bdelta=\bbeta^{(1)}-\bbeta^{(2)}.
	\end{equation*}
	Furthermore, by definition, we can write $\hat{\sigma}^2_{\epsilon}-\sigma_{\epsilon}^2$ as the following 8 parts:
	\begin{equation}\label{equation: eight parts}
		\hat{\sigma}^2_{\epsilon}-\sigma_{\epsilon}^2=I+II+III+IV+V+VI+VII+VIII,
	\end{equation}
	where $I-VIII$ are defined as
	\begin{equation}\label{inequality: 8 parts}
		\begin{array}{ll}
			I=\dfrac{1}{n}\sum\limits_{i=1}^n(\epsilon_i^2-\sigma_{\epsilon}^2),  \quad\quad\quad\quad \quad II=\dfrac{1}{n}\Big\|\Xb_{(0,\hat{t}_0)}\big(\hat{\bbeta}^{(0,\hat{t}_0)}-\bbeta^{(1)}\big)\Big\|_2^2,\\
			III=\dfrac{2}{n}(\bepsilon_{(0,\hat{t}_0)})^\top\Xb_{(0,\hat{t}_0) }\big(\bbeta^{(1)}-\hat{\bbeta}^{(0,\hat{t}_0)}\big),\quad\quad IV=\dfrac{1}{n}\Big\|\Xb_{(\hat{t}_0,1)}\big(\hat{\bbeta}^{(\hat{t}_0,1)}-\bbeta^{(2)}\big)\Big\|_2^2,\\
			V=\dfrac{2}{n}(\bepsilon_{(\hat{t}_0,1) })^\top\Xb_{(\hat{t}_0,1) }\big(\bbeta^{(2)}-\hat{\bbeta}^{(\hat{t}_0,1)}\big), \quad\quad VI=\dfrac{2}{n} (\bepsilon_{(\hat{t}_0,t_0)})^\top\Xb_{(\hat{t}_0,t_0)}(\bbeta^{(1)}-\bbeta^{(2)}),\\
			VII=\dfrac{1}{n}(\bbeta^{(1)}-\bbeta^{(2)})^\top(\Xb_{(\hat{t}_0,t_0)})^\top\Xb_{(\hat{t}_0,t_0)}(\bbeta^{(1)}-\bbeta^{(2)}),\\
			VIII=\dfrac{1}{n}(\bbeta^{(1)}-\bbeta^{(2)})^\top(\Xb_{(\hat{t}_0,t_0)})^\top\Xb_{(\hat{t}_0,t_0)}(\bbeta^{(2)}-\hat{\bbeta}^{(\hat{t}_0,1)}).
		\end{array}
	\end{equation}
	By	(\ref{equation: eight parts}), we need to bound the eight parts on the RHS of (\ref{equation: eight parts}), respectively. For the rest of the proof, we assume the event $\cap_{t\in[\tau_0,1-\tau_0]}\big\{\cA(t)\cap\cB(t)\big\} $ holds.
	For $I$, using (\ref{inequality: upper bounds for residual term}) in Theorem \ref{theorem: variance estimator under H0}, we have $I=o(1)$ as $n,p\rightarrow\infty$. For $II$, by Lemma \ref{lemma: esimation error bound}, we have $II\leq Cs^{(1)}\dfrac{\log(p)}{n}=o(1)$ as $n,p\rightarrow\infty$. For $III$, by Lemma \ref{lemma: esimation error bound} and Assumption $\mathbf{(A.4)}$, we have
	\begin{equation*}\label{inequality: III}
		\begin{array}{ll}
			III&\leq C\dfrac{\lfloor n\hat{t}_0 \rfloor}{n}\sqrt{\dfrac{\log(p)}{\lfloor n\hat{t}_0 \rfloor}}\|\hat{\bbeta}^{(0,\hat{t}_0)}-\bbeta^{(1)}\|_1,\\
			&\leq Cs^{(1)}\dfrac{\log(p)}{\lfloor n\hat{t}_0 \rfloor}\leq Cs^{(1)}\dfrac{\log(p)}{\lfloor n\tau_0 \rfloor}=o_p(1).
		\end{array}
	\end{equation*}
	Recall $s=s^{(1)}\vee s^{(2)}$.	For $IV$, by Lemma \ref{lemma: esimation error bound} and Assumption $\mathbf{(A.4)}$, we have
	\begin{equation*}\label{inequality: IV}
		\begin{array}{ll}
			IV&\leq C\Big( \dfrac{\lfloor nt_0 \rfloor -\lfloor n\hat{t}_0 \rfloor}{\lfloor n\hat{t}_0 \rfloor^*}\Big)^2 \big\|\bbeta^{(2)}-\bbeta^{(1)}\big\|_2^2,\\
			&\leq C\Big(\dfrac{\lfloor nt_0 \rfloor -\lfloor n\hat{t}_0 \rfloor}{n}\Big)^2 \big\|\bbeta^{(2)}-\bbeta^{(1)}\big\|_2^2=O_p(\epsilon_n^2s\|\bdelta\|^2_{\infty}).
		\end{array}
	\end{equation*}	
	For $V$, by Lemma \ref{lemma: esimation error bound} and Assumption $\mathbf{(A.4)}$, we have
	\begin{equation*}\label{inequality: V}
		\begin{array}{ll}
			|V|&\leq C\dfrac{\lfloor n\hat{t}_0 \rfloor^*}{n}\sqrt{\dfrac{\log(p)}{\lfloor n\hat{t}_0 \rfloor^*}}\big\|\hat{\bbeta}^{(\hat{t}_0,1)}-\bbeta^{(2)}\big\|_1,\\
			&\leq C\dfrac{\lfloor n\hat{t}_0 \rfloor^*}{n}\sqrt{\dfrac{\log(p)}{\lfloor n\hat{t}_0 \rfloor^*}}\times\dfrac{\lfloor nt_0 \rfloor -\lfloor n\hat{t}_0 \rfloor}{\lfloor n\hat{t}_0 \rfloor^*} \big\|\bbeta^{(2)}-\bbeta^{(1)}\big\|_1,\\
			&\leq C\sqrt{\dfrac{\log(p)}{\lfloor n\hat{t}_0 \rfloor^*}}\dfrac{\lfloor nt_0 \rfloor -\lfloor n\hat{t}_0 \rfloor}{n} \big\|\bbeta^{(2)}-\bbeta^{(1)}\big\|_1=O_p(\epsilon_ns\sqrt{\dfrac{\log(p)}{n}}\|\bdelta\|_{\infty}).
		\end{array}
	\end{equation*}
	For $VI$, by Assumptions $\mathbf{(A.1)}$ and  $\mathbf{(A.3)}$, we have
	\begin{equation*}\label{inequality: VI}
		\begin{array}{ll}
			|VI|&\leq C\dfrac{\lfloor nt_0 \rfloor -\lfloor n\hat{t}_0 \rfloor}{n}\sqrt{\dfrac{\log(p)}{\lfloor nt_0 \rfloor-\lfloor n\hat{t}_0 \rfloor}}\big\|\bbeta^{(2)}-\bbeta^{(1)}\big\|_1,\\
			&\leq C\sqrt{\dfrac{\lfloor nt_0 \rfloor -\lfloor n\hat{t}_0 \rfloor}{n}}\sqrt{\dfrac{\log(p)}{n}}\big\|\bbeta^{(2)}-\bbeta^{(1)}\big\|_1=O_p(s\sqrt{\epsilon_n\dfrac{\log(p)}{n}}\|\bdelta\|_{\infty}).
		\end{array}
	\end{equation*}
	For $VII$, using the fact that $\bx^\top \Ab \bx\leq \|\Ab\|_{\infty}\|\bx\|_1^2$, we have
	\begin{equation*}
		\begin{array}{ll}
			VII&=_{(1)}	\dfrac{\floor{nt_0}-\floor{n\hat{t}_0}}{n}(\bbeta^{(1)}-\bbeta^{(2)})^\top \hat{\bSigma}_{(\hat{t}_0,t_0)} (\bbeta^{(1)}-\bbeta^{(2)})\\
			&=_{(2)}	\dfrac{\floor{nt_0}-\floor{n\hat{t}_0}}{n}(\bbeta^{(1)}-\bbeta^{(2)})^\top (\hat{\bSigma}_{(\hat{t}_0,t_0)}-\bSigma) (\bbeta^{(1)}-\bbeta^{(2)})\\
			&\quad\quad	+	\dfrac{\floor{nt_0}-\floor{n\hat{t}_0}}{n}(\bbeta^{(1)}-\bbeta^{(2)})^\top \bSigma (\bbeta^{(1)}-\bbeta^{(2)})\\
			&\leq_{(3)} C_1\sqrt{\epsilon_n\dfrac{\log(p)}{n}}\|\bbeta^{(2)}-\bbeta^{(1)}\big\|_1^2+C_2\epsilon_n\|\bbeta^{(2)}-\bbeta^{(1)}\big\|_2^2\\
			&=_{(4)}O_p(s^2\sqrt{\epsilon_n\dfrac{\log(p)}{n}}\|\bdelta\|^2_{\infty}+\epsilon_ns\|\bdelta\|^2_{\infty}),
		\end{array}
	\end{equation*}
	where $(3)$ comes from the concentration inequality for $\|\hat{\bSigma}_{(\hat{t}_0,t_0)}-\bSigma\|_{\infty}$ and by Assumption ${\bf{(A.3)}}$ that $\Sigma_{j,j}=O(1)$. Lastly, for $VIII$,  by Lemma \ref{lemma: esimation error bound}, and similar to $VII$, we have 
	\begin{equation*}\label{inequality: VIII}
		\begin{array}{ll}
			VIII=O_p(\epsilon_n^2s^2\|\bdelta\|_{\infty}^2+s^2\sqrt{\epsilon^3_n\dfrac{\log(p)}{n}}\|\bdelta\|^2_{\infty}).
		\end{array}
	\end{equation*}
	Combining the obtained upper bounds of $I,\ldots,VIII$, we have
	\begin{equation}\label{inequality: variance difference}
		\begin{array}{ll}
			|\hat{\sigma}^2_{\epsilon}-\sigma_{\epsilon}^2|=o_p(1)+O_p(\epsilon_n^2s\|\bdelta\|^2_{\infty})+O_p(\epsilon_ns\sqrt{\dfrac{\log(p)}{n}}\|\bdelta\|_{\infty})+O_p(s\sqrt{\epsilon_n\dfrac{\log(p)}{n}}\|\bdelta\|_{\infty})\\
			\quad+O_p(\epsilon_ns\|\bdelta\|^2_{\infty}+s^2\sqrt{\epsilon_n\dfrac{\log(p)}{n}}\|\bdelta\|^2_{\infty})+O_p(\epsilon_n^2s^2\|\bdelta\|_{\infty}^2+s^2\sqrt{\epsilon^3_n\dfrac{\log(p)}{n}}\|\bdelta\|^2_{\infty}).
		\end{array}
	\end{equation}
	By (\ref{inequality: variance difference}), to bound $|\hat{\sigma}^2_{\epsilon}-\sigma_{\epsilon}^2|$, we consider the following two cases:\\
	$\mathbf{Case 1:}$ The signal satisfies $\|\bdelta\|_{\infty}\gg \sqrt{\log(p)/n}$. In this case, by Theorem \ref{theorem: cpt estimation results}, we have $\epsilon_n=o_p(1)$. Moreover, by Assumption ${\bf{(A.6)}}$, we have $s\|\bdelta\|_{\infty}=O(1)$ and $\|\bdelta\|_{\infty}=o(1)$. Combining (\ref{inequality: variance difference}), we have $	|\hat{\sigma}^2_{\epsilon}-\sigma_{\epsilon}^2|=o_p(1)$.\\
	$\mathbf{Case 2:}$ The signal satisfies $\|\bdelta\|_{\infty}=O (\sqrt{\log(p)/n})$. In this case, we can not obtain a consistent change point estimator. In other words, we only have $\epsilon_{n}=O_p(1)$. Moreover, we can show that 
	\begin{equation}\label{inequality: variance difference2}
		|\hat{\sigma}^2_{\epsilon}-\sigma_{\epsilon}^2|=O_p(s^2\|\bdelta\|^2_{\infty}).
	\end{equation}
	Considering (\ref{inequality: variance difference2}), and  by the assumption that $s\sqrt{\log(p)/n}=o(1)$, we have $	|\hat{\sigma}^2_{\epsilon}-\sigma_{\epsilon}^2|=o_p(1)$, which finishes the proof.
\end{proof}

\subsection{Proof of Theorem \ref{theorem: power results}}	
\begin{proof}
	As a very mild technical assumption, throughout this section, we assume 
	\begin{equation*}
		{s\sqrt{\log(p)/n\tau_0}\|\bdelta\|_{\infty}/\|\bdelta\|_{\cG,\infty}=o(1).}
	\end{equation*}
	The proof of Theorem \ref{theorem: power results} proceeds in two steps. In Step 1, we obtain the upper bound of $c_{T_{\cG}^b}(1-\alpha)$, where $c_{T_{\cG}^b}(1-\alpha)$ is  the $1-\alpha$ quantile of $T^{b}_{\cG}$, which is defined as
	\begin{equation}
		c_{T_{\cG}^b}(1-\alpha):=\inf\big\{t:\P(T_{\cG}^b\leq t|\cX)\geq 1-\alpha  \big\}.
	\end{equation}
	In Step 2, using the obtained upper bound, we get the lower bound of $\P\big(T_{\cG}\geq c_{T_{\cG}^b}(1-\alpha)\big)$ and prove
	\begin{equation}
		\P\big(T_{\cG}\geq c_{T_{\cG}^b}(1-\alpha)\big)\rightarrow1, ~\text{as}~n,p\rightarrow\infty.
	\end{equation}
	Note that $\{\Phi_{\cG,\alpha}=1\}\Leftrightarrow\{T_{\cG}\geq \hat{c}_{T_{\cG}^b}(1-\alpha) \}$, where 
	\begin{equation}
		\hat{c}_{T_{\cG}^b}(1-\alpha):=\inf\Big\{t:
		\dfrac{1}{B+1}\sum_{b=1}^{B}\mathbf{1}\{T_{\cG}^{b}\leq t|\cX\}\geq 1-\alpha\Big\}.
	\end{equation}
	Finally, using the fact that $\hat{c}_{T_{\cG}^b}(1-\alpha)$ is the estimation for $c_{T_{\cG}^b}(1-\alpha)$ based on the bootstrap samples, we complete the proof. Now, we consider the two steps in detail.  \\
	\textbf{Step~1}: In this step, we aim to obtain the upper bound for $c_{T_{\cG}^b}(1-\alpha)$. Define $\hat{\xi}_{i,j}^b=\hat{\bTheta}_j^\top\bX_i\epsilon_{i}^b$ for $1\leq i\leq n$ and $1\leq j\leq p$. Recall $C^b_j(\lfloor nt \rfloor)$ in (\ref{equation: bootstraped partial sum process for each coordinate }) and the decomposition in (\ref{equation: bootstraped leading and residual term of partial sum process}). By the definition of $T_{\cG}^b$ and using the fact that $\dfrac{\lfloor nt \rfloor}{n}\dfrac{\lfloor nt \rfloor^* }{n}\leq 1$ with $t\in[\tau_0,1-\tau_0]$, we have
	\begin{equation}\label{inequality: upper bound for the bootstrap based test statistics}
		\begin{array}{ll}
			T_{\cG}^b&\leq \max\limits_{t\in[\tau_0,1-\tau_0]}\max\limits_{j\in\cG}|C^{b,\rm I}_j(\lfloor nt \rfloor)|+\max\limits_{t\in[\tau_0,1-\tau_0]}\max\limits_{j\in\cG}|C^{b,\rm II}_j(\lfloor nt \rfloor)|\\\\
			&\leq W_{\cG}^b+\max\limits_{t\in[\tau_0,1-\tau_0]}\max\limits_{j\in\cG}|C^{b,\rm II}_j(\lfloor nt \rfloor)|,
		\end{array}
	\end{equation}
	where 
	\begin{equation}\label{equation: WGb}
		W_{\cG}^b:=\max\limits_{t\in[\tau_0,1-\tau_0]}\max\limits_{j\in\cG}\underbrace{\sqrt{n}\sqrt{\dfrac{\lfloor nt \rfloor}{n}\dfrac{\lfloor nt \rfloor^* }{n}}\dfrac{\Big|\dfrac{1}{\lfloor nt \rfloor}\sum\limits_{i=1}^{\lfloor nt \rfloor}\hat{\xi}_{i,j}^b-\dfrac{1}{\lfloor nt \rfloor^*}\sum\limits_{i=\lfloor nt \rfloor+1}^{n}\hat{\xi}_{i,j}^b\Big|}{\sqrt{\hat{\sigma}_{\epsilon}^2\hat{\omega}_{j,j }}}}_{D_{j}^b(\lfloor nt \rfloor)}.
	\end{equation}
	By (\ref{inequality: upper bound for the bootstrap based test statistics}), we have $c_{T_{\cG}^b}(1-\alpha)\leq c_{W_{\cG}^b}(1-\alpha)+\max\limits_{t\in[\tau_0,1-\tau_0]}\max\limits_{j\in\cG}|C^{b,\rm II}_j(\lfloor nt \rfloor)|$, where $c_{W_{\cG}^b}(1-\alpha)$ is the $1-\alpha$ quantile of $W_{\cG}^b$. Hence, to obtain the upper bound of $c_{T_{\cG}^b}(1-\alpha)$, it is sufficient to get the upper bound of $c_{W_{\cG}^b}(1-\alpha)$ and $\max\limits_{t\in[\tau_0,1-\tau_0]}\max\limits_{j\in\cG}|C^{b,\rm II}_j(\lfloor nt \rfloor)|$, respectively. 
	
	We first consider $c_{W_{\cG}^b}(1-\alpha)$. By the definition of $D_{j}^b(\lfloor nt \rfloor)$ in (\ref{equation: WGb}), conditional on $\cX$, some basic calculations show that 
	\begin{equation}\label{equation: norm distribution}
		D_{j}^b(\lfloor nt \rfloor)\sim N(0,\sigma_{j}^2(t)),~\text{with}~t\in[\tau_0,1-\tau_0] ~\text{and}~1\leq j\leq p,
	\end{equation}
	where 
	\begin{equation}
		\sigma_{j}^2(t):=\dfrac{\hat{\bTheta}_j^\top\Big(\dfrac{\lfloor nt \rfloor^*}{n}\sum\limits_{i=1}^{\lfloor nt \rfloor}\bX_i\bX_i^\top+\dfrac{\lfloor nt \rfloor}{n}\sum\limits_{i=\lfloor nt \rfloor+1}^{n}\bX_i\bX_i^\top\Big) \hat{\bTheta}_j  }{\hat{\bTheta}_j^\top\big(\dfrac{1}{n}\sum\limits_{i=1}^{n}\bX_i\bX_i^\top\big) \hat{\bTheta}_j }.
	\end{equation}
	Under Assumptions $\mathbf{(A.1)}$ - $\mathbf{(A.5)}$, we can prove that as $n,p\rightarrow\infty$
	\begin{equation}\label{equation: large deviation for bootstraped variance}
		\max_{t\in[\tau_0,1-\tau_0]}\max_{1\leq j\leq p}|\sigma_{j}^2(t)-1|=o_p(1).
	\end{equation}
	Let $q'=|\cG|(n-2\lfloor n\tau_0\rfloor+1)$. Combining (\ref{equation: norm distribution}) and (\ref{equation: large deviation for bootstraped variance}), and using Lemma \ref{lemma:maximum inequality}, for any $t>0$, we have
	\begin{equation}\label{inequality: maximum inequality}
		\E \Big(\max_{t\in[\tau_0,1-\tau_0]}\max_{j\in\cG} |D_{j}^b(\lfloor nt \rfloor)|\Big)\leq \dfrac{\log(2p')}{t}+\dfrac{tA_0^2}{2},~\text{with}~A^2_0:=\dfrac{3}{2}.
	\end{equation}
	Furthermore, taking $t=A_0^{-1}\sqrt{2\log(q')}$ in (\ref{inequality: maximum inequality}), we have
	\begin{equation}\label{inequality: maximum inequality2}
		\E \Big(\max_{t\in[\tau_0,1-\tau_0]}\max_{j\in\cG} |D_{j}^b(\lfloor nt \rfloor)|\Big)\leq A_0\sqrt{2\log(q')}\big(1+\dfrac{1}{2\log q'}\big).
	\end{equation}
	By Theorem 5.8 in \cite{boucheron2013concentration}, we have
	\begin{equation}\label{inequality: upper bound expectation}
		\P\Big( \max\limits_{\tau_0\leq t\leq 1-\tau_0\atop j\in \cG}|D_{j}^b(\lfloor nt \rfloor)|\geq \E\Big[ \max\limits_{\tau_0\leq t\leq 1-\tau_0\atop j\in \cG}|D_{j}^b(\lfloor nt \rfloor)|\Big]+z\Big|\mathcal{X}\Big)\leq \exp\Big(-\dfrac{z^2}{2A_0^2}\Big).
	\end{equation}
	Based on (\ref{inequality: maximum inequality2}), and taking $z=A_0\sqrt{2\log(\alpha^{-1})}$ in (\ref{inequality: upper bound expectation}), we have 
	\begin{equation}\label{inequality: upper bound for 1-alpha quantile1}
		c_{W_{\cG}^b}(1-\alpha)\leq A_0\sqrt{2\log(q')}\big(1+\dfrac{1}{2\log q'}\big)+A_0\sqrt{2\log(\alpha^{-1})}.
	\end{equation}
	After obtaining the upper bound of $c_{W_{\cG}^b}(1-\alpha)$ in (\ref{inequality: upper bound for 1-alpha quantile1}), we next consider the upper bound of  $\max\limits_{t\in[\tau_0,1-\tau_0]}\max\limits_{j\in\cG}|C^{b,\rm II}_j(\lfloor nt \rfloor)|$. To this end, we define 
	\begin{equation}\label{equation: upper bound and low bound of estimated variances}
		\cE'=\big\{\min_{1\leq j\leq p}\hat{\sigma}_{\epsilon}^2\hat{\omega}_{j,j}\geq c_{\epsilon}\kappa_1^{-1} /2, ~~\max_{1\leq j\leq p}\hat{\sigma}_{\epsilon}^2\hat{\omega}_{j,j}\leq 2C_{\epsilon}\kappa_2\big\}.
	\end{equation}
	By Theorem \ref{theorem: variance estimation under H1} and Assumptions $\mathbf{(A.2)}$ and $\mathbf{(A.3)}$, we have $\P(\cE')\rightarrow1$ as $n,p\rightarrow\infty$. Under $\cE'$, by the definition of $C^{b,\rm II}_j(\lfloor nt \rfloor)$ in (\ref{equation: bootstraped leading and residual term of partial sum process}), we have
	\begin{equation}\label{inequality: delta1+delta2}
		\begin{array}{ll}
			\max\limits_{t\in[\tau_0,1-\tau_0]}\max\limits_{j\in\cG}|C^{b,\rm II}_j(\lfloor nt \rfloor)|\\
			\leq \underbrace{C_1\max\limits_{t\in[\tau_0,1-\tau_0]}\sqrt{n}\dfrac{\lfloor nt \rfloor}{n}\dfrac{\lfloor nt \rfloor^* }{n}\|\bDelta^{b,(0,t)}\|_{\cG,\infty}}_{\bDelta_1}\\
			\quad\quad\quad+\underbrace{C_1\max\limits_{t\in[\tau_0,1-\tau_0]}\sqrt{n}\dfrac{\lfloor nt \rfloor}{n}\dfrac{\lfloor nt \rfloor^* }{n}\|\bDelta^{b,(t,1)}\|_{\cG,\infty}}_{\bDelta_2},
		\end{array}
	\end{equation}
	where $C_1:=\sqrt{C_{\epsilon}\kappa_1^{-1} /2}$, $\bDelta^{b,(0,t)}$ and $\bDelta^{b,(t,1)}$ are defined in (\ref{equation: bootstraped residual term}). Next, we consider $\bDelta_1$ and $\bDelta_2$, respectively. Without loss of generality, we assume $\hat{t}_{0,\cG}\in[\tau_0,t_0]$.
	
	{\bf{Control of $\bDelta_1$}.}	For $\bDelta_1$, by the definition of $\bDelta^{b,(0,t)}$ in (\ref{equation: bootstraped residual term}), we have
	\begin{equation}\label{inequality: delta1}
		\small
		\bDelta_1\leq C_1\Big(\underbrace{\max\limits_{t\in[\tau_0,\hat{t}_{0,\cG}]}\sqrt{n}\dfrac{\lfloor nt \rfloor}{n}\dfrac{\lfloor nt \rfloor^* }{n}\|\bDelta^{b,(0,t),\rm I}\|_{\cG,\infty}}_{\bDelta_{1,1}}\vee \underbrace{\max\limits_{t\in[\hat{t}_{0,\cG},1-\tau_0]}\sqrt{n}\dfrac{\lfloor nt \rfloor}{n}\dfrac{\lfloor nt \rfloor^* }{n}\|\bDelta^{b,(0,t),\rm II}\|_{\cG,\infty}}_{\bDelta_{1,2}}\Big).
	\end{equation}
	
	{\bf{Control of $\bDelta_{1,1}$}.}	For $\bDelta_{1,1}$, consider $\bDelta^{b,(0,t),\rm I}$ in (\ref{equation: bootstraped residual before cpt1}) with $t\in[\tau_0,\hat{t}_{0,\cG}]$. Conditional on $\cX$, using concentration inequalities and by Lemma \ref{lemma: esimation error bound}, we have
	\begin{equation}\label{inequality: upper bound of bootstrap residual1}
		\begin{array}{ll}
			\|\bDelta^{b,(0,t),\rm I}\|_{\cG,\infty}\\
			\leq C \sqrt{\dfrac{\log(pn)}{\lfloor nt \rfloor}}\big\|\hat{\bbeta}^{b,(0,t)}-\hat{\bbeta}^{(0,\hat{t}_{0,\cG})}\big\|_1\\
			\leq C s\big(\hat{\bbeta}^{(0,\hat{t}_{0,\cG})}\big) \dfrac{\log(pn)}{\lfloor nt \rfloor},\\
			\leq Cs^{(1)}\dfrac{\log(pn)}{\lfloor nt \rfloor}(\text{by}~\text{Lemma}~\text{\ref{lemma: esimation error bound}}).
		\end{array}
	\end{equation}
	Hence, by (\ref{inequality: upper bound of bootstrap residual1}), we have 
	\begin{equation}\label{inequality: final upper bound for bootstraped residual1}
		\begin{array}{ll}
			\bDelta_{1,1}&=\max\limits_{t\in[\tau_0,\hat{t}_{0,\cG}]}\sqrt{n}\dfrac{\lfloor nt \rfloor}{n}\dfrac{\lfloor nt \rfloor^* }{n}\|\bDelta^{b,(0,t),\rm I}\|_{\cG,\infty},\\
			&\leq Cs\dfrac{\log(pn)}{\sqrt{n}}=o\big(\sqrt{\log(|\cG|n)}\big),
		\end{array}
	\end{equation}
	where the last equation of (\ref{inequality: final upper bound for bootstraped residual1}) comes from the assumption that ${s\sqrt{\log(pn)/n}=o(1)}$ with $s:=s^{(1)}\vee s^{(2)}$ and $|\cG|=p^{\gamma}$ for $\gamma\in(0,1]$.
	
	{\bf{Control of $\bDelta_{1,2}$}.} For $\bDelta_{1,2}$, considering  $\bDelta^{b,(0,t),\rm II}$  in (\ref{equation: bootstraped residual before+after cpt2}) with $t\in[\hat{t}_{0,\cG},1-\tau_0]$, we have
	\begin{equation}\label{inequality: upper bound for delta12}
		\|\bDelta^{b,(0,t),\rm II}\|_{\cG,\infty}\leq \|\bDelta_1^{b,(0,t),\rm II}\|_{\cG,\infty}+\|\bDelta_2^{b,(0,t),\rm II}\|_{\cG,\infty},
	\end{equation}
	where 
	\begin{equation}\label{inequality: two parts of bootstrap residual term2 }
		\begin{array}{ll}
			\bDelta^{b,(0,t), \rm II}_1&=-\big(\hat{\bTheta}\hat{\bSigma}_{(0,t)}-\Ib\big)\big(\hat{\bbeta}^{b,(0,t)}-\hat{\bbeta}^{(0,\hat{t}_{0,\cG})} \big),\\
			\bDelta^{b,(0,t), \rm II}_2&=-\dfrac{\lfloor nt \rfloor -\lfloor n\hat{t}_{0,\cG}\rfloor}{\lfloor nt \rfloor }\big(\hat{\bTheta}\hat{\bSigma}_{(\hat{t}_{0,\cG},t)}-\Ib\big)\big(\hat{\bbeta}^{(0,\hat{t}_{0,\cG})}-\hat{\bbeta}^{(\hat{t}_{0,\cG},1)} \big).\\
		\end{array}
	\end{equation}
	Hence, by (\ref{inequality: two parts of bootstrap residual term2 }), we need to consider $\bDelta^{b,(0,t), \rm II}_1$ and $\bDelta^{b,(0,t), \rm II}_2$, respectively. 
	
	{\bf{Control of $\bDelta^{b,(0,t), \rm II}_1$}.}	For $\bDelta^{b,(0,t), \rm II}_1$, using Lemma \ref{lemma: esimation error bound} for the bootstrap based samples, we have
	\begin{equation}\label{inequality: upper bound for bootstraped21}
		\begin{array}{ll}
			\|\bDelta^{b,(0,t), \rm II}_1\|_{\cG,\infty}\\
			\leq C\sqrt{\dfrac{\log(pn)}{\lfloor nt \rfloor}}\big\|\hat{\bbeta}^{b,(0,t)}-\hat{\bbeta}^{(0,\hat{t}_{0,\cG})}\big\|_{1},\\
			\leq C\sqrt{\dfrac{\log(pn)}{\lfloor nt \rfloor}} \dfrac{\lfloor nt\rfloor-\lfloor n\hat{t}_{0,\cG}\rfloor}{\lfloor nt \rfloor}\|\hat{\bbeta}^{(0,\hat{t}_{0,\cG})}-\hat{\bbeta}^{(\hat{t}_{0,\cG},1)}\|_1,\\
			\leq C\sqrt{\dfrac{\log(pn)}{\lfloor nt \rfloor}}\Big(\|\hat{\bbeta}^{(0,\hat{t}_{0,\cG})}-\bbeta^{(1)}\|_1+\|\hat{\bbeta}^{(\hat{t}_{0,\cG},1)}-\bbeta^{(2)}\|_1+\|\bbeta^{(1)}-\bbeta^{(2)}\|_1\Big).
		\end{array}
	\end{equation}
	Note that we assume $\hat{t}_{0,\cG}\in[\tau_0,t_0]$. Using Lemma \ref{lemma: esimation error bound}, we have
	\begin{equation}\label{inequality: upper bound for bootstraped 21 three terms}
		\|\hat{\bbeta}^{(0,\hat{t}_{0,\cG})}-\bbeta^{(1)}\|_1\leq Cs\sqrt{\dfrac{\log(p)}{n\tau_0}},~~\|\hat{\bbeta}^{(\hat{t}_{0,\cG},1)}-\bbeta^{(2)}\|_1\leq C\dfrac{\lfloor nt_0\rfloor-\lfloor n\hat{t}_{0,\cG}\rfloor}{\lfloor n\hat{t}_{0,\cG} \rfloor^*}\|\bbeta^{(2)}-\bbeta^{(1)}\|_1.
	\end{equation}
	Combining (\ref{inequality: upper bound for bootstraped21}) and (\ref{inequality: upper bound for bootstraped 21 three terms}), and using the fact that $\|\bdelta\|_1\leq s\|\bdelta\|_{\infty}$, we have
	\begin{equation}\label{inequality: final upper bound for bootstraped 21}
		\|\bDelta^{b,(0,t), \rm II}_1\|_{\cG,\infty}\leq C_1s\dfrac{\log(pn)}{\lfloor n\tau_0 \rfloor}+C_2s\sqrt{\dfrac{\log(pn)}{\lfloor nt \rfloor}}\|\bdelta\|_{\infty}, ~\text{with}~t\in[\hat{t}_{0,\cG},1-\tau_0].
	\end{equation}
	
	{\bf{Control of $\bDelta^{b,(0,t), \rm II}_2$}.}	After bounding $\bDelta^{b,(0,t), \rm II}_1$ in (\ref{inequality: final upper bound for bootstraped 21}), we next consider $\bDelta^{b,(0,t), \rm II}_2$. Using concentration inequalities and the trianlge inequality and by Lemma \ref{lemma: esimation error bound}, we have
	\begin{equation}\label{inequality: final upper bound for bootstraped 22}
		\begin{array}{ll}
			\|\bDelta^{b,(0,t), \rm II}_2\|_{\cG,\infty}\\
			\leq C\sqrt{\dfrac{\log(pn)}{\lfloor nt \rfloor}}\big\|\hat{\bbeta}^{(0,\hat{t}_{0,\cG})}-\hat{\bbeta}^{(\hat{t}_{0,\cG},1)}\big\|_{1},\\
			\leq C\sqrt{\dfrac{\log(pn)}{\lfloor nt \rfloor}}\Big(\|\hat{\bbeta}^{(0,\hat{t}_{0,\cG})}-\bbeta^{(1)}\|_1+\|\hat{\bbeta}^{(\hat{t}_{0,\cG},1)}-\bbeta^{(2)}\|_1+\|\bbeta^{(1)}-\bbeta^{(2)}\|_1\Big),\\
			\leq C\sqrt{\dfrac{\log(pn)}{\lfloor nt \rfloor}}\Big(s\sqrt{\dfrac{\log(p)}{n\tau_0}}+\|\bbeta^{(1)}-\bbeta^{(2)}\|_1+\|\bbeta^{(1)}-\bbeta^{(2)}\|_1\Big)(\text{Lemma \ref{lemma: esimation error bound}}),\\
			\leq C_1s\dfrac{\log(pn)}{\lfloor n\tau_0 \rfloor}+C_2s\sqrt{\dfrac{\log(p)}{\lfloor nt \rfloor}}\|\bdelta\|_{\infty}.
		\end{array}
	\end{equation}
	Combining (\ref{inequality: upper bound for delta12}), (\ref{inequality: final upper bound for bootstraped 21}), and (\ref{inequality: final upper bound for bootstraped 22}), by Assumption $\mathbf{(A.4)}$, we have
	\begin{equation}\label{inequality: final upper bound for bootstraped residual2}
		\begin{array}{ll}
			\bDelta_{1,2}&=\max\limits_{t\in[\hat{t}_{0,\cG},1-\tau_0]}\sqrt{n}\dfrac{\lfloor nt \rfloor}{n}\dfrac{\lfloor nt \rfloor^* }{n}\|\bDelta^{b,(0,t),\rm II}\|_{\cG,\infty}\\
			&\leq C_1\sqrt{n}s\dfrac{\log(pn)}{{n\tau_0}}+C_2\sqrt{n}\big(s\sqrt{\dfrac{\log(p)}{n\tau_0}}\|\bdelta\|_{\infty}\big),\\
			&\leq o\big(\sqrt{\log(|\cG|n)}\big)+C_2\sqrt{n}\big(s\sqrt{\dfrac{\log(p)}{n\tau_0}}\|\bdelta\|_{\infty}\big).
		\end{array}
	\end{equation}
	Combining (\ref{inequality: delta1}), (\ref{inequality: final upper bound for bootstraped residual1}), and (\ref{inequality: final upper bound for bootstrap residual2}), we have
	\begin{equation}\label{inequality: final upper bound for delta1}
		\bDelta_1\leq o\big(\sqrt{\log(|\cG|n)}\big)+C_1\sqrt{n}\big(s\sqrt{\dfrac{\log(p)}{n\tau_0}}\|\bdelta\|_{\infty}\big).
	\end{equation}
	
	{\bf{Control of $\bDelta_2$}.}	Similarly, we can obtain the upper bound for $\bDelta_2$  as
	\begin{equation}\label{inequality: final upper bound for delta2}
		\bDelta_2\leq o\big(\sqrt{\log(|\cG|n)}\big)+C_2\sqrt{n}\big(s\sqrt{\dfrac{\log(p)}{n\tau_0}}\|\bdelta\|_{\infty}\big).
	\end{equation}
	Combining (\ref{inequality: delta1+delta2}), (\ref{inequality: final upper bound for delta1}), and (\ref{inequality: final upper bound for delta2}), we have
	\begin{equation}\label{inequality: final upper bound for bootstraped residual term}
		\max\limits_{t\in[\tau_0,1-\tau_0]}\max\limits_{j\in\cG}|C^{b,\rm II}_j(\lfloor nt \rfloor)|\leq o\big(\sqrt{\log(|\cG|n)}\big)+C_1\sqrt{n}\big(s\sqrt{\dfrac{\log(p)}{n\tau_0}}\|\bdelta\|_{\cG,\infty}\big).
	\end{equation}
	Finally, using (\ref{inequality: upper bound for the bootstrap based test statistics}), (\ref{inequality: upper bound for 1-alpha quantile1}), and (\ref{inequality: final upper bound for bootstraped residual term}), we obtain an upper bound of $c_{T_{\cG}^b}(1-\alpha)$ as
	\begin{equation}\label{inequality: final upper bound of 1-alpha quantile}
		\begin{array}{ll}
			c_{T_{\cG}^b}(1-\alpha)&\leq A_0\sqrt{2\log(q')}\big(1+\dfrac{1}{2\log q'}\big)+A_0\sqrt{2\log(\alpha^{-1})}+o\big(\sqrt{\log(|\cG|n)}\big)\\
			&\quad\quad+C_1\sqrt{n}\big(s\sqrt{\dfrac{\log(p)}{n\tau_0}}\|\bdelta\|_{\cG,\infty}\big).
		\end{array}
	\end{equation}
	\textbf{Step~2}:
	In this step, we aim to prove that $\P\big(T_{\cG}\geq c_{T_{\cG}^b}(1-\alpha)\big)\rightarrow1$ as $n,p\rightarrow\infty$. 
	Let 
	\begin{equation}\label{equation: upper bound for 1-alpha quantile}
		\begin{array}{ll}
			c^{\rm u}_{T_{\cG}^b}(1-\alpha)&=A_0\sqrt{2\log(q')}\big(1+\dfrac{1}{2\log q'}\big)+A_0\sqrt{2\log(\alpha^{-1})}\\
			&\quad\quad\quad+o\big(\sqrt{\log(|\cG|n)}\big)+ C_1\sqrt{n}\big(s\sqrt{\dfrac{\log(p)}{n\tau_0}}\|\bdelta\|_{\cG,\infty}\big).
		\end{array}
	\end{equation}
	Considering the upper bound obtained in (\ref{inequality: final upper bound of 1-alpha quantile}), it is sufficient to prove $H_1\rightarrow1$, where
	\begin{equation}\label{inequality: H1}
		H_1=\P\big(T_{\cG}\geq c^{\rm u}_{T_{\cG}^b}(1-\alpha)\big).
	\end{equation}
	By replacing $\hat{\sigma}_{\epsilon}\hat{\omega}_{j,j }$ by its true values, we define the oracle testing statistics as
	\begin{equation}\label{equation: oracle test statistics}
		\tilde{T}_\cG=\max_{t\in[\tau_0,1-\tau_0]}\max_{j\in \cG}\sqrt{n}\dfrac{\lfloor nt \rfloor}{n}\Big(1-\dfrac{\lfloor nt \rfloor }{n}\Big) \Big| \dfrac{\breve{\beta}^{(0,t)}_j-\breve{\beta}^{(t,1)}_j}{\sqrt{\sigma_{\epsilon}^2\omega_{j,j }}}\Big|.
	\end{equation} 
	Considering (\ref{inequality: H1}) and (\ref{equation: oracle test statistics}), it is sufficient to prove $H_2\rightarrow1$ as $n,p\rightarrow\infty$, where
	\begin{equation}\label{inequality: H2}
		H_2=\P\big(\tilde{T}_\cG\geq c^{\rm u}_{T_{\cG}^b}(1-\alpha)+|T_{\cG}-\tilde{T}_{\cG}|\big).
	\end{equation}
	Recall $\{Z_j(\lfloor nt \rfloor),\tau_0\leq t\leq 1-\tau_0,1\leq j\leq p\}$ defined in (\ref{equation: zj(nt)}). By definition, we have
	\begin{equation}
		\tilde{T}_{\cG}=\max\limits_{t\in[\tau_0,1-\tau_0]}\max\limits_{j\in \cG}\dfrac{|Z_j(\lfloor nt \rfloor)|}{\sqrt{\sigma_{\epsilon}^2\omega_{j,j }}}.
	\end{equation}
	Let $\bZ(\lfloor nt \rfloor)=\big(Z_1(\lfloor nt \rfloor),\ldots,Z_p(\lfloor nt \rfloor)\big)^\top$. Under $\Hb_1$, we have the following decomposition:
	\begin{equation}\label{equation: decomposition under H1-2}
		\bZ(\lfloor nt \rfloor)=\bdelta(t)+\sqrt{n}\dfrac{\lfloor nt \rfloor}{n}\dfrac{\lfloor nt \rfloor^* }{n}\Big(\dfrac{1}{\lfloor nt \rfloor}\sum\limits_{i=1}^{\lfloor nt \rfloor}\hat{\bxi}_{i}-\dfrac{1}{\lfloor nt \rfloor^*}\sum\limits_{i=\lfloor nt \rfloor+1}^{n}\hat{\bxi}_{i}+\bR^{(0,t)}-\bR^{(t,1)}\Big),
	\end{equation}
	where $\hat{\bxi}_i:=(\hat{\xi}_{i,1},\ldots,\hat{\xi}_{i,p})^\top$ with $\hat{\xi}_{i,j}=\hat{\bTheta}_j^\top\Xb_i\epsilon_i$, $\bdelta(t)=(\delta_1(t),\ldots,\delta_p(t))^\top$ is defined in  (\ref{equation: delta (nt)}), $\bR^{(0,t)}$ and $\bR^{(t,1)}$ are defined in (\ref{inequality: R(nt) and R(nt2)}).
	Using (\ref{equation: decomposition under H1-2}), under the event $\cE'$, we have
	\begin{equation}\label{inequality: R1-R3}
		\tilde{T}_{\cG}\geq \max\limits_{t\in[\tau_0,1-\tau_0]}\max\limits_{j\in \cG}\dfrac{\delta_j(t)}{\sqrt{\sigma_{\epsilon}^2\omega_{j,j }}}-(c_{\epsilon}\kappa_2^{-1} /2)^{-1/2}(\bR_1+\bR_2+\bR_3),
	\end{equation}
	with
	\begin{equation}
		\begin{array}{l}
			\bR_1=\max\limits_{t\in[\tau_0,1-\tau_0]}\sqrt{n}\dfrac{\lfloor nt \rfloor}{n}\dfrac{\lfloor nt \rfloor^* }{n}\Big\|\dfrac{1}{\lfloor nt \rfloor}\sum\limits_{i=1}^{\lfloor nt \rfloor}\hat{\bxi}_{i}-\dfrac{1}{\lfloor nt \rfloor^*}\sum\limits_{i=\lfloor nt \rfloor+1}^{n}\hat{\bxi}_{i}\Big\|_{\cG,\infty},\\
			\bR_2=\max\limits_{t\in[\tau_0,1-\tau_0]}\sqrt{n}\dfrac{\lfloor nt \rfloor}{n}\dfrac{\lfloor nt \rfloor^* }{n}\|\bR^{(0,t)}\|_{\cG,\infty},\\
			\bR_3=\max\limits_{t\in[\tau_0,1-\tau_0]}\sqrt{n}\dfrac{\lfloor nt \rfloor}{n}\dfrac{\lfloor nt \rfloor^* }{n}\|\bR^{(t,1)}\|_{\cG,\infty}.
		\end{array}
	\end{equation}
	By (\ref{inequality: H2}) and (\ref{inequality: R1-R3}), to prove $H_2\rightarrow1$, it is sufficient to prove $H_3\rightarrow1$, where
	\begin{equation}\label{inequality: H3}
		\begin{array}{ll}
			H_3&=\P\Big(\max\limits_{t\in[\tau_0,1-\tau_0]}\max\limits_{j\in \cG}\dfrac{\delta_j(t)}{\sqrt{\sigma_{\epsilon}^2\omega_{j,j }}}\geq (c_{\epsilon}\kappa_2^{-1} /2)^{-1/2}(\bR_1+\bR_2+\bR_3)\\
			&\quad\quad\quad\quad+c^{\rm u}_{T_{\cG}^b}(1-\alpha)+|T_{\cG}-\tilde{T}_{\cG}|\Big).
		\end{array}
	\end{equation}
	Next, we prove $H_3\rightarrow1$. To this end, we need to obtain the upper bound of $\bR_1$, $\bR_2$, and $\bR_3$, and $|T_{\cG}-\tilde{T}_{\cG}|$, respectively. 
	
	{\bf{Control of $\bR_1$}}. We first consider $\bR_1$. By Assumptions $\mathbf{(A.1)}$ -- $\mathbf{(A.5)}$, using basic concentration inequalities, we can prove that with probability at least $1-C_1(np)^{-C_2}$, 
	\begin{equation}\label{inequality: final upper bound for R1}
		\bR_1\leq C_2\sqrt{\log(|\cG|n)}.
	\end{equation}
	
	{\bf{Control of $\bR_2$}}.	We next bound $\bR_2$. Considering  $\bR^{(0,t)}$ in (\ref{inequality: R(nt) and R(nt2)}), we have
	\begin{equation}\label{inequality: R21+R22}
		\bR_2\leq \underbrace{\max\limits_{t\in[\tau_0,t_0]}\sqrt{n}\dfrac{\lfloor nt \rfloor}{n}\dfrac{\lfloor nt \rfloor^* }{n}\big\|\bR^{(0,t),\rm I}\big\|_{\cG,\infty}}_{\bR_{2,1}} \vee \underbrace{\max\limits_{t\in[t_0,1-\tau_0]}\sqrt{n}\dfrac{\lfloor nt \rfloor}{n}\dfrac{\lfloor nt \rfloor^* }{n}\big\|\bR^{(0,t),\rm II}\big\|_{\cG,\infty}}_{\bR_{2,2}},
	\end{equation}
	where $\bR^{(0,t),\rm I}$ and $\bR^{(0,t),\rm II}$ are defined in (\ref{equation: residual before cpt1}).
	Next, we bound $\bR_{2,1}$ and $\bR_{2,2}$, respectively. 
	
	{\bf{Control of $\bR_{2,1}$}}.	For $\bR_{2,1}$, using concentration inequalities and Lemma \ref{lemma: esimation error bound}, we have 
	\begin{equation}\label{inequality: final upper of R21}
		\begin{array}{ll}
			\bR_{2,1}&\leq C_1\max\limits_{t\in[\tau_0,t_0]}\sqrt{n}\dfrac{\lfloor nt \rfloor}{n}\dfrac{\lfloor nt \rfloor^* }{n} \sqrt{\dfrac{\log(pn)}{\lfloor nt \rfloor}}\big\|\hat{\bbeta}^{(0,t)}-\bbeta^{(1)}\big\|_{1},\\
			&\leq C_1s\dfrac{\log(pn)}{\sqrt{n}}=o\big(\log(|\cG|n)\big),
		\end{array}
	\end{equation}
	where the last equation of (\ref{inequality: final upper of R21}) comes from the assumption that {$s\sqrt{\log(pn)/n}=o(1)$} and $|\cG|=p^\gamma$ with $\gamma\in(0,1]$. 
	
	{\bf{Control of $\bR_{2,2}$}}.	For $\bR_{2,2}$, by the decomposition in (\ref{equation: residual before cpt1}), we have
	\begin{equation}\label{inequality: R221+R222}
		\bR_{2,2}\leq \bR_{2,2,1}+\bR_{2,2,2},
	\end{equation}
	where $\bR_{2,2,1}$ and $\bR_{2,2,2}$ are defined as
	\begin{equation}\label{inequality: R221}
		\begin{array}{ll}
			\bR_{2,2,1}&= \max\limits_{t\in[t_0,1-\tau_0]}\sqrt{n}\dfrac{\lfloor nt \rfloor}{n}\dfrac{\lfloor nt \rfloor^*}{n}\dfrac{\lfloor nt_0\rfloor-\lfloor nt \rfloor}{\lfloor nt \rfloor }\big\|\big(\hat{\bTheta}\hat{\bSigma}_{(t_0,t)}-\Ib\big)\big(\bbeta^{(1)}-\bbeta^{(2)} \big)\|_{\cG,\infty},\\
			&\leq C_1\sqrt{n}\sqrt{\dfrac{\log(pn)}{n\tau_0}}\|\bbeta^{(1)}-\bbeta^{(2)} \|_1,\\
			&\leq C_1\sqrt{n}s\sqrt{\dfrac{\log(pn)}{n\tau_0}}\big\|\bdelta\big\|_{\infty},
		\end{array}
	\end{equation}
	and
	\begin{equation}\label{inequality: R222}
		\begin{array}{ll}
			\bR_{2,2,2}&=\max\limits_{t\in[t_0,1-\tau_0]}\sqrt{n}\dfrac{\lfloor nt \rfloor}{n}\dfrac{\lfloor nt \rfloor^*}{n}\big\|\big(\hat{\bTheta}\hat{\bSigma}_{(0,t)}-\Ib\big)\big(\hat{\bbeta}^{(0,t)}-\bbeta^{(1) } \big)\big\|_{\cG,\infty},\\
			&\leq \max\limits_{t\in[t_0,1-\tau_0]}C_1\sqrt{n}\sqrt{\dfrac{\log(pn)}{n\tau_0}}\|\hat{\bbeta}^{(0,t)}-\bbeta^{(1) }\|_1,\\
			&\leq \max\limits_{t\in[t_0,1-\tau_0]}C_1\sqrt{n}\sqrt{\dfrac{\log(pn)}{n\tau_0}}\dfrac{\lfloor nt\rfloor-\lfloor nt_0\rfloor}{\lfloor nt \rfloor}\|\bbeta^{(2)}-\bbeta^{(1)}\|_1,~(\text{by Lemma \ref{lemma: esimation error bound}})\\
			&\leq C_1\sqrt{n}s\sqrt{\dfrac{\log(pn)}{n\tau_0}}\big\|\bdelta\big\|_{\infty}.
		\end{array}
	\end{equation}
	Combining (\ref{inequality: R21+R22}), (\ref{inequality: final upper of R21}), (\ref{inequality: R221+R222}),
	(\ref{inequality: R221}), (\ref{inequality: R222}), we have
	\begin{equation}\label{inequality: final upper bound for R2}
		\bR_2\leq o\big(\log(|\cG|n)\big)+C_1\sqrt{n}s\sqrt{\dfrac{\log(pn)}{n\tau_0}}\big\|\bdelta\big\|_{\infty}.
	\end{equation}
	
	{\bf{Control of $\bR_3$}}.	With a similar proof, we can obtain the upper bound of $\bR_3$ as 
	\begin{equation}\label{inequality: final upper bound for R3}
		\bR_3\leq o\big(\log(|\cG|n)\big)+C_1\sqrt{n}s\sqrt{\dfrac{\log(pn)}{n\tau_0}}\big\|\bdelta\big\|_{\infty}.
	\end{equation}
	
	{\bf{Control of $|T_{\cG}-\tilde{T}_{\cG}|$}}.	After bounding $\bR_1$, $\bR_2$, and $\bR_3$ in (\ref{inequality: final upper bound for R1}), (\ref{inequality: final upper bound for R2}), and (\ref{inequality: final upper bound for R3}), we next bound $|T_{\cG}-\tilde{T}_{\cG}|$. Using the fact that $|\max_i|a_i|-\max_i|b_i||\leq \max_i|a_i-b_i|$, we have
	\begin{equation}\label{inequality: difference between oracle and data-driven}
		\begin{array}{ll}
			|T_{\cG}-\tilde{T}_{\cG}|\\
			=\Big|\max\limits_{t\in[\tau_0,1-\tau_0]}\max\limits_{j\in \cG}\dfrac{|Z_j(\lfloor nt \rfloor)|}{\sqrt{\hat{\sigma}_{\epsilon}^2\hat{\omega}_{j,j }}}-\max\limits_{t\in[\tau_0,1-\tau_0]}\max\limits_{j\in \cG}\dfrac{|Z_j(\lfloor nt \rfloor)|}{\sqrt{\sigma_{\epsilon}^2\omega_{j,j }}}\Big|,\\
			\leq \max\limits_{t\in[\tau_0,1-\tau_0]}\max\limits_{j\in \cG}\Big|\dfrac{Z_j(\lfloor nt \rfloor)}{\sqrt{\hat{\sigma}_{\epsilon}^2\hat{\omega}_{j,j }}}-\dfrac{Z_j(\lfloor nt \rfloor)}{\sqrt{\sigma_{\epsilon}^2\omega_{j,j }}}\Big|,\\
			\leq \tilde{T}_{\cG}\max\limits_{j\in\cG}\Big|\dfrac{\sqrt{\sigma_{\epsilon}^2\omega_{j,j }}}{\sqrt{\hat{\sigma}_{\epsilon}^2\hat{\omega}_{j,j }}}-1\Big|.
		\end{array}
	\end{equation}
	Note that conditional on the event $\cE'$, using Theorem \ref{theorem: variance estimation under H1}, we have 
	\begin{equation}\label{inequality: large deviation for variance estimator under H1}
		\max\limits_{j\in\cG}\Big|\dfrac{\sqrt{\sigma_{\epsilon}^2\omega_{j,j }}}{\sqrt{\hat{\sigma}_{\epsilon}^2\hat{\omega}_{j,j }}}-1\Big|\leq C_1\underbrace{\max_{1\leq j\leq p}|\hat{\sigma}_{\epsilon}^2\hat{\omega}_{j,j }-\sigma_{\epsilon}^2\omega_{j,j }|}_{\epsilon_{n}'}=o_p(1).
	\end{equation}
	Considering (\ref{inequality: difference between oracle and data-driven}) and (\ref{inequality: large deviation for variance estimator under H1}), using the decomposition for $T_{\cG}$ in (\ref{equation: decomposition under H1-2}), we have
	\begin{equation}\label{inequality: final upper bound for the difference between oracle and data-driven}
		|T_{\cG}-\tilde{T}_{\cG}|\leq C_1\epsilon_{n}' \max\limits_{t\in[\tau_0,1-\tau_0]}\max\limits_{j\in \cG}\dfrac{\delta_j(t)}{\sqrt{\sigma_{\epsilon}^2\omega_{j,j }}}+C_2\epsilon_n'(\bR_1+\bR_2+\bR_3).
	\end{equation} 
	Let $\epsilon_n''=s\sqrt{\log(pn)/n\tau_0}\|\bdelta\|_{\infty}/\|\bdelta\|_{\cG,\infty}$. By the definition of $\bdelta(t)$ in (\ref{equation: delta (nt)}), we have
	\begin{equation}
		\begin{array}{ll}
			\max\limits_{t\in[\tau_0,1-\tau_0]}\max\limits_{j\in \cG}\dfrac{\delta_j(t)}{\sqrt{\sigma_{\epsilon}^2\omega_{j,j }}}&=\sqrt{n}\dfrac{\lfloor nt_0 \rfloor}{n}\dfrac{\lfloor nt_0 \rfloor^*}{n}\max\limits_{j\in\cG}\dfrac{|\beta_j^{(1)}-\beta_j^{(2)}|}{\sqrt{\sigma_{\epsilon}^2\omega_{j,j }}},\\
			&=\sqrt{n}\max\limits_{j\in\cG}\Big|\dfrac{t_0(1-t_0)(\beta_j^{(2)}-\beta_j^{(1)})}{(\sigma^2_{\epsilon}\omega_{j,j })^{1/2}}\Big|+O(\dfrac{1}{\sqrt{n}}),
		\end{array}
	\end{equation}
	where the last equation comes from the fact that $|\lfloor nt \rfloor/n-t|=O(1/n)$ as $n\rightarrow\infty$.
	
	Finally, for $H_3$ in (\ref{inequality: H3}), considering the upper bounds in (\ref{equation: upper bound for 1-alpha quantile}), (\ref{inequality: final upper bound for R1}), (\ref{inequality: final upper bound for R2}), (\ref{inequality: final upper bound for R3}), (\ref{inequality: final upper bound for the difference between oracle and data-driven}), we have
	\begin{equation}\label{inequality: final lower bound for H3}
		\begin{array}{ll}
			H_3&\geq \P\Big( \sqrt{n}\max\limits_{j\in\cG}\Big|\dfrac{t_0(1-t_0)(\beta_j^{(2)}-\beta_j^{(1)})}{(\sigma^2_{\epsilon}\omega_{j,j })^{1/2}}\Big|\geq C_1\sqrt{2\log(|\cG|n)}+C_2\sqrt{2\log(\alpha^{-1})}    \\
			&\quad\quad\quad+C_3(\epsilon_n'\vee\epsilon_n'')(\sqrt{n}\|\bdelta\|_{\cG,\infty})\Big),\\
			&\geq \P\Big(\sqrt{n}\max\limits_{j\in\cG}\big|D_j\big|\geq \dfrac{C_4}{(1-\epsilon_n'\vee\epsilon_n'')}\big(\sqrt{2\log(|\cG|n)}+\sqrt{2\log(\alpha^{-1})}\big) \Big).
		\end{array}
	\end{equation}
	Considering (\ref{inequality: final lower bound for H3}), by choosing a large enough constant in (\ref{inequality: theoretical signal strengh}), we have $H_3\rightarrow1$, which completes the proof of Theorem \ref{theorem: power results}.

\end{proof}

\section{Proofs of lemmas in Section S7}\label{section: proofs of lemmas in the main proof}
\subsection{Proof of Lemma \ref{lemma: negligible}}\label{section: proof of negligible}
\begin{proof}
	In this section, we aim to prove
	\begin{equation}\label{inequality: proof of lemma B.1}
		\P\big(\max_{\tau_0\leq t\leq 1-\tau_0} \big\|\bC(\floor{nt})-\tilde{\bC}^{\text{I}}(\floor{nt})\big\|_{\cG,\infty}\geq \epsilon \big)=o(1).
	\end{equation}
	Without loss of generality, we assume $\cG=\{1,\ldots,p\}$.	Using the triangle inequality, we have
	\begin{equation}\label{inequality: D1+D2}
		\P\big(\max_{\tau_0\leq t\leq 1-\tau_0} \big\|\bC(\floor{nt})-\tilde{\bC}^{\text{I}}(\floor{nt})\big\|_{\infty}\geq \epsilon \big)\leq D_1+D_2,
	\end{equation}
	where 
	\begin{equation}
		\begin{array}{l}
			D_1:=\P\big(\max\limits_{\tau_0\leq t\leq 1-\tau_0}\big\|\bC(\floor{nt})-\bC^{\text{I}}(\floor{nt})\big\|_{\infty}\geq \epsilon/2 \big),\\\\
			D_2:=\P\big(\max\limits_{\tau_0\leq t\leq 1-\tau_0}\big\|\bC^{\text{I}}(\floor{nt})-\tilde{\bC}^{\text{I}}(\floor{nt})\big\|_{\infty}\geq \epsilon/2 \big).
		\end{array}
	\end{equation}
	
	{\bf{Control of $D_1$}}. By (\ref{inequality: D1+D2}), to prove (\ref{inequality: proof of lemma B.1}), we need to bound $D_1$ and $D_2$, respectively. We first consider $D_1$. To this end, we define 
	\begin{equation}\label{equation: set for variance estimation}
		\cE=\big\{\min_{1\leq j\leq p}\hat{\sigma}_{\epsilon}^2\hat{\omega}_{j,j}>c_{\epsilon}\kappa_1^{-1} /2  \big\},
	\end{equation}
	where $\kappa_2$ and $c_{\epsilon}$ are defined in Assumptions $\mathbf{(A.2)}$ and $\mathbf{(A.3)}$. By introuducing $\cE$, we have
	\begin{equation}
		D_1\leq \P\big(\max\limits_{\tau_0\leq t\leq 1-\tau_0}\big\|\bC(\floor{nt})-\bC^{\text{I}}(\floor{nt})\big\|_{\infty}\geq \epsilon/2\cap\cE \big)+\P(\cE^c).
	\end{equation}
	By Theorem \ref{theorem: variance estimator under H0}, we have $\P(\cE^c)=o(1)$ as $n,p\rightarrow\infty$. Under the event $\cE$, we have
	\begin{equation}\label{inequality: basic inequality for D1}
		\begin{array}{l}
			\P\big(\max\limits_{\tau_0\leq t\leq 1-\tau_0}\big\|\bC(\floor{nt})-\bC^{\text{I}}(\floor{nt})\big\|_{\infty}\geq \epsilon/2\cap\cE \big)\\
			\leq \P\big(\max\limits_{t\in[\tau_0,1-\tau_0]}\max\limits_{1\leq j\leq p}\sqrt{n}\dfrac{\lfloor nt \rfloor}{n}\dfrac{\lfloor nt \rfloor^* }{n}\big(\hat{\sigma}_{\epsilon}^2\hat{\omega}_{j,j }\big)^{-1/2} |\Delta^{(0,t)}_j-\Delta^{(t,1)}_j|\geq \epsilon/2 \big),\\
			\leq \P\big(\max\limits_{t\in[\tau_0,1-\tau_0]}\|\bDelta^{(0,t)}-\bDelta^{(t,1)}\|_{\infty} \geq \frac{1}{2}\sqrt{c_{\epsilon}\kappa_2^{-1}/2}\epsilon n^{-1/2}  \big),\\
			\leq \P\big(\max\limits_{t\in[\tau_0,1-\tau_0]}\|\bDelta^{(0,t)}\|_{\infty} \geq C_1\epsilon n^{-1/2}\big)+\P\big(\max\limits_{t\in[\tau_0,1-\tau_0]}\|\bDelta^{(t,1)}\|_{\infty} \geq C_2\epsilon n^{-1/2}\big).
		\end{array}
	\end{equation}
	By the definitions of $\bDelta^{(0,t)}$ and $\bDelta^{(t,1)}$ in (\ref{equation: residual term before and after cpt}), we have 
	\begin{equation}
		\begin{array}{ll}
			\|\bDelta^{(0,t)}\|_{\infty}&\leq \big\|\hat{\bTheta}\hat{\bSigma}_{(0,t)}-\Ib\big\|_{\infty}\big\|\hat{\bbeta}^{(0,t)}-\bbeta^{(0)} \big\|_1.
		\end{array}
	\end{equation}
	To bound $\|\bDelta^{(0,t)}\|_{\infty}$, we need to consider $\big\|\hat{\bTheta}\hat{\bSigma}_{(0,t)}-\Ib\big\|_{\infty}$ and $\big\|\hat{\bbeta}^{(0,t)}-\bbeta^{(0)} \big\|_1$, respectively. For $\big\|\hat{\bTheta}\hat{\bSigma}_{(0,t)}-\Ib\big\|_{\infty}$,  by the triangle inequality, we have
	\begin{equation}\label{inequality: three parts for precision matrix estimation}
		\begin{array}{ll}
			\|\hat{\bTheta}\hat{\bSigma}_{(0,t)}-\Ib\big\|_{\infty}\\
			\leq \|\hat{\bTheta}\hat{\bSigma}^{n}-\Ib\big\|_{\infty}+\|(\hat{\bTheta}-\bTheta)(\hat{\bSigma}^{n}-\hat{\bSigma}_{(0,t)})\big\|_{\infty}+\|\bTheta(\hat{\bSigma}^{n}-\hat{\bSigma}_{(0,t)})\big\|_{\infty},\\
			\leq \|\hat{\bTheta}\hat{\bSigma}^{n}-\Ib\big\|_{\infty}+\|(\hat{\bTheta}-\bTheta)(\hat{\bSigma}^{n}-\hat{\bSigma}_{(0,t)})\big\|_{\infty}+\|\bTheta\hat{\bSigma}^{n}-\Ib\big\|_{\infty}+\|\bTheta\hat{\bSigma}_{(0,t)}-\Ib\big\|_{\infty},\\
			\leq \|\hat{\bTheta}\hat{\bSigma}^{n}-\Ib\big\|_{\infty}+\max\limits_{1\leq j\leq p}\|\hat{\bTheta}_j-\bTheta_j\|_1\|\hat{\bSigma}^{n}-\hat{\bSigma}_{(0,t)}\|_{\infty}+\|\bTheta\hat{\bSigma}^{n}-\Ib\big\|_{\infty}+\|\bTheta\hat{\bSigma}_{(0,t)}-\Ib\big\|_{\infty}.
		\end{array}
	\end{equation}
	To bound $\|\hat{\bTheta}\hat{\bSigma}_{(0,t)}-\Ib\big\|_{\infty}$, we consider the four parts on the  RHS of (\ref{inequality: three parts for precision matrix estimation}), respectively.\\
	For 
	$\|\hat{\bTheta}\hat{\bSigma}^{n}-\Ib\big\|_{\infty}$, by \cite{van2014asymptotically} and Assumption $\mathbf{(A.5)}$, we have 
	\begin{equation}\label{inequality: upper bound for precision matrix}
		\|\hat{\bTheta}\hat{\bSigma}^{n}-\Ib\big\|_{\infty}\leq O_{p}\Big(\max\limits_{1\leq j\leq p}\lambda_{(j)}\Big)=O_p\Big( \sqrt{\dfrac{\log(p)}{n}} \Big).
	\end{equation}
	For $\max\limits_{1\leq j\leq p}\|\hat{\bTheta}_j-\bTheta_j\|_1\|\hat{\bSigma}^{n}-\hat{\bSigma}_{(0,t)}\|_{\infty}$,  by Lemma \ref{lemma: upper bounds for precision matrix}, we have  
	\begin{equation}
		\max_{1\leq j\leq p}\|\hat{\bTheta}_j-\bTheta_j\|_1=O_p\Big(s_j\sqrt{\dfrac{\log(p)}{n}}\Big).
	\end{equation}
	Note that  we can write  $\|\hat{\bSigma}^{n}-\hat{\bSigma}_{(0,t)}\|_{\infty}$ into 
	\begin{equation}
		\|\hat{\bSigma}_{(0,t)}-\hat{\bSigma}^{n}\|_{\infty}=\max_{1\leq j,k\leq p}\Big|\dfrac{1}{\lfloor nt \rfloor}\sum_{i=1}^{\lfloor nt \rfloor}\big(X_{i,j}X_{i,k}-\E (X_{i,j}X_{i,k})\big)  - \dfrac{1}{n} \sum_{i=1}^{n}\big(X_{i,j}X_{i,k}-\E (X_{i,j}X_{i,k}) \big)  \Big|.
	\end{equation}
	Under Assumption $\mathbf{(A.1)}$, $X_{i,j}X_{i,k}-\E X_{i,j}X_{i,k}$ follows sub-exponential distributions for $1\leq j,k\leq p$ and $1\leq i\leq n$. By Bernstein's inequality, with probabilty tending to 1, we have
	\begin{equation}\label{inequality: large deviation for covariance partial sum process}
		\|\hat{\bSigma}_{(0,t)}-\hat{\bSigma}^{n}\|_{\infty}\leq C_3\sqrt{\dfrac{\log(pn)}{\lfloor nt \rfloor}}.
	\end{equation}
	For $\|\bTheta\hat{\bSigma}^{n}-\Ib\big\|_{\infty}+\|\bTheta\hat{\bSigma}_{(0,t)}-\Ib\big\|_{\infty} $, using concentration inequalities again, we have
	\begin{equation}\label{inequality: upper bound for the third part}
		\|\bTheta\hat{\bSigma}^{n}-\Ib\big\|_{\infty}+\|\bTheta\hat{\bSigma}_{(0,t)}-\Ib\big\|_{\infty}=O_p\Big(\sqrt{\dfrac{\log pn}{\lfloor nt \rfloor}}\Big).
	\end{equation}
	Combining the results in (\ref{inequality: upper bound for precision matrix}) - (\ref{inequality: upper bound for the third part}), we have 
	\begin{equation}\label{inequality: upper bound for precision matrix2}
		\max_{t\in[\tau_0,1-\tau_0]}\big\|\hat{\bTheta}\hat{\bSigma}_{(0,t)}-\Ib\big\|_{\infty}\leq O_p\Big(\sqrt{\dfrac{\log(pn)}{\lfloor n\tau_0 \rfloor}}\Big).
	\end{equation}
	Note that by Lemma \ref{lemma: esimation error bound}, under $\Hb_0$, $\sup\limits_{t\in[\tau_0,1-\tau_0]}\big\|\hat{\bbeta}^{(0,t)}-\bbeta^{(0)} \big\|_1\leq s^{(0)}\sqrt{\log(p)/\lfloor n\tau_0\rfloor}$ holds. Considering (\ref{inequality: upper bound for precision matrix2}), we have 
	\begin{equation}\label{inequality: upper bound for residual term 1}
		\max_{t\in[\tau_0,1-\tau_0]}\|\bDelta^{(0,t)}\|_{\infty}\leq O_p\Big(s^{(0)}\dfrac{\log(pn)}{\lfloor n\tau_0 \rfloor}\Big).
	\end{equation}
	With a similar proof technique , for $\max\limits_{t\in[\tau_0,1-\tau_0]}\|\bDelta^{(t,1)}\|_{\infty}$, we can obtain
	\begin{equation}\label{inequality: upper bound for residual term 2}
		\max_{t\in[\tau_0,1-\tau_0]}\|\bDelta^{(t,1)}\|_{\infty}\leq O_p\Big(s^{(0)}\dfrac{\log(pn)}{\lfloor n\tau_0 \rfloor}\Big).
	\end{equation}
	Note that 
	\begin{equation}\label{equation: value of epsilon}
		\epsilon=C\max(\max\limits_{1\leq j\leq p}s_j\dfrac{\log(pn)}{\sqrt{ n }},s\sqrt{n}\dfrac{\log(pn)}{{\lfloor n\tau_0 \rfloor}})
	\end{equation}
	holds  for some large enough constant $C>0$. Considering (\ref{inequality: basic inequality for D1}), (\ref{inequality: upper bound for residual term 1}), and (\ref{inequality: upper bound for residual term 2}), as $n,p\rightarrow \infty$, we have
	$D_1=o(1)$.  
	
	{\bf{Control of $D_2$}}.	After bounding $D_1$, we next consider $D_2$. By the definitions of $\bC^{\text{I}}(\floor{nt})$ and $\tilde{\bC}^{\text{I}}(\floor{nt})$ in (\ref{equation: leading and residual term of partial sum process}) and (\ref{equation: oracle leading term}), and 
	using the triangle inequality, we have 
	\begin{equation}\label{equation: upper bounds for D2}
		\max_{t\in[\tau_0,1-\tau_0]}\big\|\bC^{\text{I}}(\floor{nt})-\tilde{\bC}^{\text{I}}(\floor{nt})\big\|_{\infty}\leq I+II+III,
	\end{equation}
	where $I-III$ are defined as
	\begin{equation}\label{inquation: I+II+III}
		\begin{array}{l}
			I=\max\limits_{t\in[\tau_0,1-\tau_0]}\max\limits_{1\leq j\leq p}\Big|\dfrac{\sqrt{\sigma_{\epsilon}^2\omega_{j,j }}}{\sqrt{\hat{\sigma}_{\epsilon}^2\hat{\omega}_{j,j}}}-1\Big|\Big|\dfrac{\Big(\sum\limits_{i=1}^{\lfloor nt \rfloor }(\hat{\bTheta}_j^\top-\bTheta_j^\top)\Xb_i\epsilon_i-\dfrac{\lfloor nt \rfloor}{n}\sum\limits_{i=1}^{n }(\hat{\bTheta}_j^\top-\bTheta_j^\top)\Xb_i\epsilon_i\Big)}{\sqrt{n\sigma_{\epsilon}^2\omega_{j,j }}}\Big|,\\
			
			II=\max\limits_{t\in[\tau_0,1-\tau_0]}\max\limits_{1\leq j\leq p}\Big|\dfrac{\sqrt{\sigma_{\epsilon}^2\omega_{j,j }}}{\sqrt{\hat{\sigma}_{\epsilon}^2\hat{\omega}_{j,j}}}-1\Big|\Big|\dfrac{\Big(\sum\limits_{i=1}^{\lfloor nt \rfloor }\bTheta_j^\top\Xb_i\epsilon_i-\dfrac{\lfloor nt \rfloor}{n}\sum\limits_{i=1}^{n }\bTheta_j^\top\Xb_i\epsilon_i\Big)}{\sqrt{n\sigma_{\epsilon}^2\omega_{j,j }}}\Big|,\\
			III=\max\limits_{t\in[\tau_0,1-\tau_0]}\max\limits_{1\leq j\leq p}\Big|\dfrac{\Big(\sum\limits_{i=1}^{\lfloor nt \rfloor }(\hat{\bTheta}_j^\top-\bTheta_j^\top)\Xb_i\epsilon_i-\dfrac{\lfloor nt \rfloor}{n}\sum\limits_{i=1}^{n }(\hat{\bTheta}_j^\top-\bTheta_j^\top)\Xb_i\epsilon_i\Big)}{\sqrt{n\sigma_{\epsilon}^2\omega_{j,j }}}\Big|.\\
		\end{array}
	\end{equation}
	We next consider $I$, $II$, and $III$, respectively. For $I$, we have $I\leq I^{(1)}+I^{(2)}$, where 
	\begin{equation}
		\begin{array}{l}
			I^{(1)}=\max\limits_{t\in[\tau_0,1-\tau_0]}\max\limits_{1\leq j\leq p}\Big|\dfrac{\sqrt{\sigma_{\epsilon}^2\omega_{j,j }}}{\sqrt{\hat{\sigma}_{\epsilon}^2\hat{\omega}_{j,j}}}-1\Big|,\\
			I^{(2)}=\max\limits_{t\in[\tau_0,1-\tau_0]}\max\limits_{1\leq j\leq p}\Big|\dfrac{\Big(\sum\limits_{i=1}^{\lfloor nt \rfloor }(\hat{\bTheta}_j^\top-\bTheta_j^\top)\Xb_i\epsilon_i-\dfrac{\lfloor nt \rfloor}{n}\sum\limits_{i=1}^{n }(\hat{\bTheta}_j^\top-\bTheta_j^\top)\Xb_i\epsilon_i\Big)}{\sqrt{n\sigma_{\epsilon}^2\omega_{j,j }}}\Big|.
		\end{array}
	\end{equation}
	To bound $I^{(1)}$, define
	\begin{equation}
		\begin{array}{ll}
			\tilde{I}^{(1)}&=\max\limits_{t\in[\tau_0,1-\tau_0]}\max\limits_{1\leq j\leq p}\Big|\dfrac{\sqrt{\hat{\sigma}_{\epsilon}^2\hat{\omega}_{j,j}}}{\sqrt{\sigma_{\epsilon}^2\omega_{j,j }}}-1\Big|=\max\limits_{t\in[\tau_0,1-\tau_0]}\max\limits_{1\leq j\leq p}\Big|\dfrac{\sqrt{\hat{\sigma}_{\epsilon}^2\hat{\omega}_{j,j}}-\sqrt{\sigma_{\epsilon}^2\omega_{j,j }}}{\sqrt{\sigma_{\epsilon}^2\omega_{j,j }}}\Big|.
		\end{array}
	\end{equation}
	Using the fact that $a^2-b^2=(a-b)(a+b)$, we have
	\begin{equation}\label{inequality: upper bounds for inverse of variance estimation}
		\begin{array}{ll}
			\tilde{I}^{(1)}&=\max\limits_{t\in[\tau_0,1-\tau_0]}\max\limits_{1\leq j\leq p}\Big|\dfrac{\hat{\sigma}_{\epsilon}^2\hat{\omega}_{j,j}-\sigma_{\epsilon}^2\omega_{j,j }}{\sqrt{\sigma_{\epsilon}^2\omega_{j,j }}(\sqrt{\hat{\sigma}_{\epsilon}^2\hat{\omega}_{j,j}}+\sqrt{\sigma_{\epsilon}^2\omega_{j,j }})}\Big|,\\
			&\leq \max\limits_{t\in[\tau_0,1-\tau_0]}\max\limits_{1\leq j\leq p}\Big|\dfrac{\hat{\sigma}_{\epsilon}^2\hat{\omega}_{j,j}-\sigma_{\epsilon}^2\omega_{j,j }}{\sigma_{\epsilon}^2\omega_{j,j }}\Big|,\\
			& \leq C\max\limits_{t\in[\tau_0,1-\tau_0]}\max\limits_{1\leq j\leq p}\big|\hat{\sigma}_{\epsilon}^2\hat{\omega}_{j,j}-\sigma_{\epsilon}^2\omega_{j,j }\big|(\text{Assumptions}~ \mathbf{(A.2)}~\text{ and}~ \mathbf{(A.3)}),\\
			&\leq O_p\Big(\sqrt{\dfrac{\log(n)}{n}}+\max_j\lambda_{(j)}\sqrt{s_j}\Big),
		\end{array}
	\end{equation}
	where the last inequality comes from Theorem \ref{theorem: variance estimator under H0}. Using (\ref{inequality: upper bounds for inverse of variance estimation}), and by Lemma C.1 in \cite{Zhou2017An}, we have 
	\begin{equation}\label{inequality: upper bound for I11}
		I^{(1)}\leq O_p\Big(\sqrt{\dfrac{\log(n)}{n}}+ \max_j\lambda_{(j)}\sqrt{s_j}\Big).
	\end{equation}
	For $I^{(2)}$, note that for two vectors $\bx$ and $\by$, we have $\|\bx^\top\by\|_\infty\leq \|\bx\|_1\|\by\|_\infty$ . By Assumptions $\mathbf{(A.2)}$ and $\mathbf{(A.3)}$, there exists a universal positive constant $C$ such that 
	\begin{equation}\label{inequality: upper bound for I12}
		\begin{array}{ll}
			I^{(2)}&\leq C \max\limits_{1\leq j\leq p}\|\hat{\bTheta}_j^\top-\bTheta_j^\top\|_1\max\limits_{t\in[\tau_0,1-\tau_0]}\Big\|\dfrac{1}{\sqrt{n}}\Big(\sum\limits_{i=1}^{\lfloor nt \rfloor }\Xb_i\epsilon_i-\dfrac{\lfloor nt \rfloor}{n}\sum\limits_{i=1}^{n }\Xb_i\epsilon_i\Big)\Big\|_{\infty}.\\
		\end{array}
	\end{equation}
	By Lemma \ref{lemma: upper bounds for precision matrix}, we have $\max\limits_{1\leq j\leq p}\|\hat{\bTheta}_j^\top-\bTheta_j^\top\|_1\leq O_p\Big(\max\limits_{1\leq j\leq p}s_j\sqrt{\dfrac{\log(p)}{n}}\Big)=o_p(1)$. Note that Assumptions $\mathbf{(A.1)}$ and $\mathbf{(A.2)}$ imply that $\Xb_i\epsilon_i$ follows the sub-exponential distribution. Using Lemma \ref{lemma: exponential inequality for partial sum process}, we have 
	\begin{equation}
		\max\limits_{t\in[\tau_0,1-\tau_0]}\Big\|\dfrac{1}{\sqrt{n}}\Big(\sum\limits_{i=1}^{\lfloor nt \rfloor }\Xb_i\epsilon_i-\dfrac{\lfloor nt \rfloor}{n}\sum\limits_{i=1}^{n }\Xb_i\epsilon_i\Big)\Big\|_{\infty}\leq O_p\big(\sqrt{\log(pn)}\big).
	\end{equation}
	Combining (\ref{inequality: upper bound for I11}) and (\ref{inequality: upper bound for I12}), we have 
	\begin{equation}\label{inequality: final upper bound for I}
		I\leq  O_p\Big(\max_{1\leq j\leq p}s_j\dfrac{\log^{3/2}(pn)}{n} + \max\limits_{1\leq j\leq p}\dfrac{(s_j\log(pn))^{3/2}}{n}\Big).
	\end{equation}
	Similarly, for $II$ and $III$, we can obtain their upper bounds as follows:
	\begin{equation}\label{inequality: final upper bound for II+III}
		\begin{array}{ll}
			II \leq\times  O_p\Big(\dfrac{\log(pn)}{\sqrt{n}} +\max\limits_{1\leq j\leq p}\dfrac{\log(pn)\sqrt{s_j}}{\sqrt{n}}\Big),\\
			~~	III\leq O_p\Big( \max\limits_{1\leq j\leq p}s_j\dfrac{\log(pn)}{\sqrt{n}}     \Big).
		\end{array}
	\end{equation}
	Considering (\ref{equation: value of epsilon}), (\ref{equation: upper bounds for D2}), (\ref{inquation: I+II+III}), (\ref{inequality: final upper bound for I}), and (\ref{inequality: final upper bound for II+III}), as $n,p\rightarrow\infty$, we have $D_2\rightarrow0$.
	
	Finally, combining (\ref{inequality: D1+D2}), $D_1\rightarrow0$, and $D_2\rightarrow0$, we complete the proof of Lemma \ref{lemma: negligible}.
\end{proof}

\subsection{Proof of Lemma \ref{lemma: bootstraped negligible}}\label{section: proof of bootstraped negligible}
\begin{proof}
	In this section, we aim to prove 	
	\begin{equation}\label{equation: proof bootstraped negligible}
		\P\big(\max_{\tau_0\leq t\leq 1-\tau_0} \big\|\bC^b(\floor{nt})-\tilde{\bC}^{b,\text{I}}(\floor{nt})\big\|_{\cG,\infty}\geq \epsilon|\cX \big)=o(1).
	\end{equation}
	Without loss of generality, we assume $\cG=\{1,\ldots,p\}$.	Using the triangle inequality, we have
	\begin{equation}\label{inequality: E1+E2}
		\P\big(\max_{\tau_0\leq t\leq 1-\tau_0} \big\|\bC^b(\floor{nt})-\tilde{\bC}^{b,\text{I}}(\floor{nt})\big\|_{\infty}\geq \epsilon \big)\leq E_1+E_2,
	\end{equation}
	where 
	\begin{equation}
		\begin{array}{l}
			E_1:=\P\big(\max\limits_{\tau_0\leq t\leq 1-\tau_0}\big\|\bC^b(\floor{nt})-\bC^{b,\text{I}}(\floor{nt})\big\|_{\infty}\geq \epsilon/2|\cX \big),\\\\
			E_2:=\P\big(\max\limits_{\tau_0\leq t\leq 1-\tau_0}\big\|\bC^{b,\text{I}}(\floor{nt})-\tilde{\bC}^{b,\text{I}}(\floor{nt})\big\|_{\infty}\geq \epsilon/2 |\cX\big).
		\end{array}
	\end{equation}
	Hence, to prove (\ref{equation: proof bootstraped negligible}), we need to prove $E_1\rightarrow0$ and $E_2\rightarrow0$, respectively. 
	
	{\bf{Control of $E_2$}}.	We first consider $E_2$. Similar to the analysis in Section \ref{section: proof of negligible}, we can show that
	\begin{equation}\label{inequality:}
		\max\limits_{\tau_0\leq t\leq 1-\tau_0}\big\|\bC^{b,\text{I}}(\floor{nt})-\tilde{\bC}^{b,\text{I}}(\floor{nt})\big\|_{\infty}\leq O_p\Big( \max\limits_{1\leq j\leq p}s_j\dfrac{\log(pn)}{\sqrt{n}}     \Big).
	\end{equation}
	Note that 
	\begin{equation}\label{inequality: epsilon value}
		\epsilon:=Cs^{(0)}\max\limits_{1\leq j\leq p}s_j\dfrac{\log(pn)}{\sqrt{\lfloor n\tau_0 \rfloor}}.
	\end{equation}
	By choosing a large enough constant $C$ in $\epsilon$, we have $E_2\rightarrow0$ as $n,p\rightarrow\infty$.
	
	{\bf{Control of $E_1$}}.	Next, we consider $E_1$. Recall $\cE$ defined in (\ref{equation: set for variance estimation}). For $E_1$, we have 
	\begin{equation}
		E_1\leq \P\big(\max\limits_{\tau_0\leq t\leq 1-\tau_0}\big\|\bC^b(\floor{nt})-\bC^{b,\text{I}}(\floor{nt})\big\|_{\infty}\geq \epsilon/2\cap\cE \big)+\P(\cE^c).
	\end{equation}
	By Theorem \ref{theorem: variance estimator under H0}, we have $\P(\cE^c)=o(1)$ as $n,p\rightarrow\infty$. By the definitions of $\bC^b(\floor{nt})$ and $\bC^{b,\rm I}(\floor{nt})$ in (\ref{equation: bootstraped 4 vectorized process}), under the event $\cE$, we have
	\begin{equation}\label{inequality: basic inequality for E1}
		\begin{array}{l}
			\P\big(\max\limits_{\tau_0\leq t\leq 1-\tau_0}\big\|\bC^b(\floor{nt})-\bC^{\text{I}}(\floor{nt})\big\|_{\cG,\infty}\geq \epsilon/2\cap\cE|\cX \big)\\
			\leq \P\big(\max\limits_{t\in[\tau_0,1-\tau_0]}\max\limits_{1\leq j\leq p}\sqrt{n}\dfrac{\lfloor nt \rfloor}{n}\dfrac{\lfloor nt \rfloor^* }{n}\big(\hat{\sigma}_{\epsilon}^2\hat{\omega}_{j,j }\big)^{-1/2} |\Delta^{b,\lfloor nt \rfloor}_j-\Delta^{b,\lfloor nt \rfloor^*}_j|\geq \epsilon/2 |\cX\big)\\
			\leq \P\big(\max\limits_{t\in[\tau_0,1-\tau_0]}\|\bDelta^{b,(0,t)}-\bDelta^{b,(t,1)}\|_{\infty} \geq \frac{1}{2}\sqrt{c_{\epsilon}\kappa_2^{-1}/2}\epsilon n^{-1/2}  \big)\\
			\leq \P\big(\max\limits_{t\in[\tau_0,1-\tau_0]}\|\bDelta^{b,(0,t)}\|_{\infty} \geq C_1\epsilon n^{-1/2}\big)+\P\big(\max\limits_{t\in[\tau_0,1-\tau_0]}\|\bDelta^{b,(t,1)}\|_{\infty} \geq C_2\epsilon n^{-1/2}\big).
		\end{array}
	\end{equation}
	To bound (\ref{inequality: basic inequality for E1}), we need to obtain the upper bounds of $\max\limits_{t\in[\tau_0,1-\tau_0]}\|\bDelta^{b,(0,t)}\|_{\infty}$ and $\max\limits_{t\in[\tau_0,1-\tau_0]}\|\bDelta^{b,(t,1)}\|_{\infty}$. 
	
	{\bf{Control of $\max\limits_{t\in[\tau_0,1-\tau_0]}\|\bDelta^{b,(0,t)}\|_{\infty}$}}.	We first obtain the upper bound of $\max\limits_{t\in[\tau_0,1-\tau_0]}\|\bDelta^{b,(0,t)}\|_{\infty}$. To this end, we consider  two cases:\\
	$\mathbf{Case~1:}$ $t\in[\tau_0,\hat{t}_{0,\cG}]$. In this case, by the definition of $\bDelta^{b,(0,t)}$ in (\ref{equation: bootstraped residual term}), it reduces to 
	\begin{equation}
		\bDelta^{b,(0,t),\rm I}=-\big(\hat{\bTheta}\hat{\bSigma}_{(0,t)}-\Ib\big)\big(\hat{\bbeta}^{b,(0,t)}-\hat{\bbeta}^{(0,\hat{t}_{0,\cG})} \big).
	\end{equation}
	Using the fact that $\|\Ab\bx\|_{\infty}\leq \|\Ab\|_{\infty}\|\bx\|_1$, we have
	\begin{equation}
		\|\bDelta^{b,(0,t),\rm I}\|_{\infty}\leq \big\|\big(\hat{\bTheta}\hat{\bSigma}_{(0,t)}-\Ib\big)\big\|_{\infty}\big\|\big(\hat{\bbeta}^{b,(0,t)}-\hat{\bbeta}^{(0,\hat{t}_{0,\cG})} \big)\big\|_1.
	\end{equation}
	By (\ref{inequality: upper bound for precision matrix2}), we have 
	\begin{equation}\label{inequality: upper bound for precision matrix before cpt}
		\max_{t\in[\tau_0,\hat{t}_{0,\cG}]}\big\|\big(\hat{\bTheta}\hat{\bSigma}_{(0,t)}-\Ib\big)\big\|_{\infty}\leq O_p\Big(\sqrt{\dfrac{\log(pn)}{\lfloor n\tau_0 \rfloor}}\Big).
	\end{equation}
	Note that under $\Hb_0$, by Lemma \ref{lemma: esimation error bound}, the lasso estimator $\hat{\bbeta}^{(0,\hat{t}_{0,\cG})}$ has the following properties:
	\begin{equation}\label{inequality: upper bounds for lasso estimation under H0}
		\begin{array}{l}
			\|\hat{\bbeta}^{(0,\hat{t}_{0,\cG})}-\bbeta^{(0)}\|_{q}\leq O_p\Big((s^{(0)})^{\frac{1}{q}}\sqrt{\dfrac{\log(p)}{\lfloor n\hat{t}_{0,\cG}\rfloor }}\Big),~\text{for}~q=1,2,\\
			\|\hat{\bbeta}^{(\hat{t}_{0,\cG},1)}-\bbeta^{(0)}\|_{q}\leq O_p\Big((s^{(0)})^{\frac{1}{q}}\sqrt{\dfrac{\log(p)}{\lfloor n\hat{t}_{0,\cG}\rfloor^* }}\Big),~\text{for}~q=1,2,\\
			\quad\quad\text{and}~\hat{s}^{(1)},\hat{s}^{(2)}\leq O_p(s^{(0)}),
		\end{array}
	\end{equation}
	where $\hat{s}^{(1)}:=|\hat{\cS}^{(1)}|$ with $\hat{\cS}^{(1)}:=\{1\leq j\leq p: \hat{\beta}_j^{(0,\hat{t}_{0,\cG})}\neq 0\}$ and $\hat{s}^{(2)}=|\hat{\cS}^{(2)}|$ with $\hat{\cS}^{(2)}=\{1\leq j\leq p: \hat{\beta}_j^{(\hat{t}_{0,\cG},1)}\neq 0\}$. Given  $\cX$, using Lemma \ref{lemma: esimation error bound} again, for $q=1,2$, we have 
	\begin{equation}\label{inequality: upper bound for bootstraped lasso1}
		\begin{array}{ll}
			\|\hat{\bbeta}^{b,(0,t)}-\hat{\bbeta}^{(0,\hat{t}_{0,\cG})}\|_{q}&\leq O_p\Big(\big(\hat{s}^{(1)}\big)^{\frac{1}{q}}\sqrt{\dfrac{\log(p)}{\lfloor nt\rfloor }}\Big)\leq O_p\Big((s^{(0)})^{\frac{1}{q}}\sqrt{\dfrac{\log(p)}{\lfloor n\tau_0\rfloor }}\Big).\\
		\end{array}
	\end{equation}
	Combining (\ref{inequality: upper bound for precision matrix before cpt}) and (\ref{inequality: upper bound for bootstraped lasso1}), conditional on $\cX$, for the case of $t\in[\tau_0,\hat{t}_{0,\cG}]$, we have
	\begin{equation}\label{inequality: upper bound for bootstraped residual1}
		\max_{t\in[\tau_0,\hat{t}_{0,\cG}]}\|\bDelta^{b,(0,t),\rm I}\|_{\infty}\leq O_p\Big(s^{(0)}\dfrac{\log(pn)}{\lfloor n\tau_0 \rfloor}\Big).
	\end{equation}
	$\mathbf{Case~2:}$ $t\in[\hat{t}_{0,\cG},1-\tau_0]$. In this case, $\bDelta^{b,(0,t)}$  reduces to $\bDelta^{b,(0,t), \rm II}$ in (\ref{equation: bootstraped residual before+after cpt2}). By its definition, we can decompose $\bDelta^{b,(0,t), \rm II}$ into the following two terms:
	\begin{equation}\label{equation: decomposition for bootstraped residual after cpt1 }
		\bDelta^{b,(0,t), \rm II}=\bDelta^{b,(0,t), \rm II}_1+\bDelta^{b,(0,t), \rm II}_2,
	\end{equation}
	where 
	\begin{equation}
		\begin{array}{ll}
			\bDelta^{b,(0,t), \rm II}_1&=-\big(\hat{\bTheta}\hat{\bSigma}_{(0,t)}-\Ib\big)\big(\hat{\bbeta}^{b,(0,t)}-\hat{\bbeta}^{(0,\hat{t}_{0,\cG})} \big),\\
			\bDelta^{b,(0,t), \rm II}_2&=-\dfrac{\lfloor nt \rfloor -\lfloor n\hat{t}_{0,\cG}\rfloor}{\lfloor nt \rfloor }\big(\hat{\bTheta}\hat{\bSigma}_{(\hat{t}_{0,\cG},t)}-\Ib\big)\big(\hat{\bbeta}^{(0,\hat{t}_{0,\cG})}-\hat{\bbeta}^{(\hat{t}_{0,\cG},1)} \big),\\
			\hat{\bSigma}_{(\hat{t}_{0,\cG},t)}&:=\dfrac{(\Xb_{(\hat{t}_{0,\cG},t)})^\top\Xb_{(\hat{t}_{0,\cG},t)}}{\lfloor nt \rfloor-\lfloor n\hat{t}_{0,\cG} \rfloor+1}.
		\end{array}
	\end{equation}
	By (\ref{equation: decomposition for bootstraped residual after cpt1 }), for controlling $\bDelta^{b,(0,t), \rm II}$, we need to consider $\bDelta^{b,(0,t), \rm II}_1$ and $\bDelta^{b,(0,t), \rm II}_2$, respectively. 
	
	{\bf{Control of $\bDelta^{b,(0,t), \rm II}_1$}}.	We first consider $\bDelta^{b,(0,t), \rm II}_1$. Similar to the analysis of (\ref{inequality: three parts for precision matrix estimation}) - (\ref{inequality: upper bound for precision matrix2}), we can prove that 
	\begin{equation}\label{inequality: bootstrap residual 11}
		\max_{t\in[\hat{t}_{0,\cG},1-\tau_0]}\Big\|\big(\hat{\bTheta}\hat{\bSigma}_{(0,t)}-\Ib\big) \Big\|_{\infty}\leq O_p\Big(\sqrt{\dfrac{\log(pn)}{\lfloor n\hat{\tau}_{0,\cG} \rfloor}}\Big)\leq O_p\Big(\sqrt{\dfrac{\log(pn)}{\lfloor n\tau_0 \rfloor}}\Big) .
	\end{equation}
	Note that  $\hat{\bbeta}^{b,(0,t)}$ is constructed using data both before $\lfloor n\hat{\tau}_{0,\cG} \rfloor$ and after $\lfloor n\hat{\tau}_{0,\cG} \rfloor$. By Lemma \ref{lemma: esimation error bound},  conditional on $\cX$, we have 
	\begin{equation}\label{inequality: bootstrap redisual12}
		\begin{array}{ll}
			\|\hat{\bbeta}^{b,(0,t)}-\hat{\bbeta}^{(0,\hat{t}_{0,\cG})}\|_{1}\\
			\leq C_1\max\Big(\hat{s}^{(1)}\sqrt{\dfrac{\log p}{n}},\dfrac{\lfloor nt\rfloor-\lfloor n\hat{t}_{0,\cG}\rfloor}{\lfloor nt \rfloor}\|\hat{\bbeta}^{(0,\hat{t}_{0,\cG})}-\hat{\bbeta}^{(\hat{t}_{0,\cG},1)}\|_1\Big),\\
			\leq C_2 s^{(0)}\max\Big(\sqrt{\dfrac{\log(p)}{\lfloor n\hat{t}_{0,\cG}\rfloor }},\sqrt{\dfrac{\log(p)}{\lfloor n\hat{t}_{0,\cG}\rfloor^* }}\Big)\big(\text{by ~(\ref{inequality: upper bounds for lasso estimation under H0}})\big),\\
			\leq C_3 s^{(0)} \sqrt{\dfrac{\log(p)}{\lfloor n\tau_0\rfloor }}.
		\end{array}
	\end{equation}
	Combining (\ref{inequality: bootstrap residual 11}) and (\ref{inequality: bootstrap redisual12}), we have 
	\begin{equation}\label{inequality: final upper bound for bootstrap residual21}
		\max_{t\in[{\hat{t}_{0,\cG},1-\tau_0}] }\|\bDelta^{b,(0,t), \rm II}_1\|_{\infty}\leq O_p\Big(s^{(0)}\dfrac{\log(p)}{\lfloor n\tau_0\rfloor }   \Big).
	\end{equation}
	
	{\bf{Control of $\bDelta^{b,(0,t), \rm II}_2$}}.	After bounding $\max_{t\in[{\hat{t}_{0,\cG},1-\tau_0}] }\|\bDelta^{b,(0,t), \rm II}_1\|_{\infty}$, we next consider $\max_{t\in[{\hat{t}_{0,\cG},1-\tau_0}] }\|\bDelta^{b,(0,t), \rm II}_2\|_{\infty}$. Similar to the analysis of (\ref{inequality: three parts for precision matrix estimation}) - (\ref{inequality: upper bound for precision matrix2}), we have
	\begin{equation}\label{inequality: bootstrap redisual21}
		\max_{t\in[\hat{t}_{0,\cG},1-\tau_0]}\Big\|\dfrac{\lfloor nt \rfloor -\lfloor n\hat{t}_{0,\cG}\rfloor}{\lfloor nt \rfloor }\big(\hat{\bTheta}\hat{\bSigma}_{(\hat{t}_{0,\cG},t)}-\Ib\big)\Big\|_{\infty}\leq O_p\Big(\sqrt{\dfrac{\log(pn)}{\lfloor n\hat{t}_{0,\cG} \rfloor}}\Big).
	\end{equation}
	By  (\ref{inequality: upper bounds for lasso estimation under H0}), we have
	\begin{equation}\label{inequality: bootstrap redisual22}
		\begin{array}{ll}
			\|\hat{\bbeta}^{(0,\hat{t}_{0,\cG})}-\hat{\bbeta}^{(\hat{t}_{0,\cG},1)}\|_1\\
			\leq \|\hat{\bbeta}^{(0,\hat{t}_{0,\cG})}-\bbeta^{(0)}\|_1+\|\hat{\bbeta}^{(\hat{t}_{0,\cG},1)}-\bbeta^{(0)}\|_1,\\
			\leq C_1s^{(0)}\max\Big(\sqrt{\dfrac{\log(p)}{\lfloor n\hat{t}_{0,\cG}\rfloor }},\sqrt{\dfrac{\log(p)}{\lfloor n\hat{t}_{0,\cG}\rfloor^* }}  \Big),\\
			\leq C_2 s^{(0)} \sqrt{\log(p)/\lfloor n\tau_0\rfloor}.
		\end{array}
	\end{equation}
	Combining (\ref{inequality: bootstrap redisual21}) and (\ref{inequality: bootstrap redisual22}), we have
	\begin{equation}\label{inequality: final upper bound for bootstrap residual22}
		\max_{t\in[{\hat{t}_{0,\cG},1-\tau_0}] }\|\bDelta^{b,(0,t), \rm II}_2\|_{\infty}\leq O_p\Big(s^{(0)}\dfrac{\log(p)}{\lfloor n\tau_0\rfloor }   \Big).
	\end{equation}
	Considering (\ref{inequality: upper bound for bootstraped residual1}), (\ref{equation: decomposition for bootstraped residual after cpt1 }), (\ref{inequality: final upper bound for bootstrap residual21}), and (\ref{inequality: final upper bound for bootstrap residual22}), we obtain
	\begin{equation}\label{inequality: final upper bound for bootstrap residual1}
		\max_{t\in[\tau_0,1-\tau_0]}\|\bDelta^{b,(0,t)}\|_{\infty}\leq O_p\Big(s^{(0)}\dfrac{\log(pn)}{\lfloor n\tau_0 \rfloor}\Big).
	\end{equation}
	
	{\bf{Control of $\max_{t\in[\tau_0,1-\tau_0]}\|\bDelta^{b,(t,1)}\|_{\infty}$}}.	After bounding $\max_{t\in[\tau_0,1-\tau_0]}\|\bDelta^{b,(0,t)}\|_{\infty}$, we next consider $\max_{t\in[\tau_0,1-\tau_0]}\|\bDelta^{b,(t,1)}\|_{\infty}$. Using a similar proof technique, we can obtain 
	\begin{equation}\label{inequality: final upper bound for bootstrap residual2}
		\max_{t\in[\tau_0,1-\tau_0]}\|\bDelta^{b,(t,1)}\|_{\infty}\leq O_p\Big(s^{(0)}\dfrac{\log(pn)}{\lfloor n\tau_0 \rfloor}\Big).
	\end{equation}
	
	Finally, considering (\ref{inequality: epsilon value}), (\ref{inequality: basic inequality for E1}), (\ref{inequality: final upper bound for bootstrap residual1}), and (\ref{inequality: final upper bound for bootstrap residual2}), by choosing a large enough constant  $C$ in $\epsilon$, we have $E_1\rightarrow0$, which completes the proof of Lemma \ref{lemma: bootstraped negligible}.
\end{proof}

\subsection{Proof of Lemma \ref{lemma: maximum signal at the identified cpt} }\label{section: proof of maximum signal at the identified cpt}
\begin{proof}
	We give the proof by contradiction. Suppose there is a constant $c<1$ such that 
	\begin{equation*}
		\delta_{j^*}\leq c\|\bdelta\|_{\cG,\infty}.
	\end{equation*}
	On one hand, by the decomposition of $\bZ(\lfloor nt \rfloor) $ in (\ref{equation: decomposition under H1}), at time point $\hat{t}_0$, we have:
	\begin{equation*}
		\begin{array}{ll}	
			\|\bZ(\lfloor n\hat{t}_0 \rfloor)\|_{\cG,\infty}&:=Z_{j^*}(\lfloor n\hat{t}_0 \rfloor)\\
			&\leq \sqrt{n}\dfrac{\lfloor n{t}_0 \rfloor}{n}
			\dfrac{\lfloor n\hat{t}_0\rfloor^* }{n} \delta_{j^*}+C_2\sqrt{\log(|\cG|\lfloor n\tau_0\rfloor)}+o_p\big(\sqrt{n}\|\bdelta\|_{\cG,\infty}\big)\\
			&\leq \sqrt{n}\dfrac{\lfloor n{t}_0 \rfloor}{n}
			\dfrac{\lfloor n\hat{t}_0\rfloor^* }{n} c(1+o_p(1))\|\bdelta\|_{\cG,\infty}.
		\end{array}
	\end{equation*}	
	On the other hand, 	at time point ${t}_0$, we have:
	\begin{equation*}
		\begin{array}{ll}	
			\|\bZ(\lfloor n{t}_0 \rfloor)\|_{\cG,\infty}&=\max_{j\in \cG}|Z_{j}(\lfloor n{t}_0 \rfloor)|\\
			&\geq \sqrt{n}\dfrac{\lfloor n{t}_0 \rfloor}{n}
			\dfrac{\lfloor n{t}_0\rfloor^* }{n} \|\bdelta\|_{\cG,\infty}-C_2\sqrt{\log(|\cG|\lfloor n\tau_0\rfloor)}-o_p\big(\sqrt{n}\|\bdelta\|_{\cG,\infty}\big)\\
			&\geq \sqrt{n}\dfrac{\lfloor n{t}_0 \rfloor}{n}
			\dfrac{\lfloor n\hat{t}_0\rfloor^* }{n} (1-o_p(1))\|\bdelta\|_{\cG,\infty}.
		\end{array}
	\end{equation*}	
	Considering the above results, we have: $\P(\|\bZ(\lfloor n{t}_0 \rfloor)\|_{\cG,\infty}>\|\bZ(\lfloor n\hat{t}_0 \rfloor)\|_{\cG,\infty})\rightarrow 1$, which is contradicted to the fact that $\hat{t}_0$ is the maximizer of $\|\bZ(\lfloor n{t} \rfloor)\|_{\cG,\infty}$.
\end{proof}

\subsection{Proof of Lemma \ref{lemma: sign consistent}}\label{section: proof of sign consistent}
{\bf{Proof of $\cH_1$}}. Without loss of generality, we assume $\hat{t}_0\in[t_0,1-\tau_0]$. The proof proceeds in two steps. In Step 1, we prove that 
\begin{equation}
	\big|\max_{j\in\cG}Z_j(\lfloor n\hat{t}_0\rfloor)\big|\geq \big|\min_{j\in\cG}Z_j(\lfloor n\hat{t}_0\rfloor)\big|.
\end{equation}
By noting that 	$$\max\limits_{j\in\cG}|Z_j(\lfloor n\hat{t}_0\rfloor)|=
|\max\limits_{j\in\cG}Z_j(\lfloor n\hat{t}_0\rfloor)|\vee |\min_{j\in\cG}Z_j(\lfloor n\hat{t}_0\rfloor)|,$$
we have $\max\limits_{j\in\cG}|Z_j(\lfloor n\hat{t}_0\rfloor)|=
|\max\limits_{j\in\cG}Z_j(\lfloor n\hat{t}_0\rfloor)\big|$. In Step 2, we prove $\max\limits_{j\in\cG}Z_j(\lfloor n\hat{t}_0\rfloor)\geq 0$. Note that $Z_{j^*}(\lfloor n\hat{t}_0 \rfloor)=\max\limits_{j\in\cG}Z_j(\lfloor n\hat{t}_0 \rfloor)$. Combining Steps 1 and 2, we complete the proof. 

Now, we consider the two steps, respectively. By the decomposition of $\bZ(\lfloor nt \rfloor) $ in (\ref{equation: decomposition under H1}), at time point $\hat{t}_0$, we have 
\begin{equation}\label{equation: decomposotion under H1 at t0}
	\begin{array}{ll}
		\bZ(\lfloor n\hat{t}_0 \rfloor)-\bdelta(\hat{t}_0)\\
		=\sqrt{n}\dfrac{\lfloor n\hat{t}_0 \rfloor}{n}\dfrac{\lfloor n\hat{t}_0 \rfloor^* }{n}\big(\dfrac{1}{\lfloor n\hat{t}_0 \rfloor}\sum\limits_{i=1}^{\lfloor n\hat{t}_0 \rfloor}\hat{\bxi}_{i}-\dfrac{1}{\lfloor n\hat{t}_0 \rfloor^*}\sum\limits_{i=\lfloor n\hat{t}_0 \rfloor+1}^{n}\hat{\bxi}_{i}+\bR^{(0,\hat{t}_0),II}-\bR^{(\hat{t}_0,1),II}\big).
	\end{array}
\end{equation}
By Assumptions $\mathbf{(A.1)}$ -- $\mathbf{(A.3)}$, using concentration inequalities, we can prove that with probability at least $1-(np)^{-C_1}$,
\begin{equation}\label{inequality: large deviations for random noise}
	\Big\|\sqrt{n}\dfrac{\lfloor n\hat{t}_0 \rfloor}{n}\dfrac{\lfloor n\hat{t}_0 \rfloor^* }{n}\Big(\dfrac{1}{\lfloor n\hat{t}_0 \rfloor}\sum\limits_{i=1}^{\lfloor n\hat{t}_0 \rfloor}\hat{\bxi}_{i}-\dfrac{1}{\lfloor n\hat{t}_0 \rfloor^*}\sum\limits_{i=\lfloor n\hat{t}_0 \rfloor+1}^{n}\hat{\bxi}_{i}\Big)\Big\|_{\cG,\infty}\leq C_2\sqrt{\log(|\cG|\lfloor n\tau_0\rfloor)} .
\end{equation} 	
Next, we consider the control of $\|\bR^{(0,\hat{t}_0),II}\|_{\cG,\infty}$ and $\|\bR^{(\hat{t}_0,1),II}\|_{\cG,\infty}$. 

{\bf{Control of $\|\bR^{(0,\hat{t}_0),II}\|_{\cG,\infty}$}}. By the definition of $\bR^{(0,\hat{t}_0),II}$ 
in (\ref{equation: residual before cpt1}),  using the triangle inequality, we have 
\begin{equation}
	\big\|\bR^{(0,\hat{t}_0),\rm II}\big\|_{\cG,\infty}\leq \big\|\bR_1^{(0,\hat{t}_0),\rm II}\big\|_{\infty}+\big\|\bR_2^{(0,\hat{t}_0),\rm II}\big\|_{\infty},
\end{equation}
where $\bR_1^{(0,\hat{t}_0),\rm II}$ and $\bR_2^{(0,\hat{t}_0),\rm II}$ are defined as
\begin{equation}
	\begin{array}{l}
		\bR_1^{(0,\hat{t}_0),\rm II}:=-\dfrac{\lfloor n\hat{t}_0\rfloor-\lfloor nt_0 \rfloor }{\lfloor n\hat{t}_0 \rfloor }\big(\hat{\bTheta}\hat{\bSigma}_{(t_0,\hat{t}_0)}-\Ib\big)\big(\bbeta^{(1)}-\bbeta^{(2)} \big),\\
		\bR_2^{(0,\hat{t}_0),\rm II}:=-\big(\hat{\bTheta}\hat{\bSigma}_{(0,\hat{t}_0)}-\Ib\big)\big(\hat{\bbeta}^{(0,\hat{t}_0)}-\bbeta^{(1) } \big).
	\end{array}
\end{equation}
Using the fact that $\|\Ab\bx\|_{\infty}\leq \|\Ab\|_{\infty}\|\bx\|_1$ and by concentration inequalities, we have,
\begin{equation*}
	\begin{array}{ll}
		\|\bR_1^{(0,\hat{t}_0),\rm II}\|_{\infty}&\leq C_1\sqrt{\dfrac{\log(p)}{\floor{n\hat{t}_0}}}\|\bbeta^{(1)}-\bbeta^{(2)}\|_1\leq C_1 s \sqrt{\dfrac{\log(p)}{\floor{n\hat{t}_0}}}\|\bbeta^{(1)}-\bbeta^{(2)}\|_{\infty}=o(\|\bdelta\|_{\cG,\infty}),
	\end{array}
\end{equation*}	
where the last equation comes from the assumption that 	$s\sqrt{\log(pn)/n\tau_0}\|\bdelta\|_{\infty}/\|\bdelta\|_{\cG,\infty}=o(1)$. For $\|\bR_2^{(0,\hat{t}_0),\rm II}\|_{\infty}$, using Lemma \ref{lemma: esimation error bound} and concentration inequalities, we have 
\begin{equation}
	\begin{array}{ll}
		\|\bR_2^{(0,\hat{t}_0),\rm II}\|_{\infty}
		&\leq C_1\sqrt{\dfrac{\log(pn)}{\lfloor n\hat{t}_0\rfloor}}\Big(\dfrac{\lfloor n\hat{t}_0\rfloor-\lfloor nt_0\rfloor}{\lfloor n\hat{t}_0 \rfloor}\big\|\bbeta^{(2)}-\bbeta^{(1)}\big\|_1\Big),\\
		&\leq C_2\sqrt{\dfrac{\log(pn)}{n\tau_0}}s\big\|\bbeta^{(1)}-\bbeta^{(2)}\big\|_{\infty}:=o(\|\bdelta\|_{\cG,\infty}).\\
	\end{array}
\end{equation}

{\bf{Control of $\|\bR^{(\hat{t}_0,1),II}\|_{\cG,\infty}$}}. By the definition of  $\bR^{(\hat{t}_0,1),II}$ in (\ref{equation: residual before cpt1}), and using Lemma \ref{lemma: esimation error bound}, we have
\begin{equation}\label{inequalities: upper bounds for residual term at t0 }
	\begin{array}{l}
		\|\bR^{(\hat{t}_0,1)}\|_{\cG,\infty}\leq 	\|\bR^{(\hat{t}_0,1)}\|_{\infty}\leq O_p\Big(\sqrt{\dfrac{\log(pn)}{\lfloor n\hat{t}_0\rfloor^*}}\Big)\Big\|\hat{\bbeta}^{(\hat{t}_0,1)}-\bbeta^{(2)}\Big\|_1\leq O_p\big(s^{(2)}\dfrac{\log(pn)}{\lfloor n\hat{t}_0\rfloor^*}\big).
	\end{array}
\end{equation}
Note that we assume $s\sqrt{\log(pn)/n\tau_0}=o(1)$ with $s:=s^{(1)}\vee s^{(2)}$ and $|\cG|=p^\gamma$ with $\gamma\in(0,1]$. Considering the above results, we have
\begin{equation}\label{inequality: final upper bound for the residual term at t0}
		\sqrt{n}	\|\bR^{(0,\hat{t}_0)}\|_{\cG,\infty}\leq o_p\big(\sqrt{n}\|\bdelta\|_{\cG,\infty}\big), ~~\sqrt{n}	\|\bR^{(\hat{t}_0,1)}\|_{\cG,\infty}\leq o_p\big(\log(|\cG|n)\big).
\end{equation}
Combining (\ref{equation: decomposotion under H1 at t0}) -- (\ref{inequality: final upper bound for the residual term at t0}), with probability at least $1-(np)^{-C_1}$, we have
\begin{equation}\label{inequality: upper bounds for centered vectors}
	\|\bZ(\lfloor n\hat{t}_0 \rfloor)-\bdelta(\hat{t}_0)\|_{\cG,\infty}:=\max\limits_{j\in\cG}|Z_j(\lfloor n\hat{t}_0 \rfloor)-\delta_j(\hat{t}_0)|\leq K^*,
\end{equation}
where $K^*:=C_2\sqrt{\log(|\cG|n)}+o\big(\sqrt{n}\|\bdelta\|_{\cG,\infty}\big)$. Note that $$\delta_j(\hat{t}_0):=\sqrt{n}\dfrac{\lfloor n{t}_0 \rfloor}{n}
\dfrac{\lfloor n\hat{t}_0\rfloor^* }{n}(\beta_j^{(1)}-\beta_j^{(2)})\geq 0.$$
By (\ref{inequality: upper bounds for centered vectors}), using the fact $|\max_ia_i-\max_ib_i|\leq\max_{i}|a_i-b_i|$ for two sequences $\{a_i\}$ and $\{b_i\}$, we have
\begin{equation}\label{inequality: minimum and maximum upper bound}
	\begin{array}{l}
		\min\limits_{j\in\cG}Z_j(\lfloor n\hat{t}_0 \rfloor)\geq -K^*,\\
		\max\limits_{j\in\cG}Z_j(\lfloor n\hat{t}_0 \rfloor)\geq \max\limits_{j\in\cG}\delta_j(\lfloor n\hat{t}_0 \rfloor)-K^*\geq K^*,
	\end{array}
\end{equation}
where the last inequality in (\ref{inequality: minimum and maximum upper bound}) comes from the assumption $\|\bdelta\|_{\cG,\infty}\gg \sqrt{\log(pn)/n\tau_0}$.
By (\ref{inequality: minimum and maximum upper bound}), we have 	$|\max\limits_{j\in\cG}Z_j(\lfloor n\hat{t}_0\rfloor)\big|\geq \big|\min\limits_{j\in\cG}Z_j(\lfloor n\hat{t}_0\rfloor)|$ and $\max\limits_{j\in\cG}Z_j(\lfloor n\hat{t}_0\rfloor)\geq 0$, which finishes the proof of $\cH_1$ in  Lemma \ref{lemma: sign consistent}.

{\bf{Proof of $\cH_2$}}. Note that the proof of $\cH_2$ is similar and easier, to save space, we omit the details. 


\subsection{Proof of Lemma \ref{lemma: three terms for the residual terms of cpt}}\label{section: proof of three terms for the residual terms of cpt}
\begin{proof}
	We first bound $\bR^{(0,\hat{t}_0),\rm II}$. By its definition in (\ref{equation: residual before cpt1}), we have
	\begin{equation}
		\begin{array}{ll}
			\sqrt{n}	\dfrac{\lfloor n\hat{t}_0 \rfloor}{n}\dfrac{\lfloor n\hat{t}_0\rfloor^* }{n}\|\bR^{(0,\hat{t}_0),\rm II}\|_{\cG,\infty}\leq \underbrace{\sqrt{n}\dfrac{\lfloor n\hat{t}_0 \rfloor-\lfloor nt_0\rfloor}{n}\big\|\big(\hat{\bTheta}\hat{\bSigma}_{(t_0,\hat{t}_0)}-\Ib\big)\big(\bbeta^{(1)}-\bbeta^{(2)} \big)\big\|_{\infty}}_{I}\\
			\quad\quad+\underbrace{\sqrt{n}\dfrac{\lfloor n\hat{t}_0 \rfloor}{n}\dfrac{\lfloor n\hat{t}_0 \rfloor^* }{n}\big\|\big(\hat{\bTheta}\hat{\bSigma}_{(0,\hat{t}_0)}-\Ib\big)\big(\hat{\bbeta}^{(0,\hat{t}_0)}-\bbeta^{(1) } \big)\big\|_{\infty}}_{II}.
		\end{array}
	\end{equation}
	For $I$,  using concentration inequalities, we have
	\begin{equation}\label{inequality: upper bound for term11}
		\begin{array}{ll}
			I&\leq C_1\sqrt{n}\dfrac{\lfloor n\hat{t}_0 \rfloor-\lfloor nt_0\rfloor}{n}\sqrt{\dfrac{\log(pn)}{\lfloor n\hat{t}_0\rfloor-\lfloor nt_0 \rfloor}}\|\bdelta\|_{1},\\
			&\leq C_1\sqrt{{\log(|\cG|n)}\dfrac{\lfloor n\hat{t}_0 \rfloor-\lfloor nt_0\rfloor}{n}}s\|\bdelta\|_{\infty}.\\	
		\end{array}
	\end{equation}	
	For $II$, using concentration inequalities and by Lemma \ref{lemma: esimation error bound}, we have
	\begin{equation}\label{inequality: upper bound for term12}
		\begin{array}{ll}
			II&\leq C_2\sqrt{n}\sqrt{\dfrac{\log(pn)}{\lfloor n\hat{t}_0\rfloor}}\big\|\hat{\bbeta}^{(0,\hat{t}_0)}-\bbeta^{(1) }\big\|_{1}\\
			&\leq C_2\sqrt{n}\sqrt{\dfrac{\log(pn)}{\lfloor n\hat{t}_0\rfloor}}\dfrac{\lfloor n\hat{t}_0 \rfloor-\lfloor nt_0\rfloor}{n}s\big\|\bdelta\big\|_{\infty},\\
			&\leq C_2\sqrt{n}\dfrac{\lfloor n\hat{t}_0 \rfloor-\lfloor nt_0\rfloor}{n}{\sqrt{\dfrac{\log(pn)}{n\tau_0}}s}\|\bdelta\|_{\infty}\\
			&=o\Big(\sqrt{n}\dfrac{\lfloor nt_0 \rfloor}{n}\dfrac{\lfloor n\hat{t}_0\rfloor -\lfloor nt_0\rfloor}{n}\|\bdelta\|_{\cG,\infty}\Big).
		\end{array}
	\end{equation}
	After bounding $\bR^{(0,\hat{t}_0),\rm II}$ in (\ref{inequality: upper bound for term11}) and (\ref{inequality: upper bound for term12}), we next consider $\bR^{(\hat{t}_0,1),\rm II}$. Using concentration inequalities and the upper bound of estimation error of $\hat{\bbeta}^{(\hat{t}_0,1)}$ for $\bbeta^{(2)}$, we have
	\begin{equation}\label{inequality: upper bound for the second term}
		\sqrt{n}\dfrac{\lfloor n\hat{t}_0 \rfloor}{n}\dfrac{\lfloor n\hat{t}_0 \rfloor^*}{n}\|\bR^{(\hat{t}_0,1),\rm II}\|_{\cG,\infty}\leq \sqrt{n}\dfrac{\lfloor n\hat{t}_0 \rfloor}{n}\dfrac{\lfloor n\hat{t}_0 \rfloor^*}{n}\|\bR^{(\hat{t}_0,1),\rm II}\|_{\infty}\leq C\sqrt{n}s\dfrac{\log(|\cG|n)}{{n}},
	\end{equation}
	where the last inequality comes from the assumption that $|\cG|=p^{\gamma}$ with $\gamma\in(0,1]$.
	
	We next consider $\bR^{(0,t_0),\rm II}$ and $\bR^{(t_0,1),\rm II}$. Using concentration inequalities and the upper bounds of estimation errors of $\hat{\bbeta}^{(0,t_0)}$ and $\hat{\bbeta}^{(t_0,1)}$ (see Lemma \ref{lemma: esimation error bound}), we have
	\begin{equation}\label{inequality: upper bound for term4}
		\sqrt{n}\dfrac{\lfloor nt_0 \rfloor}{n}\dfrac{\lfloor nt_0 \rfloor^*}{n}\big(\|\bR^{(0,t_0),\rm II}\|_{\cG,\infty}\vee\|\bR^{(t_0,1),\rm II}\|_{\cG,\infty}\big)\leq C\sqrt{n}s\dfrac{\log(|\cG|n)}{{n}}.
	\end{equation}	
	Finally, combining (\ref{inequality: upper bound for term11}) -- (\ref{inequality: upper bound for term4}), we complete the proof.
\end{proof}

\section{Proofs of useful lemmas}\label{section: proof of useful lemmas}
\subsection{Proof of Lemma \ref{lemma: exponential inequality for partial sum process}}
\begin{proof}
	The result of Lemma	\ref{lemma: exponential inequality for partial sum process} can be obtained by using Bernstein's inequality for sub-exponential distributions. To save space, we omit the details  here.
\end{proof}
\subsection{Proof of Lemma \ref{lemma: basic inequality}}\label{section: proof of basic inequality}
\begin{proof}
	We only consider the proof of (\ref{inequality: basic inequality1}). The proof of (\ref{inequality: basic inequality2}) is similar.   Using some straightforward calculations,  we have
	\begin{equation}\label{inequality: lasso+basic1}
		\begin{array}{ll}
			&\dfrac{1}{2\lfloor nt \rfloor}\big\|\bY_{(0,t)}-\Xb_{(0,t)}\hat{\bbeta}^{(0,t)}\big\|_2^2-\dfrac{1}{2\lfloor nt \rfloor}\big\|\bY_{(0,t)}-\Xb_{(0,t)}\bbeta^{(0,t)}\big\|_2^2,\\
			&\quad=\dfrac{1}{2\lfloor nt \rfloor}\big\|\Xb_{(0,t)}(\bbeta^{(0,t)}-\hat{\bbeta}^{(0,t)})\big\|_2^2-\dfrac{1}{\floor{nt}}(\hat{\bbeta}^{(0,t)}-\bbeta^{(0,t)})^\top\Xb_{(0,t)}^\top(\bY_{(0,t)}-\Xb_{(0,t)}\bbeta^{(0,t)}).
		\end{array}
	\end{equation}
	By noting that  $\hat{\bbeta}^{(0,t)}$ is the minimizer of 
	(\ref{equation: lasso estimation before and after cpt}), we have
	\begin{equation}\label{inequality: property of lasso}
		\begin{array}{ll}
			\dfrac{1}{2\lfloor nt \rfloor}\big\|\bY_{(0,t)}-\Xb_{(0,t)}\hat{\bbeta}^{(0,t)}\big\|_2^2-\dfrac{1}{2\lfloor nt \rfloor}\big\|\bY_{(0,t)}-\Xb_{(0,t)}\bbeta^{(0,t)}\big\|_2^2	\leq \lambda_1(t)\|\bbeta^{(0,t)}\|_1-\lambda_1(t)\|\hat{\bbeta}^{(0,t)}\|_1.
		\end{array}
	\end{equation}
	Note that $|\bx^\top\by|\leq \|\bx\|_{\infty}\|\by\|_1$ for two vectors $\bx$ and $\by$. Combining (\ref{inequality: lasso+basic1}) and (\ref{inequality: property of lasso}), under the event $\cA^{(1)}(t)$, taking $\lambda_1(t)=2\lambda^{(1)}$ as defined in (\ref{equation: four basic sets}) , we have
	\begin{equation}\label{inequality: lasso basic2}
		\begin{array}{ll}
			\dfrac{1}{2\lfloor nt \rfloor}\big\|\Xb_{(0,t)}(\bbeta^{(0,t)}-\hat{\bbeta}^{(0,t)})\big\|_2^2\\
			\leq \lambda_1(t)\big\|\bbeta^{(0,t)}\big\|_1-\lambda_1(t)\big\|\hat{\bbeta}^{(0,t)}\big\|_1+\dfrac{\lambda_1(t)}{2}\big\|\bbeta^{(0,t)}-\hat{\bbeta}^{(0,t)} \big\|_1,\\
			\leq \lambda_1(t)\big\|\bbeta^{(0,t)}\big\|_1-\lambda_1(t)\big\|\hat{\bbeta}^{(0,t)}\big\|_1+\dfrac{\lambda_1(t)}{2}\big\|\bbeta^{(0,t)}-\hat{\bbeta}^{(0,t)} \big\|_1.
		\end{array}
	\end{equation}
	Note that 
	\begin{equation}\label{equation: lasso basic3}
		|\beta^{(0,t)}_j|-|\hat{\beta}^{(0,t)}_j|+|\beta^{(0,t)}_j-\hat{\beta}^{(0,t)}_j|=0,~~ \text{if}~~j\in J^c(\bbeta^{(0,t)}).
	\end{equation}
	Adding $2^{-1}\|\bbeta^{(0,t)}-\hat{\bbeta}^{(0,t)} \|_1$ on both sides of (\ref{inequality: lasso basic2}), and considering  (\ref{equation: lasso basic3}), we have
	\begin{equation}
		\begin{array}{ll}
			\dfrac{1}{2\lfloor nt \rfloor}\big\|\Xb_{(0,t)}(\bbeta^{(0,t)}-\hat{\bbeta}^{(0,t)})\big\|_2^2+\dfrac{\lambda_1(t)}{2}\big\|\bbeta^{(0,t)}-\hat{\bbeta}^{(0,t)} \big\|_1,\\
			\leq \lambda_1(t)\big\|\bbeta^{(1)}_{J(\bbeta^{(0,t)})}\big\|_1-\lambda_1(t)\big\|\hat{\bbeta}^{(0,t)}_{J(\bbeta^{(1)})}\big\|_1+\lambda_1(t)\big\|(\bbeta^{(0,t)}-\hat{\bbeta}^{(0,t)})_{J(\bbeta^{(0,t)})} \big\|_1,\\
			\leq 2\lambda_1(t)\big\|(\bbeta^{(0,t)}-\hat{\bbeta}^{(0,t)})_{J(\bbeta^{(0,t)})} \big\|_1,
		\end{array}
	\end{equation}
	which completes the proof of (\ref{inequality: basic inequality1}).
\end{proof}
\subsection{Proof of Lemma \ref{lemma: key estimation error bound for lasso}}\label{section: proof of key  estimation error bound}
\begin{proof}
	Note that by Lemma \ref{lemma: basic inequality}, and the URE conditions in Assumption (A.3),
	under the event $\{\cA(t)\cap \cB(t)\}$, using standard analysis of lasso estimation (see Pages 1728 -- 1729 in \cite{bickel2009simultaneous}), one can  prove that (\ref{inequality: estimation error bound}) holds. To save space, we omit the details.
\end{proof}

\subsection{Proof of Lemma \ref{lemma: large probabilty for intersection}}\label{section: proof of intersection}
\begin{proof}
	By definitions of $\cA(t)$ and $\cB(t)$, to prove (\ref{inequality: large probability for intersection}), we need to bound $\P(\cA^c(t))$ and $\P(\cB^c(t))$, respectively. We first consider $\P(\cA^c(t))$. Before that, we need some notations. We denote $\bbeta_i$ as the regression coefficients for the $i$-th observation. By definition, we have
	\begin{equation*}
		\bbeta_i=\bbeta^{(1)}\mathbf{1}\{i\leq \floor{nt_0}\}+\bbeta^{(2)}\mathbf{1}\{i> \floor{nt_0}\}.
	\end{equation*}
	Recall $\bbeta^{(0,t)}$ as ${\bbeta}^{(0,t)}=\argmin\limits_{\bbeta\in \RR^p}\E\big\|\bY_{(0,t)}-\Xb_{(0,t)}\bbeta\big\|_2^2$. By the first order condition, we have:
	\begin{equation*}
		\E\big[\Xb_{(0,t)}^\top(\bY_{(0,t)}-\Xb_{(0,t)}\bbeta^{(0,t)})\big]=  \E\big[\sum_{i=1}^{\floor{nt}}\bX_i(Y_i-\bX_i^\top\bbeta^{(0,t)})\big]=  \mathbf{0}.
	\end{equation*}
	Hence, by the first order condition, we have:
	\begin{equation*}
		\Big\|\dfrac{1}{\lfloor nt \rfloor}\big(\Xb_{(0,t)})^\top(\bY_{(0,t)}-\Xb_{(0,t)}\bbeta^{(0,t)}\big)\Big\|_{\infty}=\Big\|\dfrac{1}{\floor{nt}}\sum_{i=1}^{\floor{nt}}\big(\bX_i(Y_i-\bX_i^\top\bbeta^{(0,t)})-\E[\bX_i(Y_i-\bX_i^\top\bbeta^{(0,t)})]\big)\Big\|_{\infty}.
	\end{equation*}
	By noting that $Y_i=\bX_i^\top\bbeta_i+\epsilon_i$, $\E{\epsilon_i}=0$, and the independence between $\epsilon_{i}$ and $\bX_i$, we have:
	\begin{equation*}
		\begin{array}{ll}
			\dfrac{1}{\lfloor nt \rfloor}\big(\Xb_{(0,t)})^\top(\bY_{(0,t)}-\Xb_{(0,t)}\bbeta^{(0,t)})\\
			=\dfrac{1}{\floor{nt}}\sum\limits_{i=1}^{\floor{nt}}\bX_i\epsilon_{i}+\dfrac{1}{\floor{nt}}\sum\limits_{i=1}^{\floor{nt}}\big(\bX_i\bX_i^\top(\bbeta_i-\bbeta^{(0,t)})-\E[\bX_i\bX_i^\top(\bbeta_i-\bbeta^{(0,t)})]\big).
		\end{array}
	\end{equation*}
	Based on the above decomposition, we have:
	\begin{equation*}
		\Big\|\dfrac{1}{\lfloor nt \rfloor}\big(\Xb_{(0,t)})^\top(\bY_{(0,t)}-\Xb_{(0,t)}\bbeta^{(0,t)}\big)\Big\|_{\infty}\leq I+II,
	\end{equation*}
	where 
	\begin{equation*}
		\begin{array}{ll}
			I=\Big\|\dfrac{1}{\lfloor nt \rfloor}\bepsilon_{(0,t)}^\top\Xb_{(0,t)}\Big\|_{\infty},
			II=\Big\|\dfrac{1}{\floor{nt}}\sum\limits_{i=1}^{\floor{nt}}\big(\bX_i\bX_i^\top(\bbeta_i-\bbeta^{(0,t)})-\E[\bX_i\bX_i^\top(\bbeta_i-\bbeta^{(0,t)})]\big)\Big\|_{\infty}.
		\end{array}
	\end{equation*}
	
	{\bf{Control of $I$}}. We first consider $I$. By Assumptions $\mathbf{(A.1)}$ and $\mathbf{(A.2)}$, for $1\leq i\leq n$ and $1\leq j\leq p$, $\epsilon_iX_{i,j}$ follows the sub-exponential distribution. By Bernstein's inequality, for each $x>0$, we have
	\begin{equation}\label{inequality: A1}
		\begin{array}{ll}
			\P\big(\big\|\dfrac{1}{\lfloor nt \rfloor}\bepsilon_{(0,t)}^\top\Xb_{(0,t)}\big\|_{\infty}\geq x\big)\\
			\quad=\P\Big(\bigcup\limits_{1\leq j\leq p}\Big|\dfrac{1}{\lfloor nt \rfloor}\sum\limits_{i=1}^{\lfloor nt \rfloor}\epsilon_iX_{i,j}\Big|\geq x\Big),\\
			\quad\leq p\max\limits_{1\leq j\leq p}\P\Big(\Big|\dfrac{1}{\lfloor nt \rfloor}\sum\limits_{i=1}^{\lfloor nt \rfloor}\epsilon_iX_{i,j}\Big|\geq x \Big),\\
			\quad\leq C_1p\exp(-C_2\lfloor nt \rfloor x^2).\\
		\end{array}
	\end{equation}
	By (\ref{inequality: A1}), taking $\lambda^{(1)}=K_1\sqrt{\log(pn)/\lfloor nt \rfloor}$ for some big enough constant $K_1>0$, we have
	\begin{equation}\label{inequality: final A1}
		\P\big(\big\|\dfrac{1}{\lfloor nt \rfloor}\bepsilon_{(0,t)})^\top\Xb_{(0,t)}\big\|_{\infty}\geq \lambda^{(1)}\Big)\leq C_3(pn)^{-C_4},
	\end{equation} 
	where $C_3$ and $C_4$ are some big enough constants. 
	
	{\bf{Control of $II$}}. Next, we consider $II$. Note that for $t\in[\tau_0,t_0]$, $II=0$. Hence, in what follows, we consider the non-trivial case that $t\in[t_0,1-\tau_0]$. Let
	\begin{equation*}
		Z_i=\bX_i^\top(\bbeta_i-\bbeta^{(0,t)})/\|\bbeta_i-\bbeta^{(0,t)}\|_2,~~W_i=\|\bbeta_i-\bbeta^{(0,t)}\|_2.
	\end{equation*}
	By Assumption $\mathbf{(A.1)}$, $Z_i$ follows the sub-Gaussian distributions. Moreover, By Assumptions $\mathbf{(A.1)}$ and $\mathbf{(A.2)}$, for $1\leq i\leq n$ and $1\leq j\leq p$, $Z_iX_{i,j}$ follows the sub-exponential distribution. Hence, for $II$, we have
	\begin{equation*}
		\begin{array}{ll}
			
			\Big\|\dfrac{1}{\floor{nt}}\sum\limits_{i=1}^{\floor{nt}}\big(\bX_i\bX_i^\top(\bbeta_i-\bbeta^{(0,t)})-\E[\bX_i\bX_i^\top(\bbeta_i-\bbeta^{(0,t)})]\big)\Big\|_{\infty}\\
			\quad=\max\limits_{j}\Big|\dfrac{1}{\floor{nt}}\sum\limits_{i=1}^{\floor{nt}}W_i\big(X_{i,j}Z_i-\E[X_{i,j}Z_i]\big)\Big|.
		\end{array}
	\end{equation*}
	For each fixed $j$, using concentration inequality for weighted sub-exponential sums, we have:
	\begin{equation*}
		\begin{array}{ll}
			\P(\max_{j}|\dfrac{1}{\floor{nt}}\sum\limits_{i=1}^{\floor{nt}}W_i\big(X_{i,j}Z_i-\E[X_{i,j}Z_i]\big)|\geq x)\\
			\quad\leq p\max_{j}\P(|\dfrac{1}{\floor{nt}}\sum\limits_{i=1}^{\floor{nt}}W_i\big(X_{i,j}Z_i-\E[X_{i,j}Z_i]\big)|\geq x)\\
			\quad \leq 2p\exp(\dfrac{-C_1\floor{nt}^2x^2}{\|\bW\|^2}).
		\end{array}
	\end{equation*}
	Note that by definition, we have $\|\bW\|=\sqrt{\floor{nt_0}(1-\floor{nt_0}/\floor{nt})}\|\bbeta^{(2)}-\bbeta^{(1)}\|_2\leq \sqrt{\floor{nt_0}}C_{\bDelta}$. Hence, taking $\lambda^{(1)}=K_2\sqrt{\log(p)}\|\bW\|/\floor{nt}=K'_2\sqrt{\log(pn)/\lfloor nt \rfloor}$, we have:
	\begin{equation*}
		\P\Big\{\Big\|\dfrac{1}{\floor{nt}}\sum\limits_{i=1}^{\floor{nt}}\big(\bX_i\bX_i^\top(\bbeta_i-\bbeta^{(0,t)})-\E[\bX_i\bX_i^\top(\bbeta_i-\bbeta^{(0,t)})]\big)\Big\|_{\infty}\geq \lambda^{(1)}\Big\}\leq C_3(p)^{-C_4}.
	\end{equation*}
	for some big enough $C_3,C_4>0$. Combining the above two cases, taking $\lambda^{(1)}=K_1\sqrt{\log(pn)/\lfloor nt \rfloor}$ for some big enough constant $K_1>0$, we have:
	\begin{equation*}
		\P\Big\{\big\|\dfrac{1}{\lfloor nt \rfloor}\big(\Xb_{(0,t)})^\top(\bY_{(0,t)}-\Xb_{(0,t)}\bbeta^{(0,t)}\big\|_{\infty}\geq \lambda^{(1)}\Big\}\leq C_3(pn)^{-C_4}.
	\end{equation*}
	With similar arguments as above, we can also prove that 
	\begin{equation}\label{inequality: final B1+B2}
		\P\Big(\cB^c(t)\Big)\leq C_3(np)^{-C_4}.
	\end{equation}
	Finally, combining (\ref{inequality: final A1}) and (\ref{inequality: final B1+B2}),
	and noting that 
	\begin{equation}
		\begin{array}{ll}
			\P\big(\bigcap\limits_{t\in[\tau_0,1-\tau_0]}\big\{\cA(t)\cap\cB(t)\big\} \\
			\quad=1-  \P\big(\bigcup\limits_{t\in[\tau_0,1-\tau_0]}\big\{\cA^c(t)\cup\cB^c(t)\big\},\\
			\quad \geq 1-\sum\limits_{t\in[\tau_0,1-\tau_0]}\big(\P(\cA^c(t))+\P(\cB^c(t))\big),\\
			\quad \geq 1-C_1(np)^{-C_2},
		\end{array}
	\end{equation}
	we complete the proof of Lemma \ref{lemma: large probabilty for intersection}.
\end{proof}

\end{document}